\documentclass[english]{article}
\usepackage[T1]{fontenc}
\usepackage{geometry}
\usepackage{amsmath}
\usepackage{amssymb}
\usepackage{graphicx}
\usepackage{esint}

\usepackage[numbers,sort&compress]{natbib}
\usepackage{hyperref}
\usepackage{mciteplus}

\newcommand{\lyxdot}{.}

\usepackage{color}
\usepackage[makeroom]{cancel}
\usepackage[normalem]{ulem}
\definecolor{pkcolor}{rgb}{0,0.1,0.7}
\definecolor{ascolor}{rgb}{1,0,1}
\definecolor{mscolor}{rgb}{1,0,0}

\newcommand\pkout{\marginpar{\color{pkcolor}$\clubsuit$}\bgroup\markoverwith{\color{pkcolor}{\rule[0.4ex]{2pt}{0.8pt}}}\ULon}
\newcommand\asout{\marginpar{\color{ascolor}$\heartsuit$}\bgroup\markoverwith{\color{ascolor}{\rule[0.4ex]{2pt}{0.8pt}}}\ULon}
\newcommand\msout{\marginpar{\color{mscolor}$\diamondsuit$}\bgroup\markoverwith{\color{mscolor}{\rule[0.4ex]{2pt}{0.8pt}}}\ULon}

\begin{document}

\title{Exploring minijets beyond the leading power \date{}}

\author{P. Kotko, A.M. Stasto, M. Strikman \vspace{5pt}\\{\it \small  Department of Physics,   The Pennsylvania State University} \\{\it \small   University Park, PA 16802, United States}}
\maketitle
\begin{abstract}

 The crucial parameter of the current Monte Carlo models of high energy hadron-hadron interaction is the transverse momentum cutoff $p_{T0}$ for parton-parton interactions which slowly grows  with  energy and regularizes the cross section. This modification of the collinear factorization formula goes beyond the leading power and thus a natural question arises if such cutoff can be extracted from a formalism which takes into account power corrections. In this work, we consider the High Energy Factorization (HEF) valid at small $x$ and a new model, based on a similar principle to HEF, which in addition has a limit respecting the Dokshitzer-Dyakonov-Troyan formula for the dijet momentum disbalance spectrum. Minijet cross section and its suppression is then analyzed in two ways. First, we study minijets directly in the low-$p_T$ region, and demonstrate that higher twist corrections do generate suppression of the inclusive jet production cross section though these effects are not leading to the increase of the cutoff with incident energy.   Second, we consider hard inclusive dijet production where Multi Parton Interactions (MPIs)  with minijets produce power corrections. We introduce an observable constructed from differential cross section in the ratio $\tau$ of dijet disbalance to the average dijet $p_T$  and demonstrate that the $\tau>1$ region  is  sensitive to the cutoff $p_{T0}$ in the MPI minijet models. The energy dependence of the cutoff is reflected in the energy dependence of the bimodality coefficient $b$ of the $\tau$ distribution. We compare $b$ calculated from $\mathsf{pythia}$, where one can conveniently control MPIs by the program parameters, and HEF for a few unintegrated gluon distributions (UGDs). 
We find that the energy dependence of $b$ is very sensitive to the particular choice of UGD and in some models it  resembles predictions of the Monte Carlo models.
 \end{abstract}

%%%%%%%%%%%%%%%%%%%%%%%%%%%%%%%%%%%%%%%%%%%%
\section{Introduction}

The rise of the total cross section in hadron-hadron collisions with
energy is driven by \textit{minijets}, i.e. jets with relatively low
transverse momenta $p_{T}$, of the order of a few GeV. From the QCD
point of view, this growth is attributed to the rise of the parton
density inside a hadron with decreasing value of longitudinal momentum $x$ (or increasing CM energy of the collision). At leading order (LO) the colliding partons
(mostly gluons at high energies) produce two final state partons and
give rise to two jets. The problem is, however, that the resulting
QCD expression is divergent when $p_{T}\rightarrow0$. This is of
course not a paradox, simply the very low $p_{T}$ region is out of
the applicability of the formalism operating on partons (i.e. collinear
factorization theorem \citep{Collins:2011zzd}, see Section~\ref{sec:MinijetsColl}).
Thus one has to introduce a cutoff, $p_{T\,\mathrm{min}}$, above
which the formula makes sense \citep{Sarcevic1989}. This is the starting
point for so-called minijet models and  models including  
Multi Parton Interactions (MPIs)
which are at the heart of modern event generators like $\mathsf{pythia}$
\citep{Sjostrand:2006za} or $\mathsf{herwig++}$ \citep{Bahr2008}.
The basic idea is that since the minijet cross section $\sigma_{\mathrm{minijet}}\left(p_{T\,\mathrm{min}}\right)$
can easily exceed the total cross section $\sigma_{\mathrm{tot}}$
(for low values of $p_{T\,\mathrm{min}}$), the ratio $\sigma_{\mathrm{minijet}}/\sigma_{\mathrm{ND}}$,
with $\sigma_{\mathrm{ND}}$ being a non-diffractive inelastic component
of the total cross section, gives an average number of hard binary
collisions per event, i.e. MPI events \citep{Soediono1989b}.
The $p_{T\,\mathrm{min}}$ is a free parameter of the model. Typically, one does not implement the sharp cutoff but rather a smooth
transition regulated by another parameter $p_{T0}$.
 Comparison of the models with MPI with the  data indicates that hadron production at small impact parameters grows in these models too fast with increase of $\sqrt{S}$. Also the cross section of the interaction at large impact parameters grows faster than indicated by the data on profile function of the $pp$ interaction leading to cross section much larger than the experimental one
 \citep{Rogers2008,Rogers2010}.
 The typical resolution is
to let the $p_{T0}$ parameter to be energy dependent $p_{T0}=p_{T0}\left(S\right)$,
slowly growing with $S$. 

We see that there are two general features
of the minijet models: (i) an existence of a scale $p_{T0}$ above
which perturbative collinear factorization applies and (ii) the MPI-type
events. Let us note, that in a typical minijet model these features
are related in the sense that the MPI models require the property
(i), which in turn, on itself, can be viewed as a consequence of color
confinement \citep{Soediono1989b} and is independent on MPIs.
However at the LHC energies one needs a cutoff on the scale of $3\, \mathrm{GeV}$ and growing with $S$, making it unlikely that  the cutoff could be  solely non-perturbative effect. On
the other hand, the MPIs became a separate branch of high energy physics,
not necessarily related to minijets. For example one of the typical
direct MPI signals is expected to be a four-jet hard event with back-to-back
dijets \citep{Blok2011}. 
On the theory side the MPI physics is a very complicated subject and
most often is restricted to the double parton scattering (DPS), see \citep{Diehl2011} for a comprehensive review. 
So far no proof exists of the QCD factorization theorem for DPS, although
recently  a progress has been made towards the proof of DPS in
the double Drell-Yan process \citep{Diehl:2015bca}.

In this work we have undertaken an attempt to understand the origin
of the  cutoff and the  low $p_T$ suppression within the perturbative QCD. As we will discuss later, the
application of the cutoff to the collinear factorization formula extends
it beyond the leading power. Thus, any approach which aims to explain
the cutoff has to incorporate higher twists. Non-negligible power
corrections may be generated by large transverse momenta of incoming
partons entering the hard collision, as compared to the hard scale of
the process. These features are naturally incorporated in the High Energy Factorization
(HEF) (or $k_{T}$-factorization) approaches \citep{Gribov1983,Catani:1990eg,Catani:1990xk,Catani:1994sq,Collins1991}.
There, the transverse momentum of the dijet pair is no longer zero,
but equals to the sum of the transverse momenta of the incoming off-shell
gluons. The distribution of these gluons in longitudinal and transverse
 momenta is given by so-called Unintegrated Gluon Distribution (UGD).
Thus, in principle, the cutoff on the jet $p_{T}$ is related to the
behavior of UGDs in transverse  momentum which, in the low $x$ limit, is given by Balitsky-Fadin-Kuraev-Lipatov
(BFKL) equation \citep{Fadin:1975cb,Kuraev:1977fs,Balitsky:1978ic}
or some BFKL-type evolution. Furthermore, the gluon emissions with small
transverse momenta are suppressed by the Sudakov form factor. In fact,
for some UGD models \citep{Kimber2000a,Kimber:2001sc} the transverse
momentum of the gluons is generated by the Sudakov form factor and
the standard Dokshitzer-Gribov-Lipatov-Altarelli-Parisi (DGLAP) evolution.
This is somewhat similar to the soft gluon resummation \citep{Parisi1979}
technique which was used in \citep{Grau1999,Fagundes2015} to build
an eikonal minijet model which does not require a cutoff (but it is
suitable only for the total cross section).

The strategy for our paper is as follows. Using the HEF for inclusive dijet production we shall perform two independent studies of the $p_{T}$ cutoff:
\begin{description}
\item [Study 1.] A direct study, where we calculate the $p_T$ spectrum for  $p_T\gtrsim 2 \, \mathrm{GeV}$ and see if there is a suppression and determine its energy dependence.
\item [Study 2.] An indirect study, where we analyze the hard dijet production with $p_T\gtrsim 25 \, \mathrm{GeV}$ and look for an observable which is sensitive to power corrections which would come from MPIs in minijet models.
\end{description}

As for the study~1, the issue of a direct access to the $p_T$ cutoff within an approach involving an internal $k_{T}$ is actually known in the literature. In \citep{Gustafson2001} it was shown that indeed such approach can
produce $p_{T}$ suppression, which has roughly the correct energy
dependence. There is however an important difference to our study~1. 
We use HEF  of \citep{Catani:1990eg,Catani:1990xk,Catani:1994sq,Collins1991} which factorizes the cross section into UGD and a genuine $2\rightarrow 2$ off-shell hard process which extends collinear minijet formula beyond leading power. In \citep{Gustafson2001} the minijet production was considered in the sense
of a chain of emissions which does not have hard $2\rightarrow 2$ process. It is rather suitable for constructing shower-like Monte Carlo program \citep{Kharraziha1998a} that can be used to study particle production \citep{Gustafson2003}. 
More precisely, in \citep{Gustafson2001} the authors considered
a modification of the Catani-Ciafaloni-Fiorani-Marchesini (CCFM) \citep{Ciafaloni:1987ur,Catani:1989yc,Catani:1989sg}
evolution, so-called linked dipole chain model \citep{Andersson1996},
in which any emission in the chain can contribute a minijet (the emissions are unordered in transverse
momenta and thus following this logic any sub-collision in the chain can be considered
as `hard'). On the contrary, in HEF, we require that the large enough
hard scale is present that distinguishes the hard $2\rightarrow 2$ process from the chain of remaining emissions. Since this hard scale is identified with the jet $p_{T}$ the
two directly emitted partons 
should be actually considered as hard jets, not the minijets.
In the first approximation hard jets are produced back-to-back and described by the leading power
collinear approach, which does not feature any suppression factor. 
We will see this feature in our calculations when we compute the $p_T$ spectra in the small $p_T$ region from HEF.
That is, we will find no suppression in the $p_T$ spectra of the type present in the minijet models.
 Nonetheless, it does
not mean that there are no minijets in HEF. In fact, HEF takes into account
additional emissions visible as the jet imbalance, and thus as power
corrections.

The above motivates the study~2, which concentrates on the indirect access to minijets in HEF. 
 We  introduce an observable related to the dijet imbalance $K_T$, which is  sensitive to minijets. Specifically, we shall consider the cross section differential in the ratio $\tau$, of $K_T$ to the dijet average $p_{T}$. We will
check actual sensitivity of this observable on minijets, in particular on
$p_{T0}$ cutoff, using $\mathsf{pythia}$ and then we shall compare
them to similar calculations in HEF models. Next, we introduce bimodality coefficient which characterizes the $\tau$ spectrum. We observe that the energy dependence of this coefficient is very sensitive to the particular minijet model. 
  We will see that some of the UGDs used in HEF give energy dependence similar to the one coming from the minijet models in $\mathsf{pythia}$. This would then indirectly confirm the statement from \citep{Gustafson2001}, but in a way that can be confirmed experimentally when such observable is measured.

Our work is organized as follows. In Section~\ref{sec:Minijets} we systematically review theory behind minijets. First, in Subsection~\ref{sec:MinijetsColl}
we review the collinear factorization for the minijet production 
and then, in Subsection~\ref{sub:Minijets_models}, we describe in details
how the cutoff is introduced.
 In Subsection~\ref{sub:HEF} we review the 
HEF and discuss its relevance to minijet cross section. In particular, we shall explain that the leading twist limit of HEF does not reproduce the result of Dokshitzer-Dyakonov-Troyan (DDT) \citep{Dokshitzer1980} for the dijet momentum disbalance.
 Therefore, in the next Subsection~\ref{sub:IDDT} we construct a model similar to HEF but
having the DDT limit.
In the following sections
we will turn to numerical simulations. First, in Section~\ref{sub:setup}
we shall describe in some details the process under consideration,
kinematic cuts, etc. in order to unambiguously define the observables.
Later, in Section~\ref{sub:DirectMinijets} we will analyze the inclusive
dijet spectra in the low $p_{T}$ region in order to see whether the
suppression is produced in HEF and the DDT-based model we constructed in Subsection~\ref{sub:IDDT} (study 1).
Finally, in Section~\ref{sub:IndirectMinijets} we will turn to hard
inclusive dijets and study the minijets as a power correction (study 2). We
will summarize and make our conclusions in Section~\ref{sec:Conclusions}.\\

%%%%%%%%%%%%%%%%%%%%%%%%%%%%%%%%%%%%%%%%%%%%
\section{Minijets in selected approaches}
\label{sec:Minijets}

\subsection{Collinear factorization and soft gluon resummation}
\label{sec:MinijetsColl}

The starting point for a typical minijet model
is the collinear factorization formula, which however has to be modified.
In this introductory section we review this issue in a more quantitative
way.

The QCD collinear factorization theorem (see e.g. \citep{Collins:2011zzd}
for a review)  expresses the cross section for \textit{hard}
dijet production as
\begin{equation}
\sigma_{2\mathrm{jet}}=\sum_{a,b}\int\frac{dx_{A}}{x_{A}}\,\frac{dx_{B}}{x_{B}}\, d\hat{\sigma}_{ab}\left(x_{A},x_{B};\mu^{2}\right)f_{a/A}\left(x_{A};\mu^{2}\right)f_{b/B}\left(x_{B};\mu^{2}\right)+\mathcal{O}\left(\frac{\mu_{0}^{2}}{\mu^{2}}\right),\label{eq:CollFact1}
\end{equation}
where $f_{a/A}$, $f_{b/B}$ are integrated parton distribution functions (PDFs)
for a parton $a,b$ inside a hadron $A,B$, and $d\hat{\sigma}_{ab}$
is a partonic, fully differential, cross section which can be calculated order by order in perturbation theory.
In general the partons $a,b$ can be quarks and gluons, including
heavy quarks. The phase space cuts necessary to define a jet cross
section (i.e. a suitable jet algorithm) are hidden inside the partonic
cross section. The hard scale $\mu$ is the largest scale in the problem
and is typically taken to be the average transverse momentum of the
jets, $P_{T}=\left(\left|\vec{p}_{T1}\right|+\left|\vec{p}_{T2}\right|\right)/2$.
The remainder, i.e. the higher `twist' corrections in (\ref{eq:CollFact1})
are suppressed by the powers of the ratio $\mu_{0}^{2}/\mu^{2}$,
where $\mu_{0}$ is the largest of some other scales present in the
problem, e.g. heavy quark masses, dijet disbalance, etc.

Since the purpose of this work is to study minijets, let us restrict
to the semi-hard jets having transverse momenta $p_{T}\gtrsim2\,\mathrm{GeV}$.
In addition, we are interested in the total CM energies being much
larger then this scale. For such regime the factorization theorem (\ref{eq:CollFact1})
starts to fail. Two major sources for this are various large logs 
 (containing ratios of very different scales) 
and power corrections which are no longer small.

Certainly, the formula (\ref{eq:CollFact1}) 
would be perfectly valid for fixed $s$ and $\mu^{2}\rightarrow\infty$,
but obviously this is not the case for minijets. In order to illustrate
the problems more quantitatively, let us consider a cross section
(\ref{eq:CollFact1}) as a function of the disbalance between the
jets, $K_{T}^{2}$, when $\mu_{0}^{2}\ll K_{T}^{2}\ll\mu^{2}$. 
To
leading logarithmic accuracy it is given by the formula due to Dokshitzer-Dyakonov-Troyan, the so-called `DDT formula'%
\footnote{More precisely, the notion `DDT formula' refers to the factorization
formula for the transverse distribution of the Drell-Yan pairs in
hadron-hadron collision. Its generalization for decorrelation of a
di-hadron system in hadron-hadron collision was given in \citep{Dokshitzer1980}.
In the present work we use the term `DDT formula' for the latter. %
} \citep{Dokshitzer1980} 
\begin{multline}
\frac{d\sigma_{2\mathrm{jet}}}{dK_{T}^{2}}=\sum_{a,b,c,d}\int\frac{dx_{A}}{x_{A}}\,\frac{dx_{B}}{x_{B}}\, d\hat{\sigma}_{ab\rightarrow cd}\left(x_{A},x_{B};\mu^{2}\right)\\
\times\frac{\partial}{\partial K_{T}^{2}}\left\{ f_{a/H}\left(x_{A};K_{T}^{2}\right)T_{a}\left(K_{T}^{2},\mu^{2}\right)f_{b/H}\left(x_{B};K_{T}^{2}\right)T_{b}\left(K_{T}^{2},\mu^{2}\right)T_{c}\left(K_{T}^{2},\mu^{2}\right)T_{d}\left(K_{T}^{2},\mu^{2}\right)\right\} \\
+\mathcal{O}\left(\frac{K_{T}^{2}}{\mu^{2}}\right)\,,\label{eq:DDT1}
\end{multline}
where $T_{a}\left(\mu_{1}^{2},\mu_{2}^{2}\right)$ is a `Sudakov' form
factor for a parton $a$ 
(for the original Sudakov's form factor in QED see \citep{Sudakov1956}).
  It can be thought of as a probability for
the parton $a$ to evolve between the scales $\mu_{1}$ and $\mu_{2}$
without any resolvable emissions. We shall give the explicit formula
later (see Subsection~\ref{sub:IDDT}, Eq.~(\ref{eq:SudakovFF})), for now let us just mention that 
\begin{equation}
T_{a}\left(\mu^{2},\mu^{2}\right)=1,\,\,\,\, \,\, T_{a}\left(\mu_{0}^{2},\mu^{2}\right)\backsimeq0\label{eq:SudakovFF1}\;, \; \; \mu \gg \mu_0 \; ,
\end{equation}
 for $\mu_{0}$ being the lowest scale in our
problem. Let us remark, that the relevant DDT formula in \citep{Dokshitzer1980}
was actually derived for a production of hadrons in hadron-hadron
collision, and thus it contained fragmentation functions which accompanied
the form factors $T_{c}$, $T_{d}$ in (\ref{eq:DDT1}). For the purpose
of this paper we have adjusted that formula for dijets by setting
the fragmentation functions to be the delta functions. Let us note,
that due to the listed properties of the Sudakov form factors, this
formula reduces to (\ref{eq:CollFact1}) when integrated over the
jet disbalance $K_{T}$. Since the appearance of the DDT formula a
lot of effort has been put into improving the accuracy of perturbative
predictions for such semi-inclusive observables. In particular so-called
Transverse Momentum Dependent (TMD) factorization theorem has been
established for certain processes \citep{Collins:2011zzd}. We shall discuss these at the end
of this section and for the purpose of the present discussion we shall
stick to the leading-log formula (\ref{eq:DDT1}).

\begin{figure}
\begin{centering}
\includegraphics[width=0.49\textwidth]{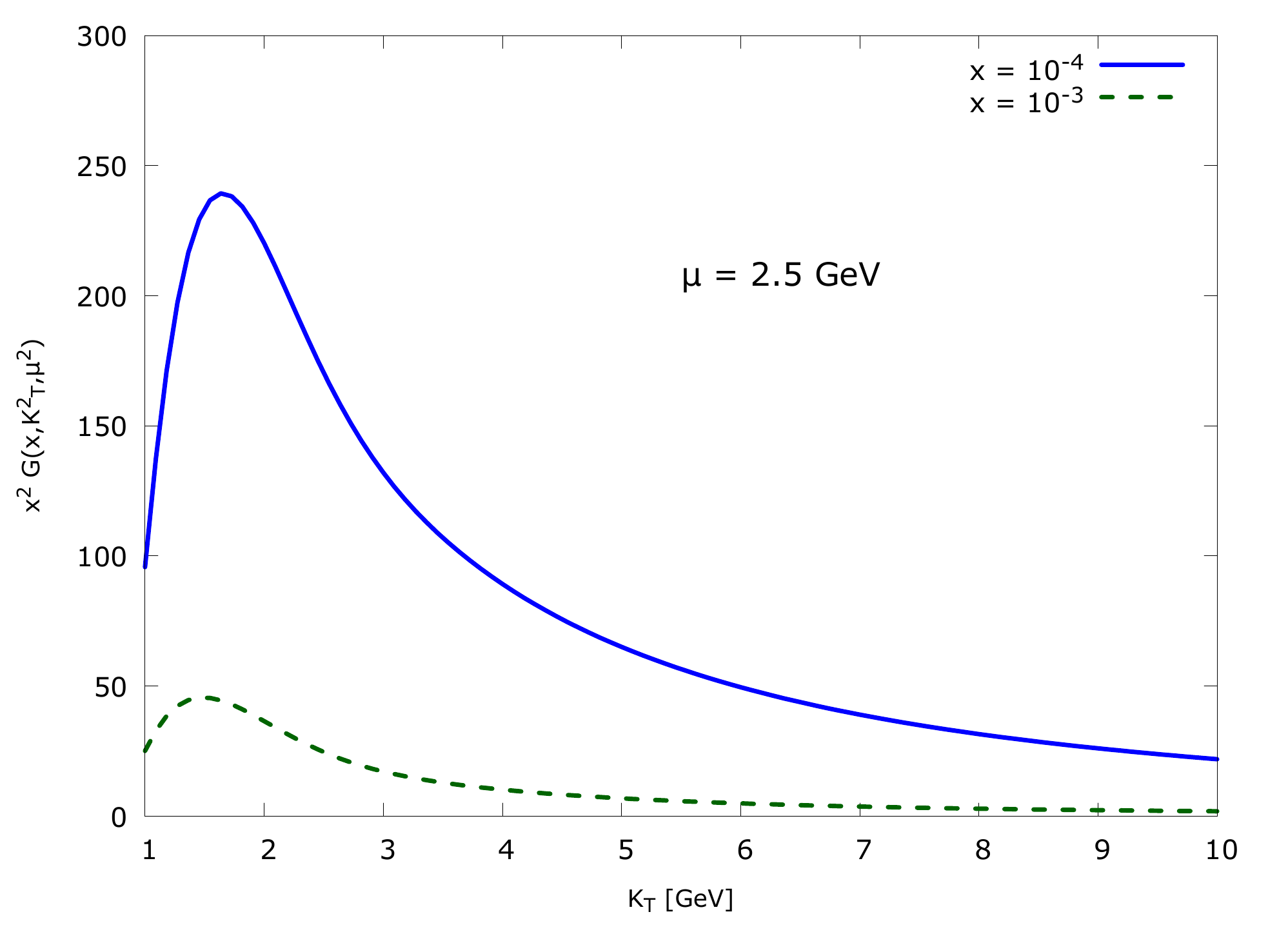}$\,\,\,\,\,\,$\includegraphics[width=0.49\textwidth]{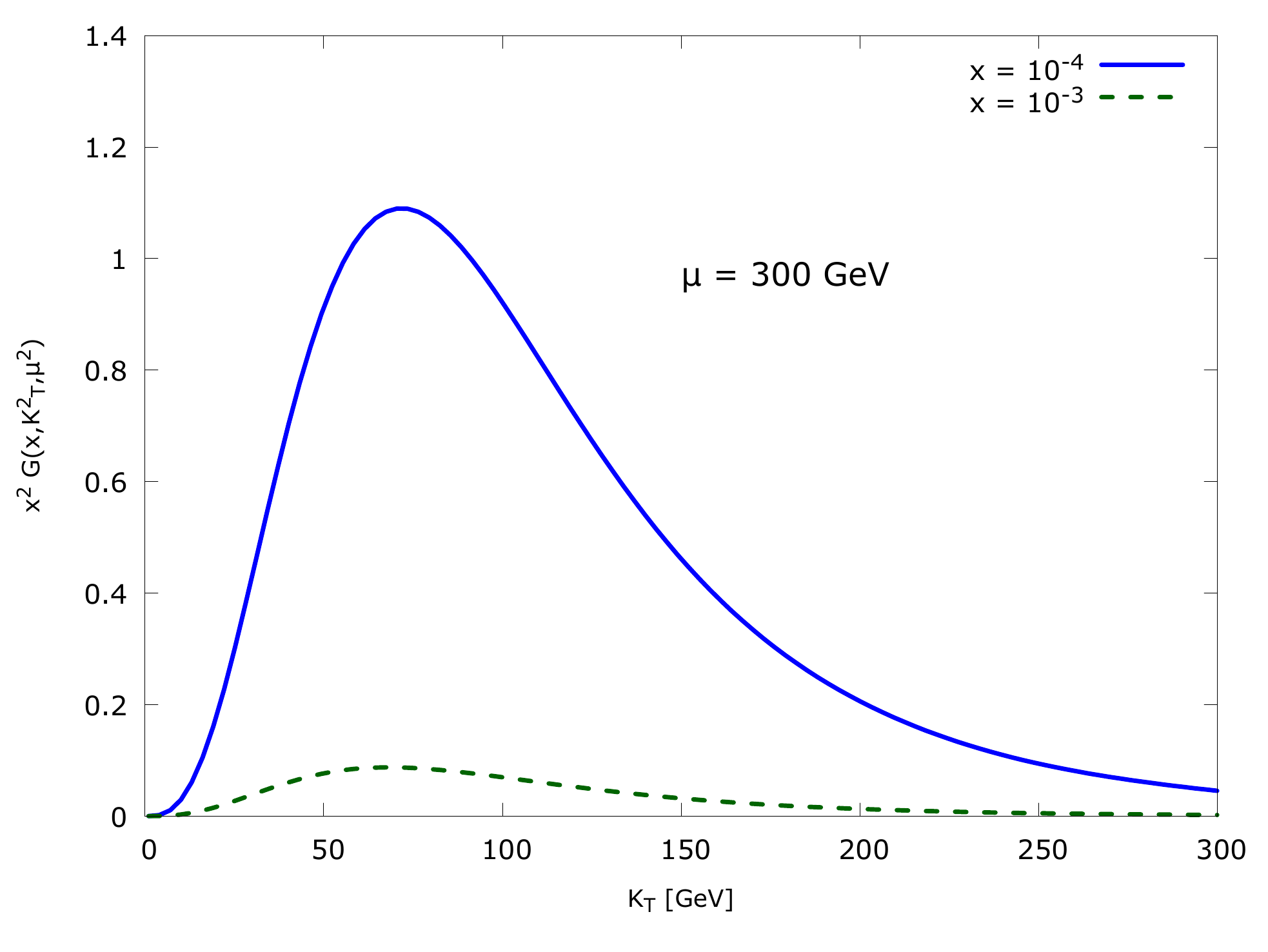}
\par\end{centering}

\caption{\label{fig:DDT_derivative} The density $G$ entering the DDT formula
for two different values of $x$ and two fixed values of $\mu=P_{T}=2.5\,\mathrm{GeV}$
(left) and $300\,\mathrm{GeV}$ (right). The PDF set used here was
GRV98 \citep{Gluck:1998xa}.}

\end{figure}

In case of minijets, the formula (\ref{eq:DDT1}) looses its accuracy
as now $K_{T}$ can be easily of the order of $\mu$ (which is the
average $p_{T}$ of the jets). This can be seen by  inspecting the
derivative in (\ref{eq:DDT1}) as a function of $K_{T}$, for example
in the pure gluonic channel: 
\begin{equation}
G\left(x,K_{T}^{2},\mu^{2}\right)=\frac{\partial}{\partial K_{T}^{2}}\left\{ f_{g/H}^{2}\left(x;K_{T}^{2}\right)T_{g}^{4}\left(K_{T}^{2},\mu^{2}\right)\right\} \,.\label{eq:G_DDT}
\end{equation}
This distribution is plotted in Fig.~\ref{fig:DDT_derivative} as a function of $K_{T}$ for fixed
$\mu=P_{T}=2.5\,\mathrm{GeV}$ and $\mu=300\,\mathrm{GeV}$, and two
values of $x$ (note that for simplicity we have used the same values
of $x$ entering both PDFs in (\ref{eq:G_DDT})). In this presentation
we use the leading order GRV98 PDF set \citep{Gluck:1998xa} (we explain
the reason for using this PDF set in Section~\ref{sub:setup}). We
see that the characteristic $K_{T}$, let us call it $K_{T0}$, generated
by the density $G$ is large comparing to the average $P_{T}$ of
minijets so that $K_{T0}/P_T\sim \mathcal{O}(1)$ (left plot in Fig.~\ref{fig:DDT_derivative}; $K_{T0}$ may
be defined for example as the value  for which the distribution
has a maximum, although median would probably be a more realistic
estimate). For comparison, we plot the same distribution for hard
jets (right plot in Fig.~\ref{fig:DDT_derivative}) with $\mu=300\,\mathrm{GeV}$.
For the latter, the ratio $K_{T0}/P_{T}$ becomes much smaller than
the unity and the situation improves with increasing scale. To summarize,
the power corrections cannot be neglected for minijets and one has
to necessarily venture beyond leading `twist' to account for minijets.
Let us remind, that the formula (\ref{eq:DDT1}) is a more `exclusive'
version of (\ref{eq:CollFact1}) and the condition that we can neglect
the power corrections is actually a condition necessary to obtain
(\ref{eq:CollFact1}) when the integral over $K_{T}$ is performed.

There is yet another source of errors in the DDT formula, namely
the sub-leading logs. Actually, in its original formulation the DDT
formula was written for processes with only two hadrons such as for
instance the Drell-Yan process \citep{Dokshitzer1978}. Assuming strong
ordering in the transverse momenta of emitted gluons one obtains a
formula similar to (\ref{eq:DDT1}) but with two Sudakov form factors
instead of four (and of course with an appropriate hard partonic cross section
relevant to Drell-Yan process). However, the strong ordering in transverse
momenta for the soft gluons is a too strong assumption and gives a
non-physical suppression in the low $K_{T}$ limit. The improved approach for Drell-Yan pairs 
was proposed in \citep{Parisi1979,Curci1979,Chiappetta1981} which
resums the soft gluons (thus the approach is often called the `soft
gluon resummation') using the impact parameter space conjugate to
transverse momenta. 
As a result one finds a flat distribution at small $K_T$ rather than an exponentially suppressed cross section. 
Soft gluon resummation extends beyond the leading
logarithm, but still more general  approach exists, so called Transverse
Momentum Dependent (TMD) factorization (see e.g. \citep{Collins:2011zzd}).
It is a rigorous factorization theorem of QCD and is valid to leading
power in the hard scale. It is important to note, that this theorem
is valid for processes with at most two hadrons. Thus the most complicated
processes are Drell-Yan process \citep{Collins1985} and semi-inclusive
deep inelastic scattering \citep{Ji2005}. The theorem is violated
when more hadrons are present \citep{Rogers:2010dm}, thus it fails
for example for jet production in hadron-hardon collision. 
However, although the TMD factorization is
not a strict leading-power theorem holding to all orders in $\alpha_s$, it has been shown in \citep{Sun2015}
that it holds to next-to-logarithmic accuracy for the latter case.

Before discussing the power corrections to (\ref{eq:DDT1}) let us
make some general comments. The twist corrections to deep inelastic
structure functions, i.e. the corrections $\mathcal{O}\left(1/Q^{2}\right)$
with $Q^{2}$ being the photon virtuality, were studied long time
ago in the context of the operator product expansion (OPE) \citep{Jaffe1982}
and using Feynman diagrams \citep{Ellis1982}. While OPE is very general,
it becomes very complicated for more exclusive processes (see e.g.
\citep{Anikin2010} and \citep{Braun2012}). As for jet production
in hadron-hadron collisions no higher twist factorization exists 
(see also a discussion of power corrections 
coming from heavy quarks in the end of this subsection). On the other hand there are approaches which take
into account all power corrections of a certain class. At very large
energies the logs of the form $\log\left(1/x\right)$, where $x$
is a fraction a hadron longitudinal momentum carried by the parton, become large
and can be resummed by means of the BFKL equation. 
Let us note, however,
that it is often arguable if such logs should be resummed at currently
achievable energies, as most of the observables measured at LHC can
be explained using collinear factorization supplemented by the DGLAP-type
parton showers. Nevertheless, the BFKL formulation leads
to HEF, which as mentioned in the Introduction resums the power corrections
of the form $K_{T}/\mu$. We shall describe HEF in more details in
Subsection~\ref{sub:HEF}.

For completeness let us discuss a special case when $K_{T}\gg\mu$.
For the case of the Drell-Yan process this kinematic region was studied
in \citep{Parisi1978,Altarelli1984}, before the DDT formula was established. The
corrections of this type can be obtained calculating explicitly additional
emission by means of $2\rightarrow3$ process, away from the singular
(soft and/or collinear) region. 
In particular, the HEF partially recovers this perturbative limit for
certain UGDs.

Finally, let us make some comments on the power corrections coming
from the heavy quark masses. Actually, they can be explicitly taken
into account in the hard cross section, order-by-order. The problem
is however, that by doing so the cross section becomes infra-red unsafe
for large $p_{T}$, i.e. we shall encounter logs of the type $\log\left(P_{T}^{2}/m_{Q}^{2}\right)$
where $m_{Q}$ is the mass of a heavy quark $Q$. This problem can
typically be addressed by so-called general-mass scheme, which supplements
the hard cross section with a proper subtraction terms (see \citep{Collins:1998rz}
for a general proof and \citep{Kotko2012} for a formulation for jets
in DIS at NLO). However, for jets in hadron-hadron collisions there
is a problem with the cancellation of soft singularities when incoming
lines are massive \citep{Doria1980} and thus the power corrections
are unlikely to be controlled using the general-mass schemes. We shall
ignore all these complications as we will be focused
on pure gluonic contributions, which should dominate at high energies.

%%%%%%%%%%%%%%%%%%%%%%%%%%%%%%%%%%%%%%%%%%%%
\subsection{Singularity $p_{T}\rightarrow0$ and soft cutoff}
\label{sub:Minijets_models}

In this section we shall discuss in detail the concept of the soft transverse momentum cutoff. We shall restrict our considerations to gluons only. This is done for two major reasons. First, the gluons dominate at high
energies and this is sufficient to illustrate all the effects we analyze in
the paper (we do not aim at giving any predictions or comparisons
with data). Second, later on we shall make  comparisons across
models including HEF, which is basically restricted only to gluons
dominating at high energies. In principle one could consider off-shell
quarks, but the subject is still poorly developed and would unnecessarily
complicate our study (see \citep{vanHameren:2013csa,Kutak2016} for
selected recent results).

Let us start by writing LO contribution to (\ref{eq:CollFact1}).
We parametrize the momenta of hadrons as
\begin{equation}
p_{A}^{\mu}=\sqrt{\frac{S}{2}}\, n_{+}^{\mu},\,\,\, \,\,\, p_{B}^{\mu}=\sqrt{\frac{S}{2}}\, n_{-}^{\mu}\,,
\end{equation}
where $n_{\pm}=\left(1,0,0,\pm1\right)$ and $S=2p_{A}\cdot p_{B}$
is the CM energy squared. The kinematics of the hard subprocess $g\left(k_{A}\right)g\left(k_{B}\right)\rightarrow g\left(p_{1}\right)g\left(p_{2}\right)$
is 
\begin{equation}
k_{A}^{\mu}=x_{A}p_{A}^{\mu},\,\,\,\,\,\, k_{B}^{\mu}=x_{B}p_{B}^{\mu}\,,
\end{equation}
\begin{equation}
p_{1}^{\mu}=z_{1}p_{A}^{\mu}+\frac{-p_{T1}^{2}}{z_{1}S}\, p_{B}^{\mu}+p_{T1}^{\mu},\,\,\,\,\,\, p_{2}^{\mu}=z_{2}p_{A}^{\mu}+\frac{-p_{T2}^{2}}{z_{2}S}\, p_{B}^{\mu}+p_{T2}^{\mu}\,,
\end{equation}
with momentum conservation $k_{A}+k_{B}=p_{1}+p_{2}$. Obviously $z_{1}$,
$z_{2}$ are directly related to rapidities $y_{1,2}$ in the following way
\begin{equation}
z_{1,2}=\frac{\left|\vec{p}_{T\,1,2}\right|}{\sqrt{S}}\, e^{\, y_{1,2}} \; ,
\end{equation}
with $-p_{T1,2}^{2}=\left|\vec{p}_{T1,2}\right|^{2}$. Due to the
transverse momentum conservation both outgoing jets have exactly the
same transverse momentum $\left|\vec{p}_{T1}\right|=\left|\vec{p}_{T2}\right|$.
In what follows we shall simply use notation $\left|\vec{p}_{T1,2}\right|\equiv p_{T}$
for brevity. In the above kinematics, the cross section can be calculated
as 
\begin{multline}
\sigma_{2\mathrm{jet}}=\frac{1}{16\pi}\,\int\frac{dp_{T}^{2}}{p_{T}^{4}}\,\int\frac{z_{1}dz_{1}\, z_{2}dz_{2}}{(z_{1}+z_{2})^{4}}\,\\
\, f_{g/H}\left(z_{1}+z_{2},\mu^{2}\right)f_{g/H}\left(\frac{p_{T}^{2}}{S}\,\frac{z_{1}+z_{2}}{z_{1}z_{2}},\mu^{2}\right)\,\frac{1}{2}\left|\overline{\mathcal{M}}\right|_{gg\rightarrow gg}^{2}\left(z_{1},z_{2}\right)\,,\label{eq:minijets1}
\end{multline}
where the amplitude squared and averaged/summed over spin and color
reads 
\begin{equation}
\left|\overline{\mathcal{M}}\right|_{gg\rightarrow gg}^{2}\left(z_{1},z_{2}\right)=g^{4}\,\frac{9}{2}\,\frac{\left(z_{1}^{2}+z_{1}z_{2}+z_{2}^{2}\right)^{3}}{z_{1}^{2}z_{2}^{2}\left(z_{1}+z_{2}\right)^{2}}\,.
\end{equation}
Typically, as the hard scale $\mu$ one chooses the $p_{T}$ of the
jets. From (\ref{eq:minijets1}) we see that the cross section diverges
like
\begin{equation}
\frac{d\sigma_{2\mathrm{jet}}}{dp_{T}^{2}}\sim\frac{\alpha_{s}^{2}\left(p_{T}^{2}\right)}{p_{T}^{4}}\,.
\end{equation}

In the pioneering work \citep{Soediono1989b} the MPI model was constructed
with $\sigma_{2\mathrm{jet}}$ modified to remove this singularity
by defining
\begin{equation}
\sigma'_{2\mathrm{jet}}=\int\frac{d\sigma_{2\mathrm{jet}}}{dp_{T}^{2}}\,\frac{p_{T}^{4}}{\left(p_{T}^{2}+p_{T0}^{2}\left(S\right)\right)^{2}}\,\frac{\alpha_{s}^{2}\left(p_{T}^{2}+p_{T0}^{2}\left(S\right)\right)}{\alpha_{s}^{2}\left(p_{T}^{2}\right)}\,,\label{eq:MinijetReg}
\end{equation}
where $p_{T0}\left(S\right)$ is the model parameter we have briefly
discussed in the Introduction. For example in version 8.1 of $\mathsf{pythia}$
it is defined as

\begin{equation}
p_{T0}\left(S\right)=2.28\,\left(\frac{\sqrt{S}}{7\,\mathrm{TeV}}\right)^{0.215}\,\mathrm{GeV}\,.\label{eq:pT0(s)}
\end{equation}
for  standard $\mathsf{pythia}$ settings (including pre-determined
PDF sets to be used by default). Let us mention, that the MPI model
and the entire event generation procedure in $\mathsf{pythia}$ is
very complex, much more then the simple Eq.~(\ref{eq:MinijetReg}).
Nevertheless Eq.~(\ref{eq:MinijetReg}) constitutes one of the core
building blocks of this powerful program.

The $p_{T}$ spectrum of minijets $d\sigma'_{2\mathrm{jet}}/dp_{T}^{2}$
within the presented model should exhibit a strong suppression for small
$p_{T}$, slowly growing with energy. It is interesting to ask if
such a suppression could be directly observed. Putting this question
aside, we will simply calculate (see Section~\ref{sub:DirectMinijets})
the inclusive dijet production in the small $p_{T}$ region using
$\mathsf{pythia}$ and compare with the minijet spectrum $d\sigma'_{2\mathrm{jet}}/dp_{T}^{2}$.
There are a few interesting features of this calculation (thought
to be more realistic than (\ref{eq:MinijetReg})) which will be discussed
later. 

%%%%%%%%%%%%%%%%%%%%%%%%%%%%%%%%%%%%%%%%%%%%
\subsection{High Energy Factorization}
\label{sub:HEF}

Let us now discuss how the power corrections in (\ref{eq:DDT1}) can
be taken into account in $k_{T}$-factorization (we use the terms
`high energy factorization' and `$k_{T}$-factorization' interchangeably
in the present work, although both terms have different origin). 

In $k_{T}$-factorization the cross section is calculated as a convolution
of so-called unintegrated gluon distributions (UGDs) and an off-shell
matrix element. UGDs depend not only on longitudinal momentum fractions $x$, but also on the transverse
momenta $k_{T}$ of the gluons -- a feature neglected in the collinear factorization
due to the power counting. For the first time $k_{T}$-factorization
was used in \citep{Gribov1983} for inclusive jet production at high
energies using basically $2\rightarrow1$ process $g^{*}g^{*}\rightarrow g$.
Let us note that the $2\to 1$ process does not exist when the incoming partons are on-shell and collinear, but it appears at lowest order   in the  $k_T$ factorization approach.
Later, a similar idea (originally called HEF) was used to compute heavy
quark production \citep{Catani:1990eg,Catani:1990xk,Catani:1994sq,Collins1991}
by means of a gauge invariant matrix element for $g^{*}g^{*}\rightarrow Q\overline{Q}$
which was extracted from the Green function utilizing suitable eikonal
projectors. The UGDs were assumed to undergo BFKL evolution. A natural
step forward was to adopt the HEF to account for jet production processes
at high energy. Thus, the HEF has been extended to all channels \citep{Leonidov:1999nc},
including gluons. At small $x$ the forward jets are especially interesting.
They can be treated in a limiting case of HEF, where one of the gluons
becomes on-shell \citep{Deak:2009xt,Deak2010,Deak:2011ga}. In this
approximation, this gluon is treated as a `large-$x$' gluon and is
assigned a standard collinear PDF. In the Color Glass Condensate (CGC)
approach \citep{Blaizot:2004wu} a similar idea was used to study
forward particle production in saturation domain and exists under
the name of the `hybrid' formalism \citep{Dumitru:2005gt}. In fact,
the hybrid version of HEF can be derived from CGC in the dilute limit
\citep{Iancu2013,Kotko:2015ura}. Several observables relevant for
LHC have been calculated within the hybrid HEF, see Refs.~\citep{VanHameren2013,vanHameren:2014ala,vanHameren:2014lna,VanHameren2015,Kutak2016}.
In the present work we are not concerned with forward jets thus we
shall not use the hybrid version of HEF, but the original one with two off-shell incoming particles.

\begin{figure}
\begin{centering}
\parbox{0.33\textwidth}{A)\\\includegraphics[width=0.33\textwidth]{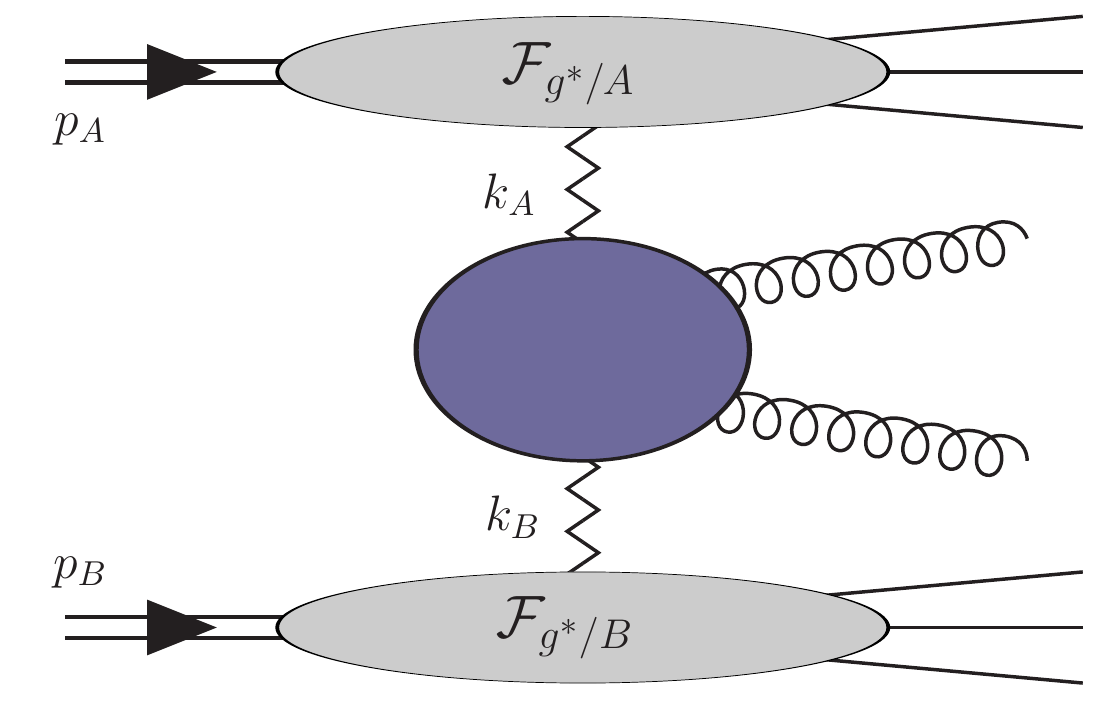}}$\,\,\,\,\,\,\,\,\,\,\,\,\,\,\,\,\,$\parbox{0.64\textwidth}{B)\\\includegraphics[width=0.64\textwidth]{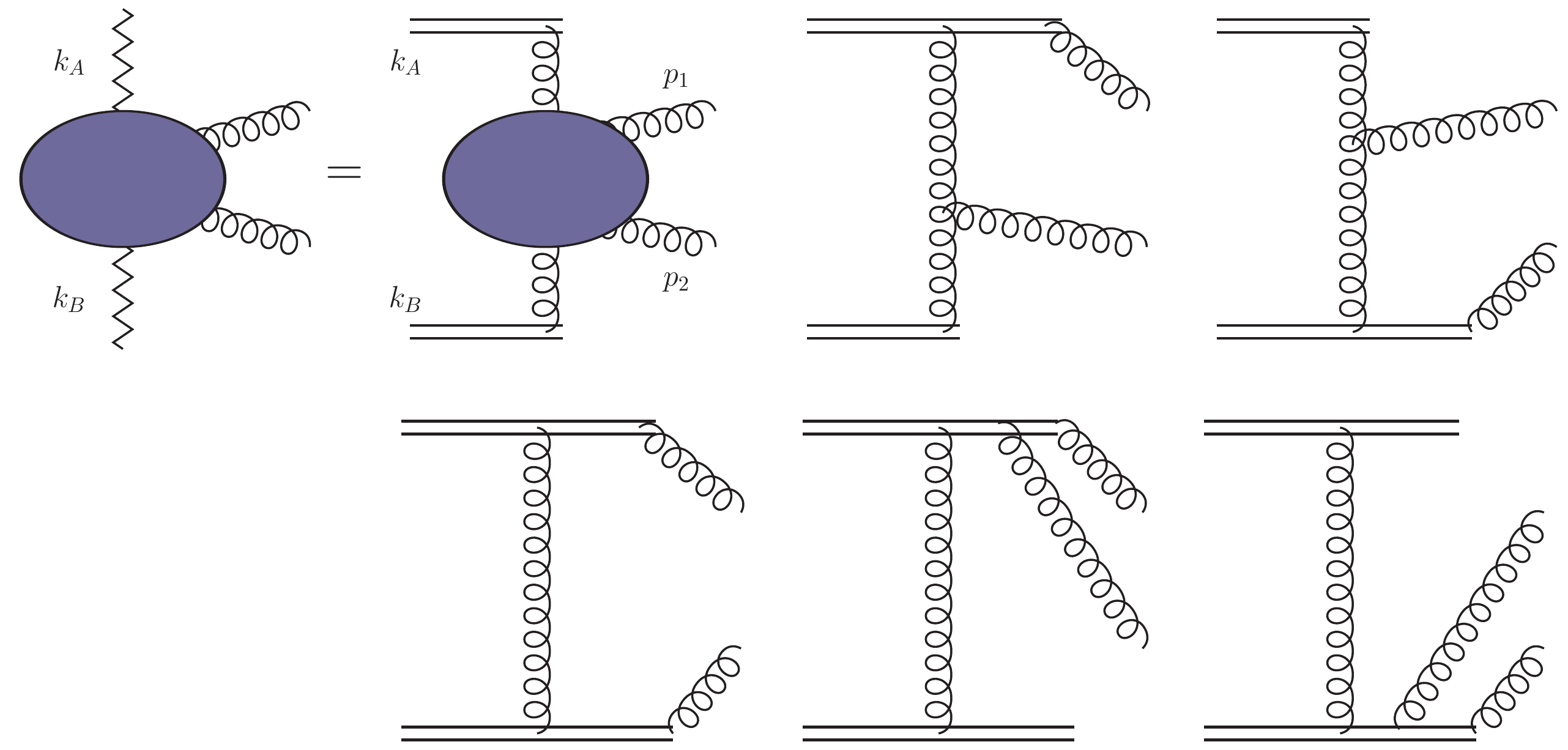}}
\par\end{centering}

\caption{A) Schematic representation of the factorization formula (\ref{eq:HEN_fact_1})
B) The hard gauge invariant tree level off-shell process expressed
in terms of a matrix element of straight infinite Wilson lines, with
the slopes being $p_{A}$ (top) and $p_{B}$ (bottom). The blue blob
on the r.h.s. denotes a standard QCD contribution with four and triple
gluon vertices. Only planar (color-ordered) diagrams are shown. \label{fig:HEF}}
\end{figure}

The factorization formula for HEF reads (including only gluons) 
\begin{multline}
d\sigma_{AB\rightarrow gg}=\int d^{2}k_{T\, A}\int\frac{dx_{A}}{x_{A}}\,\int d^{2}k_{T\, B}\int\frac{dx_{B}}{x_{B}}\,\\
\mathcal{F}_{g^{*}/A}\left(x_{A},k_{T\, A};\mu\right)\,\mathcal{F}_{g^{*}/B}\left(x_{B},k_{T\, B};\mu\right)\, d\hat{\sigma}_{g^{*}g^{*}\rightarrow gg}\left(x_{A},x_{B},k_{T\, A},k_{T\, B};\mu\right),\label{eq:HEN_fact_1}
\end{multline}
where $\mathcal{F}_{g^{*}/A}$, $\mathcal{F}_{g^{*}/B}$ are UGDs
for hadrons $A,B$ and $d\hat{\sigma}_{g^{*}g^{*}\rightarrow gg}$
is the partonic cross section build up from the gauge invariant $g^{*}g^{*}\rightarrow gg$
amplitude (Fig.~\ref{fig:HEF}A). The momenta of the off-shell gluons
have the following form relevant to the high energy approximation:
\begin{equation}
k_{A}\simeq x_{A}p_{A}+k_{TA},\,\,\,\,\,\, k_{B}\simeq x_{B}p_{B}+k_{TB}\,.\label{eq:HEF_kinem}
\end{equation}
The off-shell partonic cross section is defined by a reduction of the Green's function, where the off-shell legs $k_A$ and $k_B$ are contracted
with eikonal projectors proportional to $p_A^{\mu}$ and $p_B^{\mu}$.
Unlike the $g^{*}g^{*}\rightarrow q\overline{q}$ amplitude used in
original HEF, the gluonic off-shell hard process cannot be just calculated
from the standard Feynman diagrams in a gauge invariant way. There are
number of ways this can be done in consistency with the high energy
approximation used to define the hard process. First, one can include
the bremsstrahlung from the lines to which the hard process is attached.
At high energies those lines are eikonal. Such idea was used in \citep{Leonidov:1999nc}
to calculate $g^{*}g^{*}\rightarrow gg$ and later in \citep{VanHameren2013a}
a general method for helicity amplitudes as well as numerical algorithm
for any number of partons was developed. Second, in the approximation
used to derive (\ref{eq:HEN_fact_1}) the gauge invariant amplitude
for $g^{*}g^{*}\rightarrow gg$ is equivalent to the Lipatov's vertex
$RRPP$ \citep{Lipatov:1995pn,Antonov:2004hh} in the quasi-multi-regge
kinematics. A more general approach is to consider matrix element
of straight infinite Wilson line  operators with  the polarization of the off-shell gluon identified as the Wilson line slope \citep{Kotko2014a}.
 This method can be also used  beyond the high energy approximation \citep{Cruz-Santiago2015,Kotko2016}.
Finally, a method generalizing the BCFW recursion \citep{Britto:2004ap,Britto:2005fq}
to the off-shell case is also available \citep{vanHameren:2014iua,vanHameren:2015bba}.
Although the Lagrangian method of \citep{Lipatov:1995pn} is the most
general, in practical computations, especially for multiple external
legs, the other mentioned methods
are more efficient. For the hybrid version of (\ref{eq:HEN_fact_1})
a very efficient method of calculating helicity amplitudes for $g^{*}g\rightarrow g\dots g$
was found in \citep{VanHameren2012}.
Some other applications and different ways of calculating $g^{*}g^{*}\rightarrow gg$
were given e.g. in \citep{Leonidov:1999nc,Nefedov2013}. Moreover, many other studies have
been done using $k_{T}$-factorization, see for example \citep{Saleev:2009et,Kniehl:2011hc,Maciua2011,Saleev:2012np,Maciua2013,Maciua2013a,Kutak2016}.

The partonic cross section in (\ref{eq:HEN_fact_1}) is defined as
\begin{equation}
d\hat{\sigma}_{g^{*}g^{*}\rightarrow gg}=\frac{1}{2x_{A}x_{B}S}\,\frac{1}{2}\,\left|\overline{\mathcal{M}}\right|_{g^{*}g^{*}\rightarrow gg}^{2}d\mathrm{PS}\,,
\end{equation}
where $d\mathrm{PS}$ is the two-particle phase space while $\left|\overline{\mathcal{M}}\right|_{g^{*}g^{*}\rightarrow gg}^{2}$
is the amplitude squared for the gauge invariant off-shell process
discussed above. Using the method of \citep{Kotko2014a} it can be
calculated as follows. First, the amplitude is decomposed into the
color-ordered amplitudes \citep{Mangano:1990by}. For the one particular
ordering of the external lines the color-ordered amplitude is given
by the planar diagrams displayed in Fig.~\ref{fig:HEF}B in Feynman
gauge. The double lines on the top and the bottom correspond to the
Wilson line propagators. Calculation of these diagrams (with proper
normalization)  gives the following result for the square of the amplitude
\begin{multline}
\left|\mathcal{A}\right|^{2}\left(k_{A},p_{1},p_{2},k_{B}\right)=-\frac{g^{4}}{s^{2}t^{2}\overline{t}_{1}\overline{t}_{2}}\,\frac{1}{k_{TA}^{2}k_{TB}^{2}}\\
\Big\{ k_{TA}^{2}t\overline{t}_{2}\left[k_{TA}^{2}\left(k_{TB}^{2}s\overline{s}+t\overline{t}_{1}\overline{u}_{1}^{2}\right)+2\overline{t}_{1}\overline{u}_{1}W\right]+k_{TB}^{2}t\overline{t}_{1}\left[k_{TB}^{2}\left(k_{TA}^{2}s\overline{s}+t\overline{t}_{2}\overline{u}_{2}^{2}\right)+2\overline{t}_{2}\overline{u}_{2}W\right]\\
+k_{TA}^{2}k_{TB}^{2}t\left[t\left(s^{2}\overline{s}^{2}+2\overline{t}_{1}\overline{t}_{2}\overline{u}_{1}\overline{u}_{2}\right)+s\overline{s}\left(\overline{s}^{2}t-4\overline{t}_{1}\overline{t}_{2}\left(\overline{s}+\overline{t}_{1}+\overline{t}_{2}-t\right)\right)\right]+\overline{t}_{1}\overline{t}_{2}W^{2}\Big\}\,,\label{eq:AmpRRGG}
\end{multline}
where
\begin{equation}
W=\left[s\left(\overline{s}t+\overline{t}_{1}\overline{t}_{2}\right)-\overline{s}t\left(\overline{s}+\overline{t}_{1}+\overline{t}_{2}-t\right)\right]\,.
\end{equation}
Above we have used abbreviations $k_{TA,B}^{2}\equiv\left|\vec{k}_{TA,B}\right|^{2}$.
The standard and auxiliary Mandelstam invariants read
\begin{gather}
s=\left(k_{A}+k_{B}\right)^{2},\,\, t=\left(k_{A}-p_{1}\right)^{2},\,\, u=\left(k_{A}-p_{2}\right)^{2}\,,\label{eq:Invs1}\\
\overline{s}=\left(x_{A}p_{A}+x_{B}p_{B}\right)^{2},\,\,\,\overline{t}_{1,2}=\left(x_{A}p_{A}-p_{1,2}\right)^{2},\,\,\,\overline{u}_{1,2}=\left(x_{B}p_{B}-p_{1,2}\right)^{2}\,.\label{eq:Invs2}
\end{gather}
They satisfy $s+t+u=-k_{A}^{2}-k_{B}^{2}$ and $\overline{s}+\overline{t}_{1,2}+\overline{u}_{1,2}=0$.
The order of arguments in (\ref{eq:AmpRRGG}) corresponds to the order
of the external legs (see Fig.~\ref{fig:HEF}B). The color dressed
amplitude is obtained by summing over all noncyclic permutations of
the external legs (minus equivalent permutations due to the relations like 
$\left|\mathcal{A}\right|^{2}\left(k_{A},p_{1},p_{2},k_{B}\right)=\left|\mathcal{A}\right|^{2}\left(k_{B},p_{2},p_{1},k_{A}\right)$,
etc.)
\begin{multline}
\left|\overline{\mathcal{M}}\right|_{g^{*}g^{*}\rightarrow gg}^{2}=\frac{1}{\left(2\pi\right)^{2}}\,\frac{N_{c}^{2}}{\left(N_{c}^{2}-1\right)}\,\\
2\left[\left|\mathcal{A}\right|^{2}\left(k_{A},p_{1},p_{2},k_{B}\right)+\left|\mathcal{A}\right|^{2}\left(k_{A},p_{2},p_{1},k_{B}\right)+\left|\mathcal{A}\right|^{2}\left(k_{A},p_{1},k_{B},p_{2}\right)\right]\,.
\end{multline}
The factor $1/\left(2\pi\right)^{2}$ constitutes the helicity average
for the off-shell gluons as their `polarization' vectors can be thought of to be `continuous'. It is because one can show that these polarizations are $k_{TA}^{\mu}/k_{TA}$ and $k_{TB}^{\mu}/k_{TB}$ which depend on the transverse angle spanning between $0$ and $2\pi$.

Using the same kinematics as for the collinear case (but now with
(\ref{eq:HEF_kinem}) for initial states) we can write the cross section
as
\begin{multline}
d\sigma_{AB\rightarrow gg}=\frac{1}{32\pi^{2}}\,\int d^{2}\vec{k}_{TA}d^{2}\vec{k}_{TB}\int\frac{dp_{T}^{2}d\phi}{\left[z_{1}\left(\vec{p}_{T}-\vec{K}_{T}\right)^{2}+z_{2}p_{T}^{2}\right]^{2}}\,\int\,\frac{z_{1}dz_{1}\, z_{2}dz_{2}}{\left(z_{1}+z_{2}\right)^{2}}\\
\,\mathcal{F}_{g^{*}/A}\left(z_{1}+z_{2},k_{TA},\mu\right)\mathcal{F}_{g^{*}/B}\left(\frac{1}{z_{2}S}\left(\vec{p}_{T}-\vec{K}_{T}\right)^{2}+\frac{1}{z_{1}S}p_{T}^{2},k_{TB},\mu\right)\,\\
\frac{1}{2}\left|\overline{\mathcal{M}}\right|_{g^{*}g^{*}\rightarrow gg}^{2}\left(z_{1},z_{2},\vec{k}_{TA},\vec{k}_{TB};\mu\right)\,,\label{eq:HEF_xsec}
\end{multline}
where 
\begin{equation}
\vec{K}_{T}=\vec{k}_{TA}+\vec{k}_{TB}\,.
\end{equation}
 The invariants in (\ref{eq:AmpRRGG}) can be easily expressed in
terms of integration variables in (\ref{eq:HEF_xsec}). Comparing
this with the collinear expression (\ref{eq:minijets1}) we see that
the singularity $p_{T}^{2}\rightarrow0$ can be potentially regularized
by a nonzero $K_{T}$. Let us note, however, that $K_{T}$ can be
zero even if $k_{TA}$, $k_{TB}$ generated in UGDs are nonzero. In
fact due to the transverse momentum conservation whenever the jets
are back-to-back $K_{T}=0$ and the singularity $p_{T}^{2}\rightarrow0$
remains bare. For nonzero $k_{TA}$, $k_{TB}$ the $K_{T}$ depends
on relative orientation of the vectors $\vec{k}_{TA}$, $\vec{k}_{TB}$.
Since UGDs do not generally depend on angles, the only correlations can be hidden
inside the matrix element. 
 Moreover, the expression (\ref{eq:HEF_xsec})
has to be integrated over transverse variables to be actually compared
with the collinear expression. We shall later perform a detailed numerical
study and see whether the modification of $1/p_{T}^{4}$ factor due
to $K_{T}$ can produce a cutoff similar to minijet models. This in
principle would be possible, as one can check that the median of the
transverse momenta given by UGDs grows with decrease of $x$. Anticipating
the result, however, let us recall, that actually (\ref{eq:HEF_xsec})
should be used in the hard scattering regime, that is for $\mu\sim p_{T}$
large. This can be also understood by realizing that the main contribution
to $\left|\overline{\mathcal{M}}\right|_{g^{*}g^{*}\rightarrow gg}^{2}$
comes from the collinear region. In fact it can be shown that
\begin{multline}
\int_{0}^{2\pi}\frac{d\alpha_{1}}{2\pi}\,\frac{d\alpha_{2}}{2\pi}\left|\overline{\mathcal{M}}\right|_{g^{*}g^{*}\rightarrow gg}^{2}\left(z_{1},z_{2},\vec{k}_{TA},\vec{k}_{TB};\mu\right)=\left|\overline{\mathcal{M}}\right|_{gg\rightarrow gg}^{2}\left(z_{1},z_{2};\mu\right)\\
+\mathcal{O}\left(\frac{k_{TA}}{\mu}\right)+\mathcal{O}\left(\frac{k_{TB}}{\mu}\right)+\mathcal{O}\left(\frac{k_{TA}k_{TB}}{\mu^2}\right)\,,\label{eq:HEF_coll_lim_}
\end{multline}
where $\alpha_{1}$, $\alpha_{2}$ are the angles on the transverse
plane of the vectors $\vec{k}_{TA}$, $\vec{k}_{TB}$ and the first
term on the r.h.s. is the collinear matrix element. 
By using the above expression and expanding in powers of $k_T/\mu$ one can find systematically power expansion of the cross section. The
UGDs are typically peaked for small values of $k_{TA}$, $k_{TB}$
thus the collinear contribution is the dominant one (the leading power
contribution). Therefore one should expect that the applicability
of (\ref{eq:HEF_xsec}) is in the high $p_{T}$ domain.

Let us  make now a few   comments about the HEF. First, concerns the collinear
limit of (\ref{eq:HEF_xsec}). One would expect that for large $p_{T}$
the cross section $d\sigma/dp_{T}$ calculated converges to  the collinear
one (\ref{eq:minijets1}). Performing the expansion (\ref{eq:HEF_coll_lim_})
and retaining the first collinear contribution only we are left with
integrals in (\ref{eq:HEF_xsec}) of the type
\begin{equation}
\int^{k_{T\mathrm{max}}^{2}}dk_{TA}^{2}\,\mathcal{F}_{g^{*}/A}\left(x_{A},k_{TA},\mu\right)\equiv f_{g/A}^{\left(k_{T\mathrm{max}}\right)}\left(x_{A},\mu\right)\,,
\end{equation}
where $k_{T\mathrm{max}}$ is the upper bound on $k_{T}$ which in
practice is constrained by the grid size of the UGDs or specific kinematic
cuts. The point is that the function $f_{g/A}^{\left(k_{T\mathrm{max}}\right)}$
is in general not exactly a collinear gluon PDF, which is defined
as
\begin{equation}
f_{g/A}\left(x_{A},\mu\right)=f_{g/A}^{\left(\mu\right)}\left(x_{A},\mu\right)\,.
\end{equation}
Thus, we will overshoot the collinear result if the hard scale $\mu$
is not too large and the UGDs do not fall very rapidly with $k_{T}$.
In other words, the convergence to the collinear result for finite
$\mu$ is rather weak. The remedy could be to set $k_{T\mathrm{max}}=\mu$,
but this not inherent part of the HEF. Let us illustrate the above
with a concrete and practical example. According to the Kimber-Martin-Ryskin
(KMR) prescription \citep{Kimber2000a,Kimber:2001sc} (actually its
commonly used simplified form), an UGD can be constructed from a collinear PDF as
follows:
\begin{equation}
\mathcal{F}_{g^{*}/H}\left(x,k_{T},\mu\right)=\frac{\partial}{\partial k_{T}^{2}}\left[f_{g/H}\left(x,k_{T}\right)T_{g}\left(k_{T},\mu\right)\right]\,,\label{eq:KMR}
\end{equation}
where $T_{g}$ is the Sudakov form factor. We see that the $k_{T\mathrm{max}}$
has to be equal to $\mu$ in order to recover $f_{g/H}$ upon integration
over $k_{T}$.

In order to address another possible issue of HEF, let us consider
the cross section as a function of the jet disbalance $K_{T}$, $d\sigma/dK_{T}$.
It can be calculated within HEF using (\ref{eq:HEF_xsec}). Let us
now find  the collinear limit of $d\sigma/dK_{T}$. It is easy
to see, that it will not converge to the DDT formula (\ref{eq:DDT1}).
This is not necessarily a problem, as the natural domains of
applicability of HEF formula and DDT are very different. Nevertheless,
it would be interesting to have a formula which includes subleading
powers of $K_{T}$ while possessing the leading twist limit given
by (\ref{eq:DDT1}). We shall construct such a formula in the next
subsection.

Finally, let us mention that in practical applications it is convenient
to use Monte Carlo programs to generate various observables for jets,
instead of using the formulae like e.g. (\ref{eq:HEF_xsec}). Thus
in our study we use an implementation of HEF in a computer program
\citep{Kotko_LxJet}
 which relies on the $\mathsf{foam}$ adaptive Monte Carlo \citep{Jadach:2002kn}. 
It allows to generate partonic events (`weighted' or `unweighted'), store them and make 
further analysis in a convenient way. No parton shower or hadronization 
is done in the current version. Let us however mention that the
$k_T$ dependence of gluon distributions acts much like the  initial state parton shower
(see e.g. \citep{vanHameren:2014ala,Bury2016}).

%%%%%%%%%%%%%%%%%%%%%%%%%%%%%%%%%%%%%%%%%%%%
\subsection{Extension of DDT beyond leading power}
\label{sub:IDDT}

In order to make our analysis as complete as possible, we will construct
now a version of HEF which in the leading power limit reduces to the
DDT formula (\ref{eq:DDT1}) for the dijet disbalance spectrum. The
goal of doing this is to use  broad spectrum of models with internal
gluon $k_{T}$. In HEF described in the previous subsection
the $k_{T}$ dependent UGDs take into account two ladders of initial
state emissions for each colliding hadron; it is most transparent when
UGDs are considered within the KMR approach (\ref{eq:KMR}) (see Fig.~\ref{fig:HEF}A).
There are no final state emission ladders in HEF, whilst the DDT formula
(\ref{eq:DDT1}) has a one ladder attached to each leg of the hard
process, including the final state lines (Fig.~\ref{fig:HEF}B) \citep{Dokshitzer1980}.
Of course, the DDT formula is the leading twist expression, on the
contrary to HEF. Below, we shall construct a HEF-based model which
has a similar philosophy to the DDT, but includes power corrections.

%%%%%%%%%%%%%%%%
\begin{figure}
\begin{centering}
\parbox{0.27\textwidth}{A)\\\includegraphics[width=0.27\textwidth]{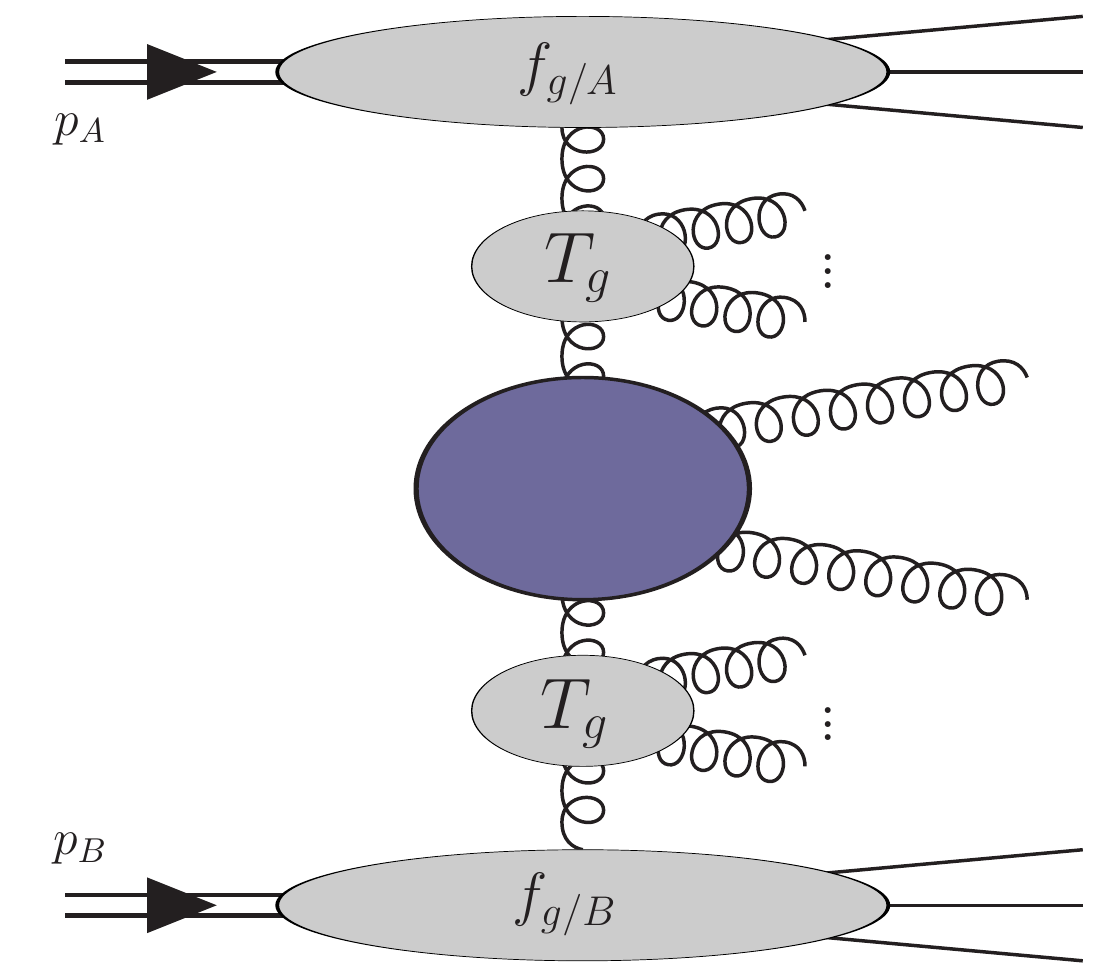}}\\ \parbox{0.92\textwidth}{B)\\\includegraphics[width=0.92\textwidth]{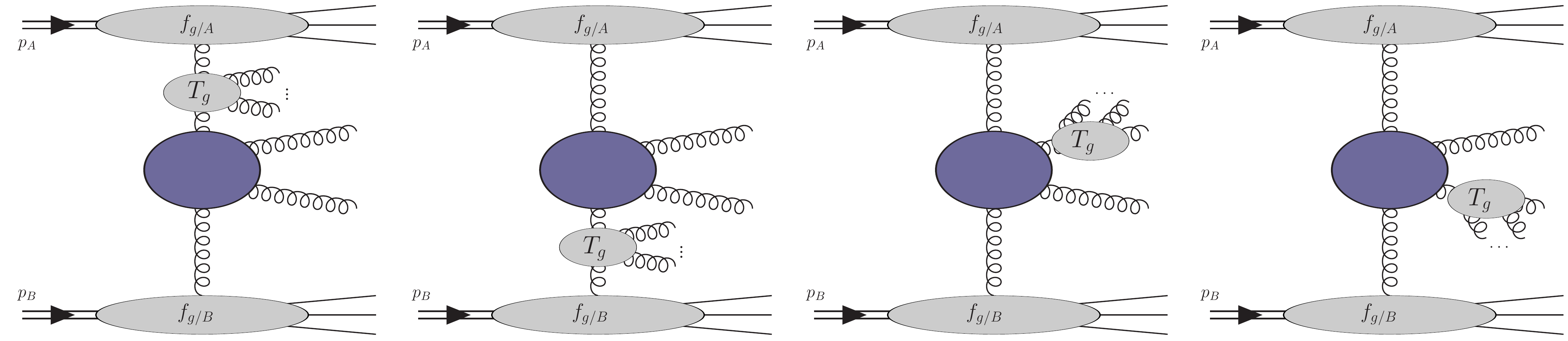}}
\par\end{centering}

\caption{A) In HEF with the KMR prescription (\ref{eq:KMR}) the $k_{T}$ of
initial state gluons on both sides is produced by the gluon PDF and
the Sudakov form factor B) In the leading twist DDT formula of Eq.~(\ref{eq:DDT1})
one ladder of emissions is associated with each leg of the hard process.
\label{fig:HEFDDT}}
\end{figure}
%%%%%%%%%%%%%%%

Let us first write (\ref{eq:DDT1}) as
\begin{multline}
\frac{d\sigma_{2\mathrm{jet}}}{dK_{T}^{2}}=2\int\frac{dx_{A}}{x_{A}}\,\frac{dx_{B}}{x_{B}}\, d\hat{\sigma}_{gg\rightarrow gg}\left(x_{A},x_{B};\mu^{2}\right)\\
\times\Bigg\{\frac{\partial}{\partial K_{T}^{2}}\left[f_{g/A}\left(x_{A};K_{T}^{2}\right)T_{g}\left(K_{T}^{2},\mu^{2}\right)\right]f_{g/B}\left(x_{B};K_{T}^{2}\right)T_{g}^{3}\left(K_{T}^{2},\mu^{2}\right)\\
+f_{g/A}\left(x_{A};K_{T}^{2}\right)f_{g/B}\left(x_{B};K_{T}^{2}\right)T_{g}^{3}\left(K_{T}^{2},\mu^{2}\right)\frac{\partial}{\partial K_{T}^{2}}T_{g}\left(K_{T}^{2},\mu^{2}\right)\Bigg\}\,,\label{eq:DDTHEF_1}
\end{multline}
where we have used the symmetry with respect to exchange of hadrons
$A\leftrightarrow B$ (this gives a factor of $2$). Basing on the
above and using (\ref{eq:KMR}) we now define
\begin{equation}
\frac{d\sigma_{2\mathrm{jet}}^{\left(\mathrm{IDDT}\right)}}{dK_{T}^{2}}=\frac{d\sigma_{2\mathrm{jet}}^{\left(\mathrm{IS}\right)}}{dK_{T}^{2}}+\frac{d\sigma_{2\mathrm{jet}}^{\left(\mathrm{FS}\right)}}{dK_{T}^{2}}\,,\label{eq:IDDT_1}
\end{equation}
where the `initial state' contribution is
\begin{multline}
\frac{d\sigma_{2\mathrm{jet}}^{\left(\mathrm{IS}\right)}}{dK_{T}^{2}}=2\int\frac{dx_{A}}{x_{A}}\,\frac{dx_{B}}{x_{B}}\int d^{2}K_{T}\, d\hat{\sigma}_{g^{*}g\rightarrow gg}\left(x_{A},x_{B},\vec{K}_{T};\mu^{2}\right)\\
\mathcal{F}_{g^{*}/A}\left(x_{A},K_{T},\mu\right)f_{g/B}\left(x_{B},K_{T}^{2}\right)T_{g}^{3}\left(K_{T}^{2},\mu^{2}\right)\,,\label{eq:IDDT_IS}
\end{multline}
while the `final state' contribution is
\begin{multline}
\frac{d\sigma_{2\mathrm{jet}}^{\left(\mathrm{FS}\right)}}{dK_{T}^{2}}=2\int\frac{dx_{A}}{x_{A}}\,\frac{dx_{B}}{x_{B}}\int d^{2}K_{T}\, d\hat{\sigma}_{gg\rightarrow gg^{*}}\left(x_{A},x_{B},\vec{K}_{T};\mu^{2}\right)\\
\times f_{g/A}\left(x_{A};K_{T}^{2}\right)f_{g/B}\left(x_{B};K_{T}^{2}\right)T_{g}^{3}\left(K_{T}^{2},\mu^{2}\right)\mathcal{T}_{g}\left(K_{T}^{2},\mu^{2}\right)\,.\label{eq:IDDT_FS}
\end{multline}
We have defined the final state transverse momentum distribution as
\begin{equation}
\mathcal{T}_{g}\left(K_{T}^{2},\mu^{2}\right)=\frac{\partial}{\partial K_{T}^{2}}T_{g}\left(K_{T}^{2},\mu^{2}\right)\,.\label{eq:FS_distribution}
\end{equation}
The Sudakov form factor we use is given by the following formula \citep{Dokshitzer1980}:
\begin{equation}
T_{g}\left(k_{T}^{2},\mu^{2}\right)=\exp\left\{-\int_{k_{T}^{2}}^{\mu^{2}}\frac{dp_{T}^{2}}{p_{T}^{2}}\,\int_{\Delta}^{1}dz\,\frac{\alpha_{s}\left(p_{T}^{2}\right)}{2\pi}\left[ (1-z)P_{gg}(z,\Delta )+N_f P_{qg}\left(z\right)\right]\right\}\,,\label{eq:SudakovFF}
\end{equation}
where
\begin{equation} 
P_{gg}(z,\Delta) = 2C_A \left( \frac{z}{1-z+\Delta} + \frac{1-z}{z} + z(1-z) \right)\,,
\end{equation}
\begin{equation}
P_{qg}(z) = \frac{1}{2}\left( z^2 + (1-z)^2 \right)\,.
\end{equation}
The cutoff parameter $\Delta$ is taken to be $\Delta=k_{T}^{2}/\mu^{2}$.
We note that there are various forms of the cutoff parameter in the literature, see for example \citep{Kimber2000a}. 
The partonic cross section $d\hat{\sigma}_{g^{*}g\rightarrow gg}$
is calculated in the exact same way as in the hybrid HEF described before,
taking into account the gauge invariant off-shell amplitude with only
one leg being off-shell
\begin{equation}
d\hat{\sigma}_{g^{*}g\rightarrow gg}=\frac{1}{2x_{A}x_{B}S}\,\frac{1}{2}\,\left|\overline{\mathcal{M}}\right|_{g^{*}g\rightarrow gg}^{2}d\mathrm{PS}\,,
\end{equation}
where $\left|\overline{\mathcal{M}}\right|_{g^{*}g\rightarrow gg}^{2}$
was calculated for instance in \citep{Deak:2009xt} and using helicity
amplitudes in \citep{VanHameren2012}. It reads
\begin{equation}
\left|\overline{\mathcal{M}}\right|_{g^{*}g\rightarrow gg}^{2}=\frac{g^{4}}{2\pi}\,\frac{N_{c}^{2}}{N_{c}^{2}-1}\,\frac{\left(\overline{s}^{4}+\overline{t}_{1}^{4}+\overline{u}_{1}^{4}\right)\left(s\overline{s}+t\overline{t}_{1}+u\overline{u}_{1}\right)}{s\overline{s}t\overline{t}_{1}u\overline{u}_{1}}\,,\label{eq:RGGG}
\end{equation}
with the invariants defined in (\ref{eq:Invs1}),(\ref{eq:Invs2}),
but now $k_{TA}\equiv K_{T}$. In the above form the on-shell limit
is visible right away: when $K_{T}\rightarrow0$ we have $\overline{s}\rightarrow s$,
$\overline{t}_{1}\rightarrow t$, $\overline{u}_{1}\rightarrow u$
and we get the known collinear result.

The partonic cross section with final state off-shell  $d\hat{\sigma}_{gg\rightarrow gg^{*}}$
is a new construction and to our knowledge does not exist in the literature.
It is constructed from the  gauge invariant off-shell amplitude with the
final state particle  taken off-shell
\begin{equation}
d\hat{\sigma}_{gg\rightarrow gg^{*}}=\frac{1}{2x_{A}x_{B}S}\,\frac{1}{2}\,\left|\overline{\mathcal{M}}\right|_{gg\rightarrow gg^{*}}^{2}d\mathrm{PS}\left(K_{T}^{2}\right)\,,
\end{equation}
where $d\mathrm{PS}\left(K_{T}^{2}\right)$ is the two-particle phase
space to produce a spacelike state with mass $K_{T}^{2}$. Let us
now explain, how the amplitude $\left|\overline{\mathcal{M}}\right|_{gg\rightarrow gg^{*}}^{2}$
is calculated, as it differs from the standard way the HEF amplitudes
are obtained. 

\begin{figure}
\begin{centering}
\parbox{0.25\textwidth}{A)\\\includegraphics[width=0.25\textwidth]{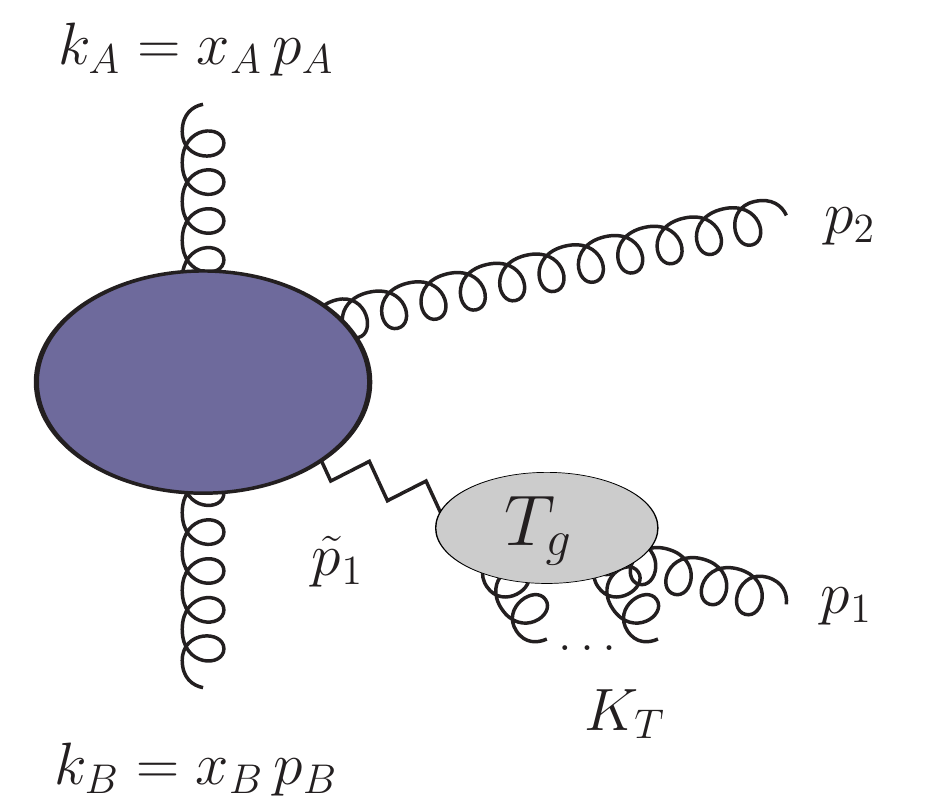}}$\,\,\,\,\,\,\,\,\,\,\,\,$\parbox{0.7\textwidth}{B)\\\includegraphics[width=0.7\textwidth]{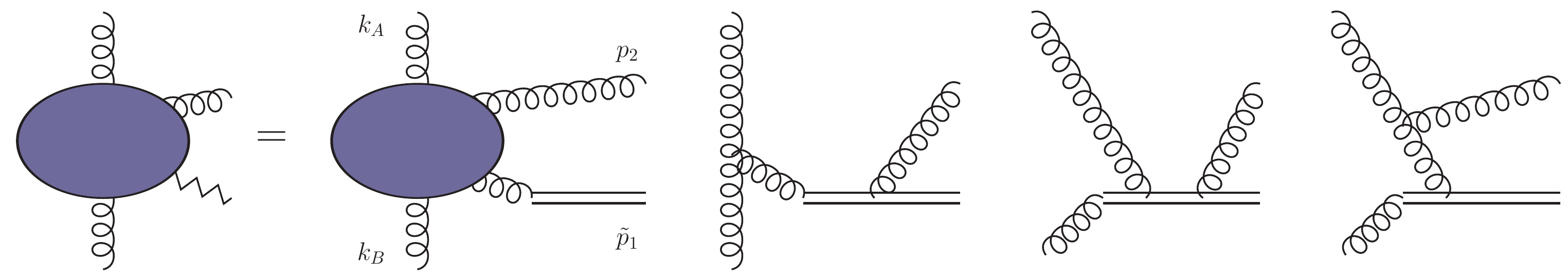}}
\par\end{centering}

\caption{A) Momentum assignment in the final state contribution to (\ref{eq:IDDT_1});
the final state momentum $\tilde{p}_{1}$ is off-shell. B) Diagrams
contributing to the gauge invariant final state off-shell process;
the Wilson line slope is given by the vector $p_{1}$ so that $\tilde{p}_{1}\cdot p_{1}=0$.
\label{fig:HEFDDT-1}}
\end{figure}

First consider the kinematics involved in (\ref{eq:IDDT_FS}), see
Fig.~\ref{fig:HEFDDT-1}A. The idea is that first the two states
are produced: an on-shell gluon $p_{2}$ and the off-shell one with
momentum $\tilde{p}_{1}$, $\tilde{p}_{1}^{2}=-K_{T}^{2}$. Next,
this off-shell dressed gluon undergoes emissions described by $\mathcal{T}_{g}$
defined in (\ref{eq:FS_distribution}) and becomes on-shell $p_{1}=\tilde{p}_{1}+K_{T}$,
$p_{1}^{2}=0$. The first stage happens via the off-shell gauge invariant
process $g\left(k_{A}\right)g\left(k_{B}\right)\rightarrow g^{*}\left(\tilde{p}_{1}\right)g\left(p_{2}\right)$
calculated from diagrams depicted in Fig.~\ref{fig:HEFDDT-1}B according
to the prescription of \citep{Kotko2014a}. As the Wilson line slope
we take here the momentum $p_{1}$ (not the eikonal vectors $p_{A,B}$,
as it was the case for HEF), so that 
\begin{equation}
\tilde{p}_{1}\cdot p_{1}=0,\,\,\,\, K_{T}\cdot p_{1}=0\,.
\end{equation}
The result reads
\begin{equation}
\left|\overline{\mathcal{M}}\right|_{gg\rightarrow gg^{*}}^{2}=\frac{g^{4}}{2}\,\frac{N_{c}^{2}}{N_{c}^{2}-1}\,\frac{\left(\tilde{s}^{4}+\tilde{t}^{4}+\tilde{u}^{4}\right)\left(s\tilde{s}+t\tilde{t}+u\tilde{u}\right)}{s\tilde{s}t\tilde{t}u\tilde{u}}\,.\label{eq:GGGR}
\end{equation}
It looks basically the same as (\ref{eq:RGGG}) but now 
\begin{equation}
\tilde{s}=\left(p_{2}+p_{1}\right)^{2},\,\,\tilde{t}=\left(x_{A}p_{A}-p_{1}\right)^{2},\,\,\tilde{u}=\left(x_{B}p_{B}-p_{1}\right)^{2}\,,
\end{equation}
\begin{equation}
s=\left(p_{2}+\tilde{p}_{1}\right)^{2},\,\, t=\left(x_{A}p_{A}-\tilde{p}_{1}\right)^{2},\,\, u=\left(x_{B}p_{B}-\tilde{p}_{1}\right)^{2}\,.
\end{equation}

It is important to mention that, by construction, the maximal allowed
value of $K_{T}$ is $K_{T\mathrm{max}}=\mu$. It is easy to see that
then, in the leading power approximation we recover both the collinear
result (\ref{eq:CollFact1}) and the DDT formula (\ref{eq:DDT1}).
In what follows we shall abbreviate the new model as IDDT (an `improved DDT').

\begin{figure}
\begin{centering}
\parbox{0.49\textwidth}{A)\\\includegraphics[width=0.49\textwidth]{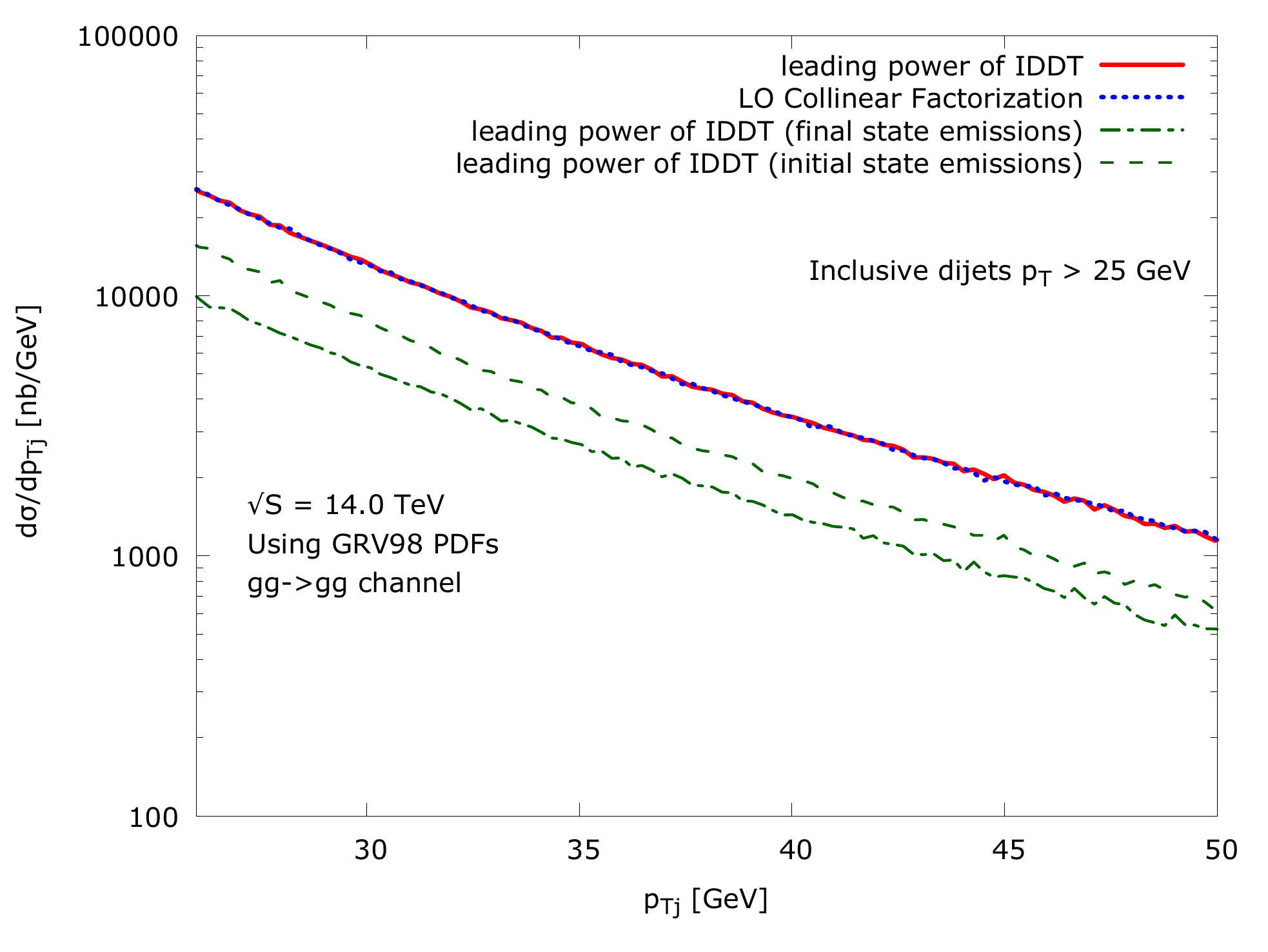}}$\,\,\,\,\,\,\,$\parbox{0.49\textwidth}{B)\\\includegraphics[width=0.49\textwidth]{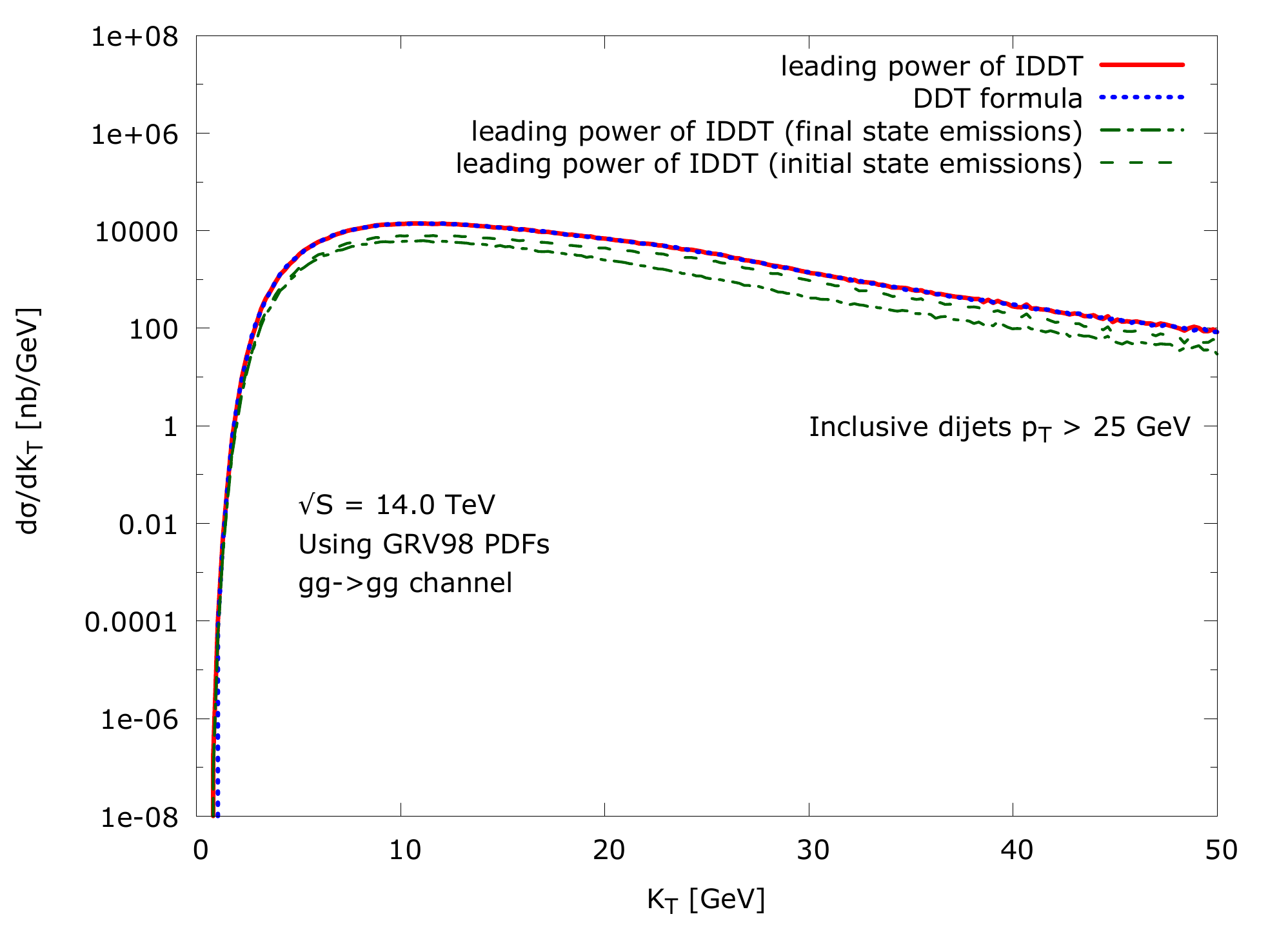}}
\par\end{centering}

\caption{A) The leading power limit of the IDDT formula (\ref{eq:IDDT_1})
for jet $p_{T}$ spectrum in comparison with the LO collinear factorization.
We see that it has the correct collinear limit as the solid line (IDDT)
is on the top of the dotted line (collinear). We show also final state
and initial state contributions of IDDT. B) The same but for the spectrum
of jet disbalance and in comparison with the DDT formula. \label{fig:HEFDDT_test}}
\end{figure}

We have implemented the IDDT approach in a computer program \citep{Kotko_LxJet}  and in  Fig.~\ref{fig:HEFDDT_test} we show the results of the consistency checks we have performed (for a detailed
description of the setup and cuts see the next section). First, we compare the
leading power limit of the IDDT  with the collinear result for the $p_{T}$
spectrum. We see (Fig.~\ref{fig:HEFDDT_test}A) that they match ideally.
We also show separately the contributions from the `initial state'
(\ref{eq:IDDT_IS}) and `final state' cross sections. Next, we compare
the spectrum in the jet disbalance $K_{T}$ with the one obtained
from the DDT formula (Fig.~\ref{fig:HEFDDT_test}B) and find a perfect agreement.
Thus we have gained exactly the properties we wanted, that is the
formula (\ref{eq:IDDT_1}) has the collinear and the DDT limits at
leading power.

In the end, let us stress that the above construction is a model of
higher twists, not a strict QCD derivation. We have neglected all
the details concerning factorization and higher order corrections.
Our aim was to catch certain properties such a formula should have
in order to study their effect on minijets. 

%%%%%%%%%%%%%%%%%%%%%%%%%%%%%%%%%%%%%%%%%%%%
\section{Setup for numerical studies}
\label{sub:setup}

Before presenting the detailed  numerical results of the different minijet formulations, we shall first define the observable we are going to calculate as well as the kinematic cuts and details of the setups of the Monte Carlo programs.

The LO collinear jet formula (\ref{eq:CollFact1}) or (\ref{eq:MinijetReg})
describes the production of exactly two jets. In more realistic simulations
we deal with multi-parton configurations and a more careful definition
of two-jet cross section is needed. This concerns simulations using
both $\mathsf{pythia}$ and Monte Carlo implementation of HEF or IDDT. In
the following, we will consider inclusive dijet cross section with
jets reconstructed using the anti-$k_{T}$ algorithm with certain
$p_{T\mathrm{min}}$ and $R=0.5$ (if not stated otherwise). We require
at least two jets to be above $p_{T\mathrm{min}}$. We do not order
the jets in their $p_{T}$, thus the spectra for both jets are identical.
We require both jets to fit within $\left[-4,4\right]$ rapidity window.

We shall use three approaches: (i) $\mathsf{pythia}$, (ii) HEF as described in Subsection~\ref{sub:HEF}, (iii) IDDT constructed in Subsection~\ref{sub:IDDT}. For reference we sometimes use also the pure collinear formula (\ref{eq:minijets1}). The approaches (i), (ii) and (iii) will be used in the direct study of minijets in Section~\ref{sub:DirectMinijets}, while the indirect minijet study shall utilize models (i) and (ii).

The $\mathsf{pythia}$ generator has two disjoint modules: `soft QCD' and `hard QCD'. The first one is used when all produced particles have transverse momenta around or slightly above $p_{T0}$ cutoff. The second is suitable for high-$p_T$ particles. From the point of view of minijet model, they differ by the fact that in the `hard QCD' module the hardest binary collision does not have the suppression factor as in (\ref{eq:MinijetReg}).

Whenever we use $\mathsf{pythia}$ we use some non-standard settings
in order to make clean comparisons. First, we use only gluonic channel.
Second, we use LO GRV98 \citep{Gluck:1998xa} PDFs with matching LO
$\alpha_{s}$. The reason we use GRV98 instead of some more up-to-date
sets in that we will compare $\mathsf{pythia}$ calculations to HEF
with KMR in the low $p_{T}$ region. This requires that the PDF used
to construct KMR has to be defined for small enough scale, smaller
than $1\,\mathrm{GeV}$. As of today, this requirement is satisfied
only by GRV98 distribution. 

Above are the generic settings. The other settings concerning MPIs
or parton showers and hadronization will be determined when necessary.
In the description of the plots we shall use the following abbreviations: PS for
final state and initial state parton showers and HAD for hadronization.
In order to comply with the minijet formula we choose the hard scale
to be the average $p_{T}$ of jets.

In our analysis within HEF we will use several UGDs: (i) the KMR gluon
distribution \citep{Kimber2000a,Kimber:2001sc} given by (\ref{eq:KMR})
based on the GRV98 collinear PDF. Note, this is actually a prescription
of DDT; the genuine KMR prescription is much more complicated, but
traditionally (\ref{eq:KMR}) functions as KMR in the literature.
(ii) The Kwiecinski-Martin-Stasto (KMS) \citep{Kwiecinski:1997ee}
gluon distribution which supplements the BFKL equation with the DGLAP
corrections. More precisely it incorporates the kinematic constraint
to maintain the energy conservation and the nonsingular parts of the gluon-gluon
splitting function. This gluon distribution has been fitted to HERA $F_{2}$
data in \citep{Kutak:2012rf} and we will call this set KMS-HERA.
In \citep{Kotko:2015ksa} fits have been performed to the jet LHC
data (using however only gluonic part of the KMS equation). We shall
call this set KMS-LHC in what follows. (iii) The CCFM equation \citep{Ciafaloni:1987ur,Catani:1989yc,Catani:1989sg}
taken from \citep{Hautmann2014} and based on the computer code \citep{Hautmann:2014uua}.
We note that various CCFM sets differ between each other and thus
we are not making any conclusions regarding CCFM from our work. The
important point of the CCFM equation is that it encodes both the BFKL
and DGLAP limits through the angular ordering constraint. A very important difference between KMS and CCFM
evolution equations is that KMS does not depend on the hard scale
of the process. We shall see, that this feature is important for jet
studies. Similar to CCFM, the KMR approach does encode the hard scale
dependence through the Sudakov form factor. In fact, in certain limit the CCFM gluon
distribution can be reduced to the one of KMR \citep{Kwiecinski2002a}.

All numerical simulations for HEF are performed using the extension of the C++ program \citep{Kotko_LxJet} briefly described in the end of Section~\ref{sub:HEF}.

%%%%%%%%%%%%%%%%%%%%%%%%%%%%%%%%%%%%%%%%%%%%
\section{Direct study of minijet suppression}
\label{sub:DirectMinijets}

In the present section we directly study $p_{T}$ spectra of minijets,
i.e. inclusive dijets with $p_{T}\gtrsim 2\,\mathrm{GeV}$ using $\mathsf{pythia}$ and 
HEF/IDDT in the small $p_{T}$ region. In particular, we will
check whether the internal gluon $k_{T}$ can give a jet suppression
compliant with the minijet formula (\ref{eq:MinijetReg}). We have
already anticipated the result: the main contribution in  HEF/IDDT 
at low $p_T$ comes from the collinear region of small
$k_T$ which does not have the suppression factor built in.
We shall check this  through numerical analysis of the $p_T$ spectra.

\begin{figure}
\begin{centering}
\includegraphics[width=0.65\textwidth]{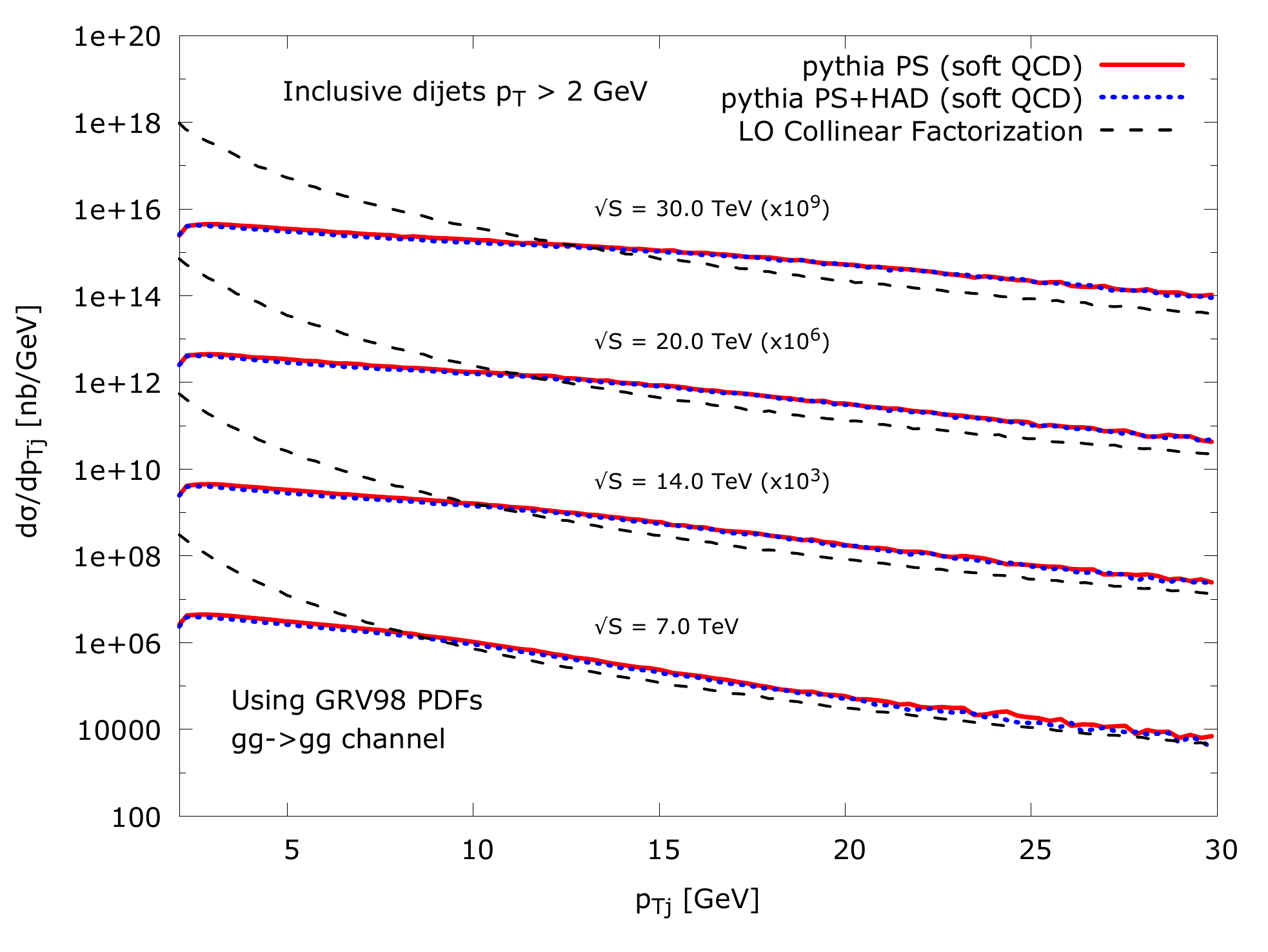}
\par\end{centering}

\caption{The suppression of minijet $p_{T}$ spectrum comparing to the LO collinear
factorization calculated with $\mathsf{pythia}$ using 'soft QCD'
simulation with parton shower and with/without hadronization. \label{fig:pythia_minijets_1}}
\end{figure}

Before we explore the  HEF, let us ask a question how the suppression
of the minijet spectrum encoded in (\ref{eq:MinijetReg}) looks in
a realistic model which implements it. To this end we use the
`soft QCD'  module of $\mathsf{pythia}$ suitable for small $p_T$ and calculate the inclusive
dijet spectra as described in the Section~\ref{sub:setup}. The result
is presented in Fig.~\ref{fig:pythia_minijets_1}. First, we indeed
see the suppression which means that every binary collision is generated
with the proper suppression factor (this is the feature of the `soft QCD'
module of $\mathsf{pythia}$; the `hard QCD' does not have this property).
Second, we observe an enhancement of the spectrum as compared to the  naive
$d\sigma'_{2\mathrm{jet}}/dp_{T}$ spectrum. This feature depends
on the jet cluster parameter $R$ and comes from the partons which originate 
in different hard collisions being clustered into one jet. This feature
does survive the hadronization, as seen in the figure. We shall further
discuss this in the next section (see Fig.~\ref{fig:events}). The
growing suppression with the CM energy is also visible.

\begin{figure}
\begin{centering}
\parbox{0.65\textwidth}{A)\\\includegraphics[width=0.65\textwidth]{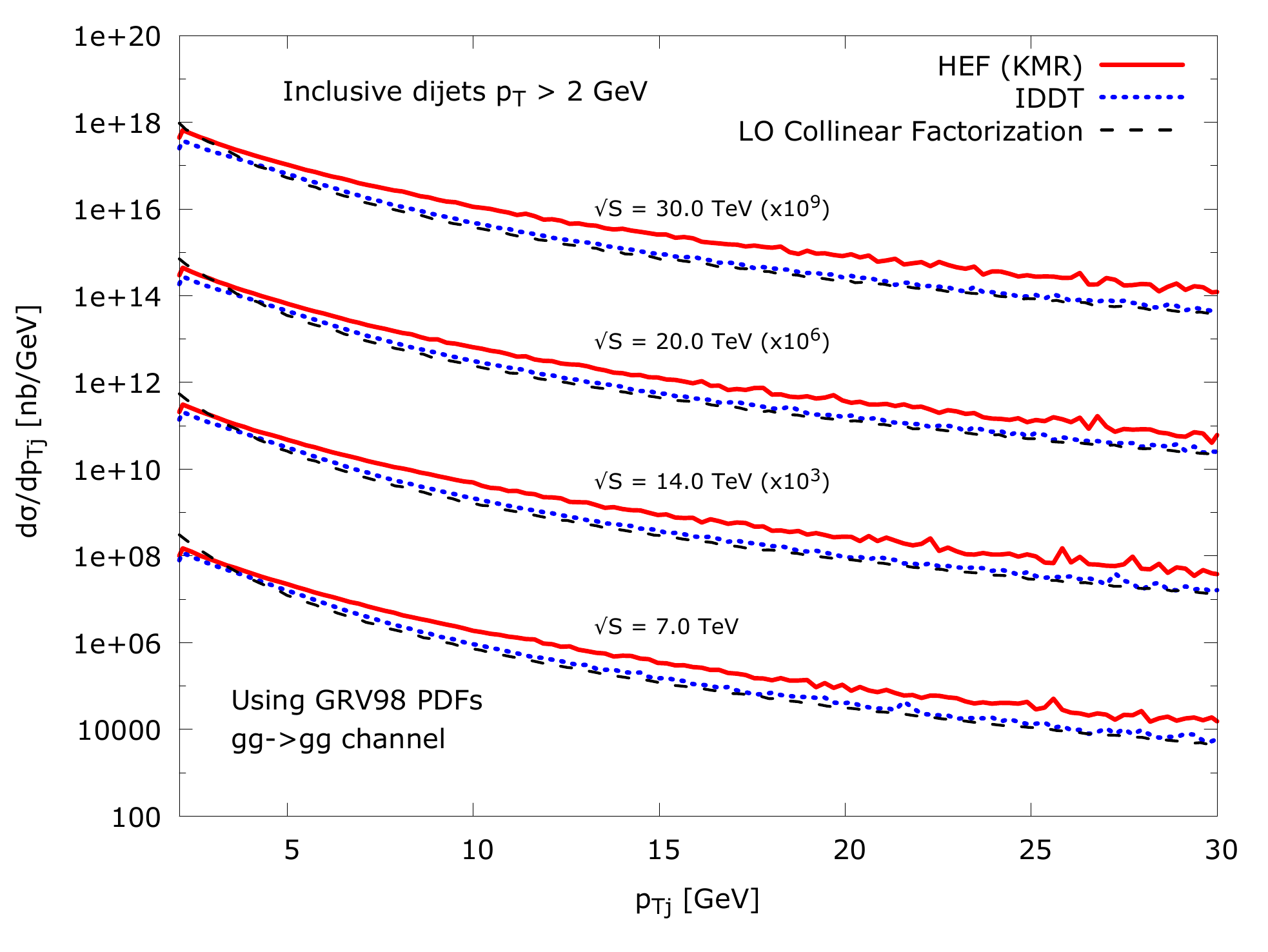}}\\ \parbox{0.65\textwidth}{B)\\\includegraphics[width=0.65\textwidth]{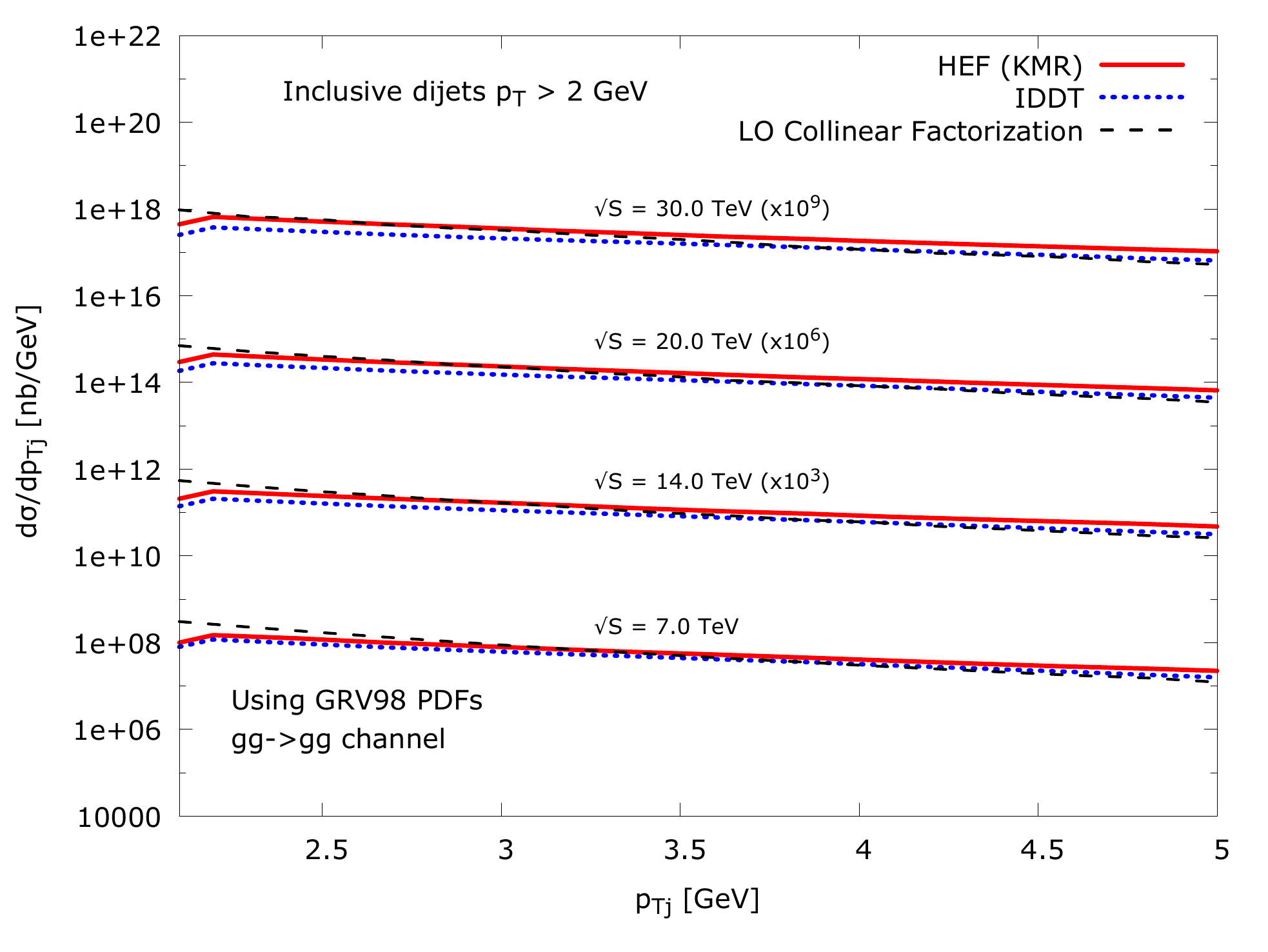}}
\par\end{centering}

\caption{A) The same as in Fig.~\ref{fig:pythia_minijets_1} but for HEF and
IDDT approaches. B) The zoom into the low-$p_{T}$ region of the top plot.
\label{fig:HEF_IDDT_minijets}}
\end{figure}

\begin{figure}
\begin{centering}
\parbox{0.65\textwidth}{A)\\\includegraphics[width=0.65\textwidth]{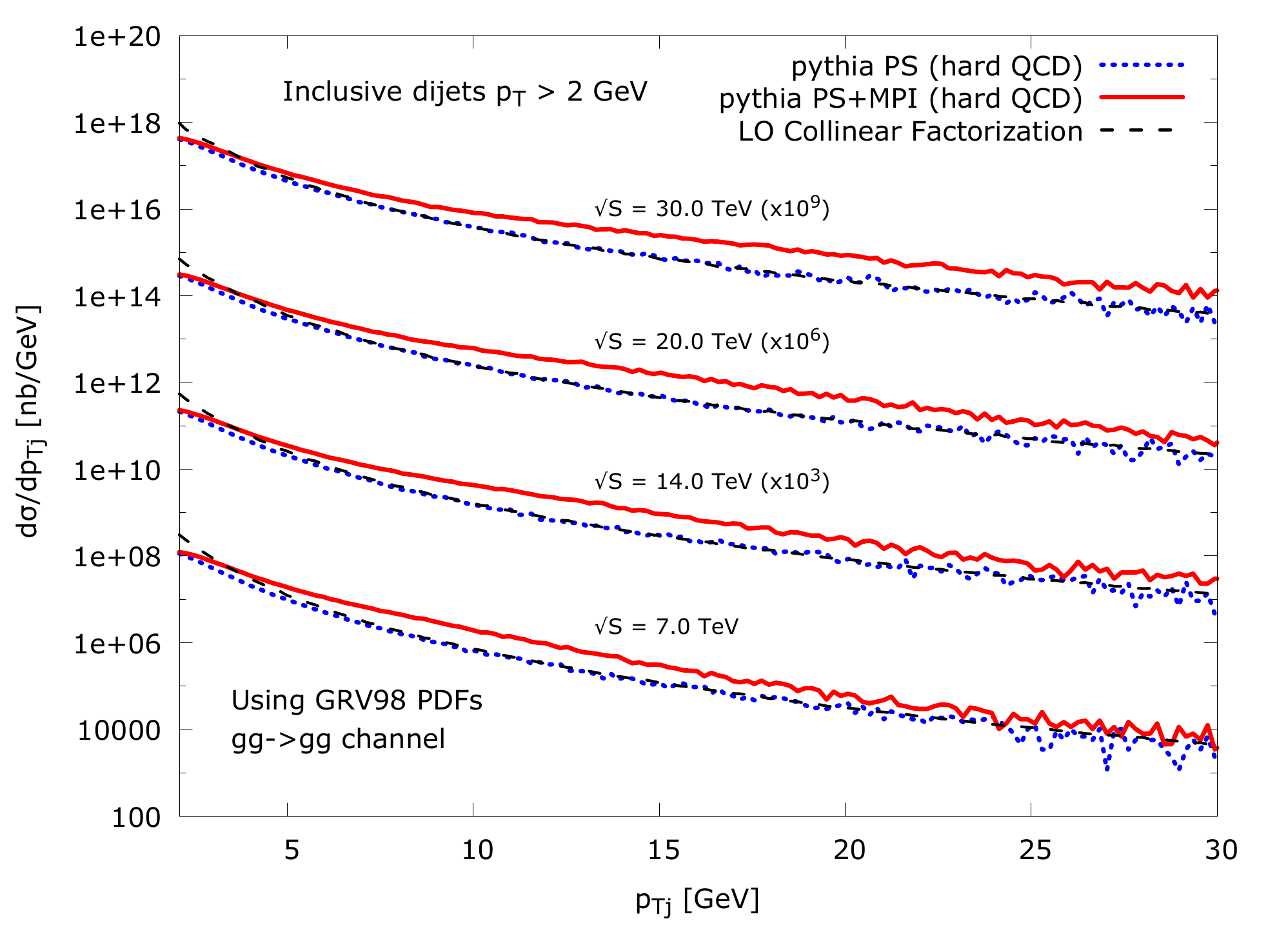}}\\ \parbox{0.65\textwidth}{B)\\\includegraphics[width=0.65\textwidth]{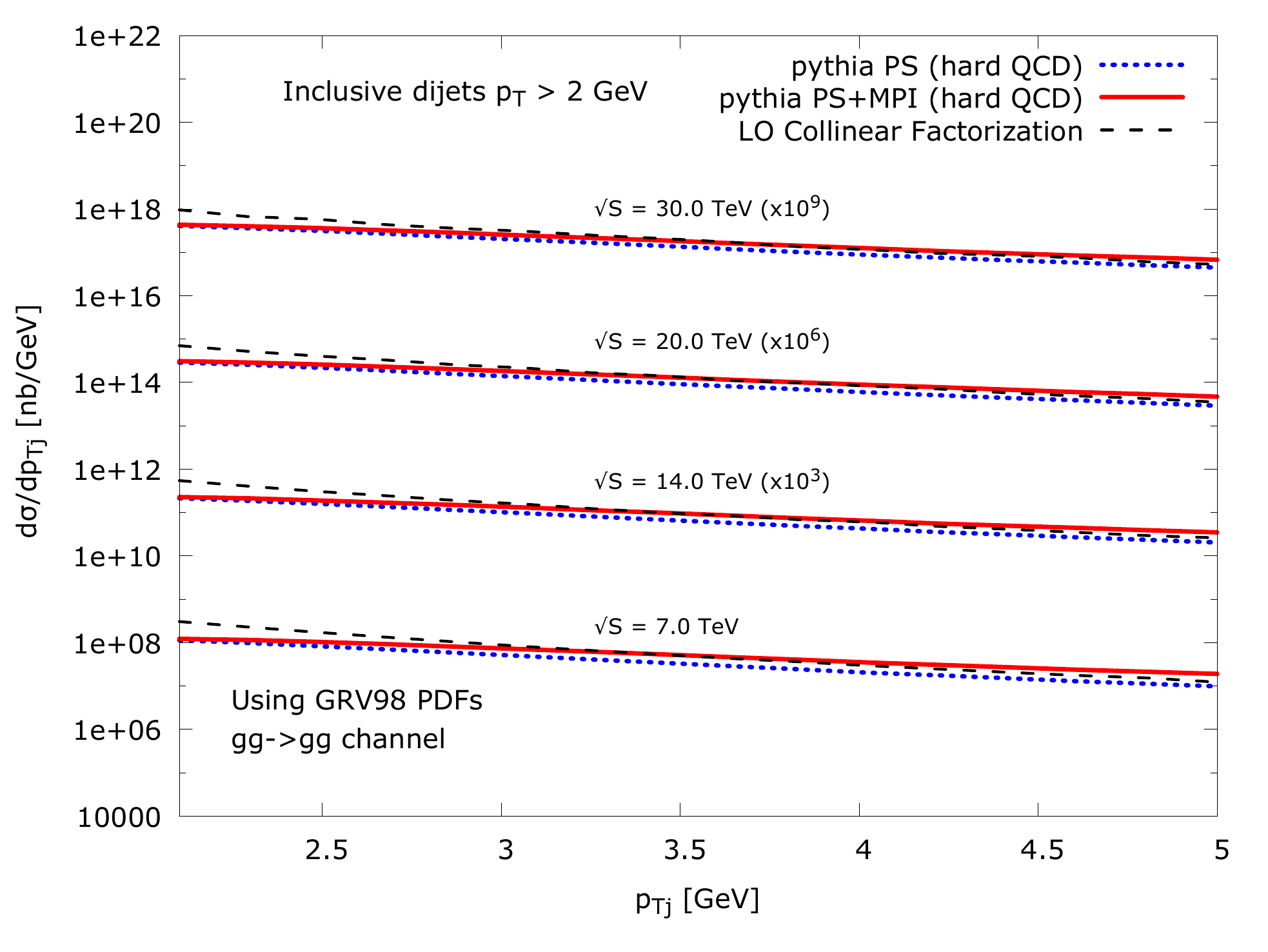}}
\par\end{centering}

\caption{A) The same as in Fig.~\ref{fig:pythia_minijets_1} but using $\mathsf{pythia}$
with `hard QCD' module. It formally is not applicable for low-$p_{T}$
but we use it on purpose to compare with HEF in this region (see the
main text for details). B) The zoom into the low-$p_{T}$ region of the top plot.
\label{fig:pythia_minijets}}
\end{figure}

Now we do the same  for HEF and IDDT models. Here we are interested
in a suppression with respect to the LO collinear factorization so we
have to use the consistent gluon distributions. Thus we use the KMR
based on GRV98 in HEF and we use the GRV98 itself in collinear factorization.
We show the results in Fig.~\ref{fig:HEF_IDDT_minijets}. First,
we see that the direct suppression of the spectrum due to the internal
gluon $k_{T}$ is very small comparing to the minijet model (the top
plot has the same horizontal scale as Fig.~\ref{fig:pythia_minijets_1},
while in the bottom plot we zoom the low-$p_{T}$ region). Second,
the suppression has the opposite energy dependence than (\ref{eq:pT0(s)}),
i.e. it becomes weaker when the energy is higher. This feature is present
in both models HEF and IDDT and is qualitatively the same. It is interesting
to compare this calculation to similar calculation made with $\mathsf{pythia}$
with the `hard QCD' module. Obviously, the `hard QCD' cannot formally be
used in the low-$p_{T}$ region, but technically it can be done and
this sheds some light on the interpretation of the results from Fig.~\ref{fig:HEF_IDDT_minijets}.
Namely, in the `hard QCD' module the hard process does not have any
$p_{T}$ regularization factor and we expect the results to exhibit
similar behavior to Fig.~\ref{fig:HEF_IDDT_minijets}. We show these
results in Fig.~\ref{fig:pythia_minijets}. By comparing them to Fig.~\ref{fig:HEF_IDDT_minijets} (in particular the bottom plots) we see that, qualitatively,
the behaviour is very similar meaning that indeed the hard process
in HEF (or IDDT) does not have the suppression of the kind (\ref{eq:MinijetReg}).

The HEF has however not only the leading power unsuppressed contribution,
but it has also power corrections which may exhibit a different behaviour.
These come from the tails in the transverse momentum of the UGDs.
Actually, one can think of a collision within HEF as  `multiple collisions'
weighted by a distribution of internal transverse momentum. This distribution
is peaked at small transverse momentum, thus the collinear contribution
(one of the multiple collision bundle) is dominant (the leading power
contribution), and it is not suppressed. Further `collisions' for
larger internal transverse momenta are less important as the internal transverse momentum
distribution falls off quickly. However, those sub-leading power corrections
may exhibit different energy behavior, as they are made of many soft
emissions. We shall investigate this point in the next section.

%%%%%%%%%%%%%%%%%%%%%%%%%%%%%%%%%%%%%%%%%%%%
\section{Indirect study of minijets}
\label{sub:IndirectMinijets}

In the previous section we saw that the internal $k_T$ flowing into the hard process as in HEF/IDDT approaches does
not give the suppression of dijet production compliant with the minijet model (\ref{eq:MinijetReg}). This is simply because the off-shell $2\rightarrow 2$ hard process is dominated by the leading power contribution for which the effect of internal gluon $k_T$ is negligible.
Thus, the next question we ask is about the relation of power corrections created in minijet model with MPIs and power corrections rendered in HEF. We stress that we consider here inclusive \textit{dijet} production. The inclusive jet cross section would not be affected by MPIs.

In order to support the above statement that minijet model with MPI can generate similar corrections to HEF let us have again a look at the $p_T$ spectra.
We can see from Fig.~\ref{fig:HEF_IDDT_minijets}
and Fig.~\ref{fig:pythia_minijets} that within HEF and $\mathsf{pythia}$
with MPIs there is an enhancement for larger $p_{T}$'s comparing
to the collinear result. Both IDDT and $\mathsf{pythia}$ without
MPIs do not have this feature. The reason for the IDDT model to
be quickly convergent to the collinear result is because, by construction,
we do not allow $K_{T}$ to be bigger than the hard scale $\mu$.
On the contrary, in HEF $K_{T}$ may be anything allowed by the jet
kinematics. Thus we may draw a conclusion that both MPI corrections
to the hard process in $\mathsf{pythia}$ and power corrections in
HEF may have similar components. In this section we will study this
point.
Because IDDT does not allow for sizeable power corrections we will not consider it in this section anymore.

\begin{figure}
\begin{centering}
\parbox{0.49\textwidth}{A)\\\includegraphics[width=0.49\textwidth]{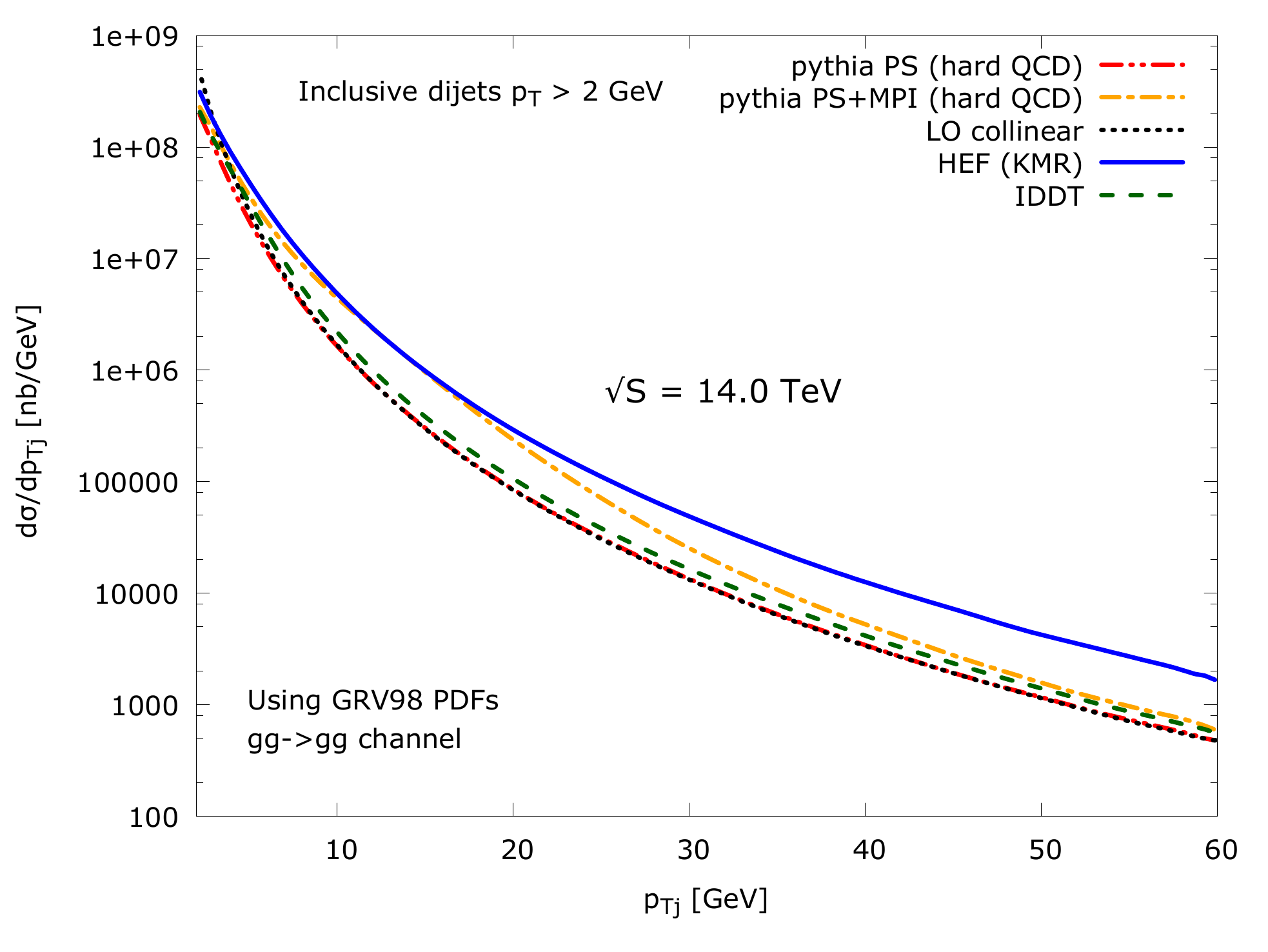}}$\,\,\,\,\,\,\,$\parbox{0.49\textwidth}{B)\\\includegraphics[width=0.49\textwidth]{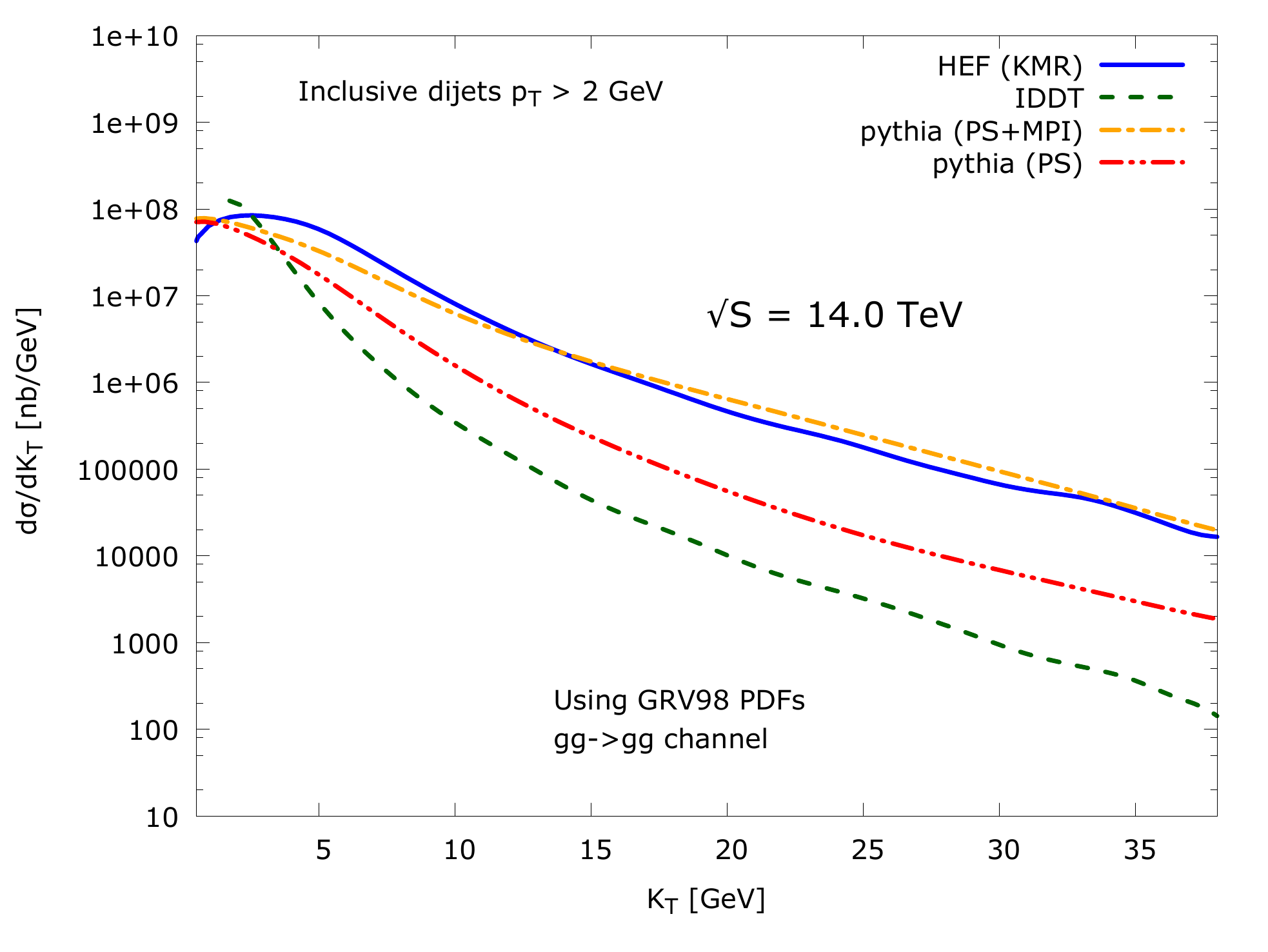}}
\par\end{centering}

\begin{centering}
\parbox{0.49\textwidth}{C)\\\includegraphics[width=0.49\textwidth]{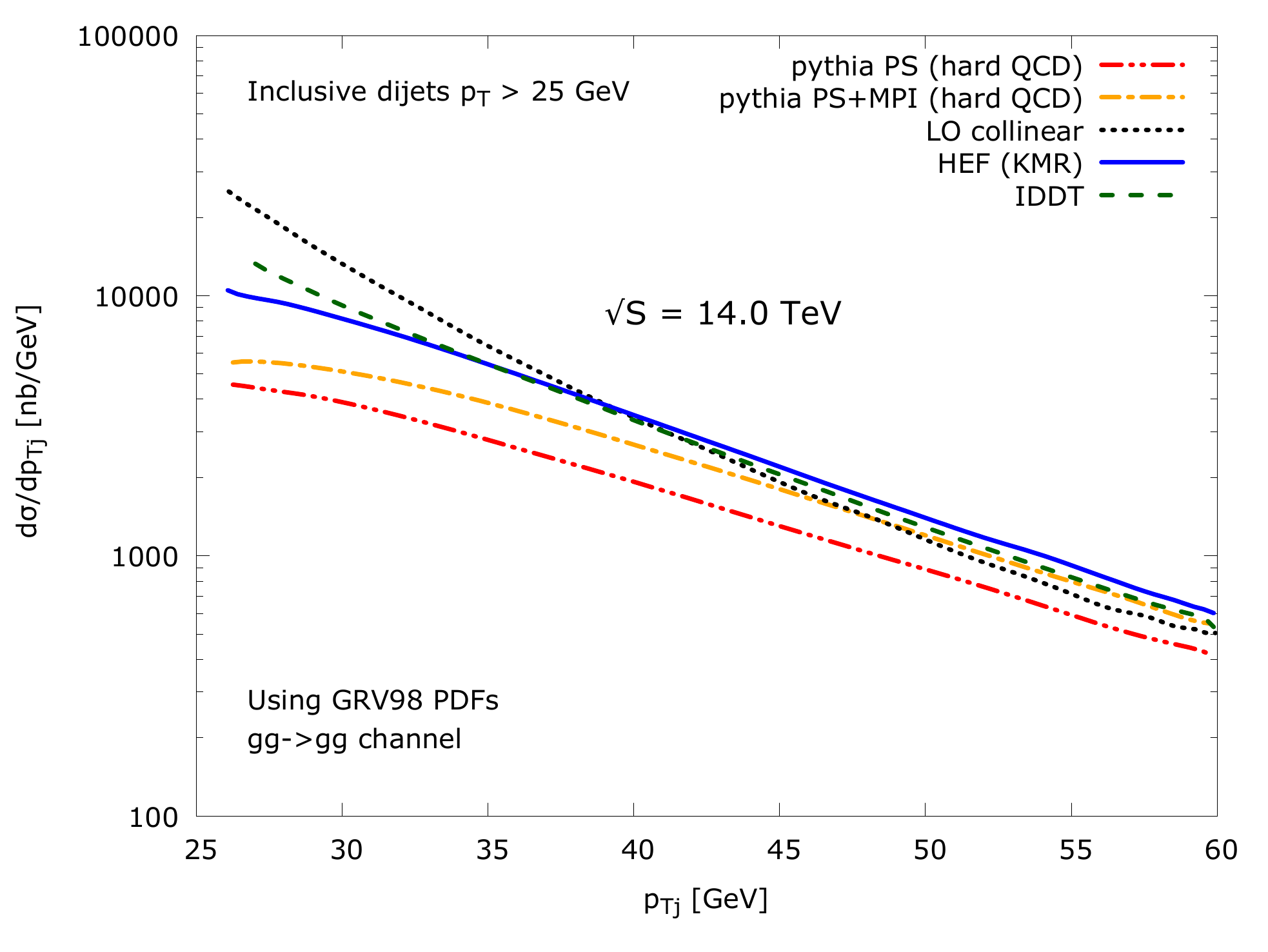}}$\,\,\,\,\,\,\,$\parbox{0.49\textwidth}{D)\\\includegraphics[width=0.49\textwidth]{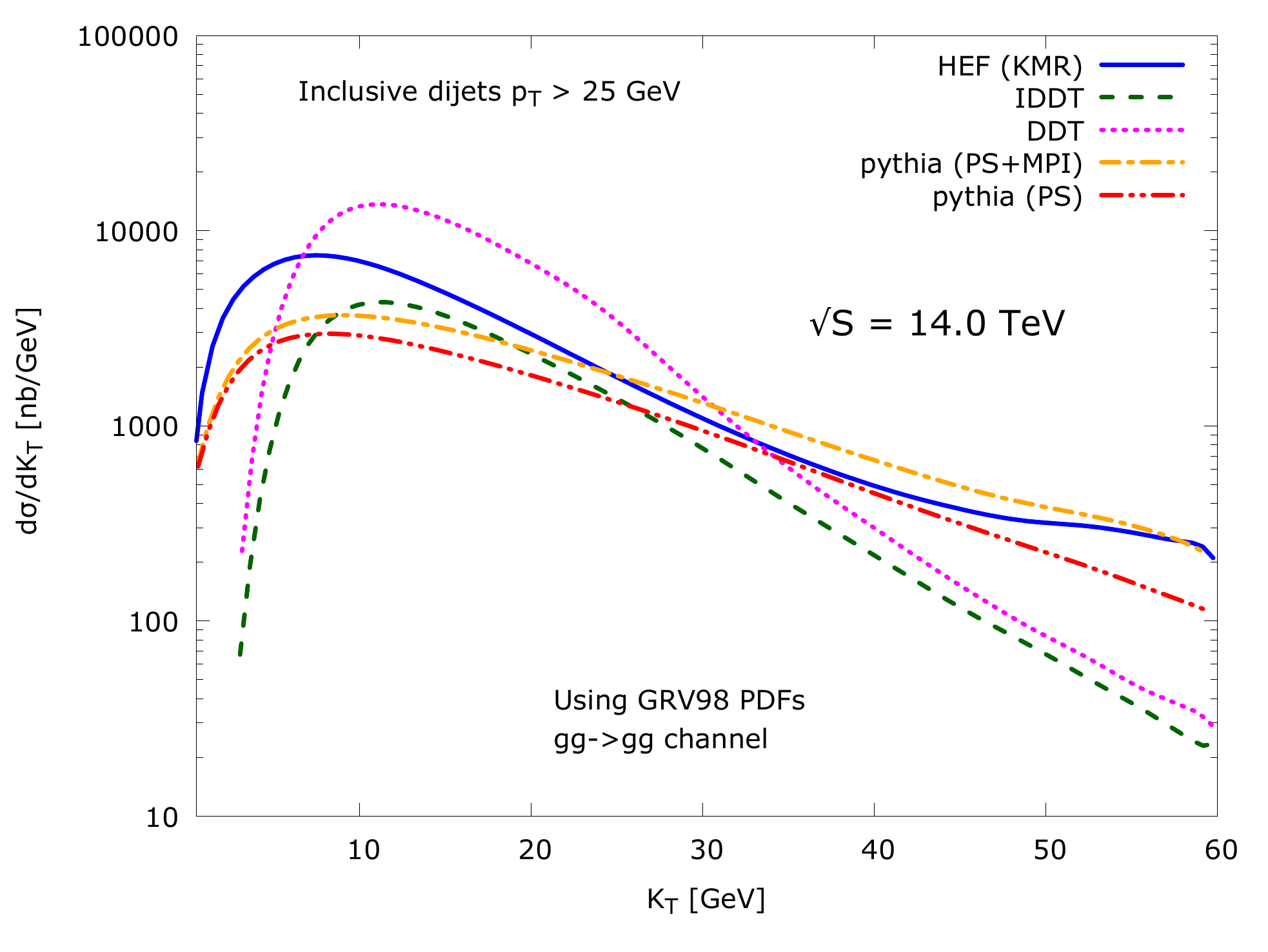}}
\par\end{centering}

\caption{Comparison of the $p_{T}$ spectra (left column) from various approaches
and the spectrum of the dijet disbalance $K_{T}$ (right column).
The top row is for actual minijets, i.e. $p_{T}>2\,\mathrm{GeV}$
while the bottom row is for $p_{T}>25\,\mathrm{GeV}$. \label{fig:pT_cmp_1}}
\end{figure}

First we take a closer look at the direct comparison of Figs.~\ref{fig:HEF_IDDT_minijets}-\ref{fig:pythia_minijets}
for certain CM energy, say $\sqrt{S}=14\,\mathrm{TeV}$ (Fig.~\ref{fig:pT_cmp_1}A).
We see, that for larger $p_{T}$'s both $\mathsf{pythia}$ with MPIs and
HEF start to exhibit indeed a similar enhancement comparing to the collinear result, but further in
$p_{T}$ the $\mathsf{pythia}$ spectrum converges to the collinear
one while HEF converges much more slowly. We calculate also the spectra
of the dijet momentum disbalance $K_{T}$ (Fig.~\ref{fig:pT_cmp_1}B).
We see that the MPIs in $\mathsf{pythia}$ model produce higher tail
of the $K_{T}$ spectrum, which in addition is close to the one from
HEF. These calculations are interesting, but as discussed before they
are extrapolated beyond the natural domain of the applicability of
the models used; both $\mathsf{pythia}$ with the `hard QCD' module and
HEF require rather high $p_{T}$ to be present. 
Let us remind that we used $\mathsf{pythia}$ with the `hard QCD' module in order to
enforce the statement that HEF is dominated by the hard process which does not have the suppression.

 Thus we make another
set of calculations, now requiring $p_{T}>25\,\mathrm{GeV}$ to get
rid of the range in $p_{T}$ which normally would be strongly suppressed.
This is the domain of applicability of both $\mathsf{pythia}$ `hard QCD'  and HEF.
The results are presented in Fig.~\ref{fig:pT_cmp_1}C-D. As for
the $p_{T}$ spectrum, the situation does not change comparing to
the smaller $p_{T}$ cut. 
For the disbalance $K_{T}$ spectrum, we see that at
first the HEF tail drops below $\mathsf{pythia}$ with MPI, but later
it again rises toward the model with MPIs. The IDDT model has a similar
(unphysical) low-$K_{T}$ behavior as the genuine DDT obtained from
(\ref{eq:DDT1}). It is important to stress, that in HEF we have been
using the KMR based on GRV98 gluon distribution so far, and the features
discussed above will be different for different UGDs. Similar, the
shape of the $\mathsf{pythia}$'s enhancement will depend on the MPI
model parameters, in particular on the $p_{T0}$ parameter, as we
will see below. 

In order to better access the power corrections and study their energy
dependence we propose the following observable. We investigate a differential
cross section for inclusive dijets in the following variable:
\begin{equation}
\tau=\frac{K_{T}}{\mu}=\frac{2K_{T}}{p_{T1}+p_{T2}}\,.
\end{equation}
It can be thought as being a measure of the  `twist content' in the approach.
That is, the small $\tau \sim 0$ corresponds to the leading power, while $\tau>1$ region is  sensitive to
higher power corrections. We expect that in HEF we will observe sizeable contribution to $\tau>1$
region. On the other hand, in $\mathsf{pythia}$ generator the small
momentum disbalance is generated by the parton shower, but it does not give significant contribution to $\tau>1$. 
However, we expect that $\tau>1$ could be generated by MPIs, because partons originating in different hard
collisions can be identified with distinctive jets in the jet pair we tag. We shall come back to this later in this section.

We will be concerned with the shape of the $\tau$ distribution only. Since
various UGDs often have different normalizations
we shall divide the differential cross sections by the total cross
section. We shall investigate this observable within the different 
approaches with a known minijet implementation for various $p_{T0}$ settings (we mean $\mathsf{pythia}$
here) and HEF. 

\begin{figure}
\begin{centering}
\parbox{0.49\textwidth}{A)\\\includegraphics[width=0.49\textwidth]{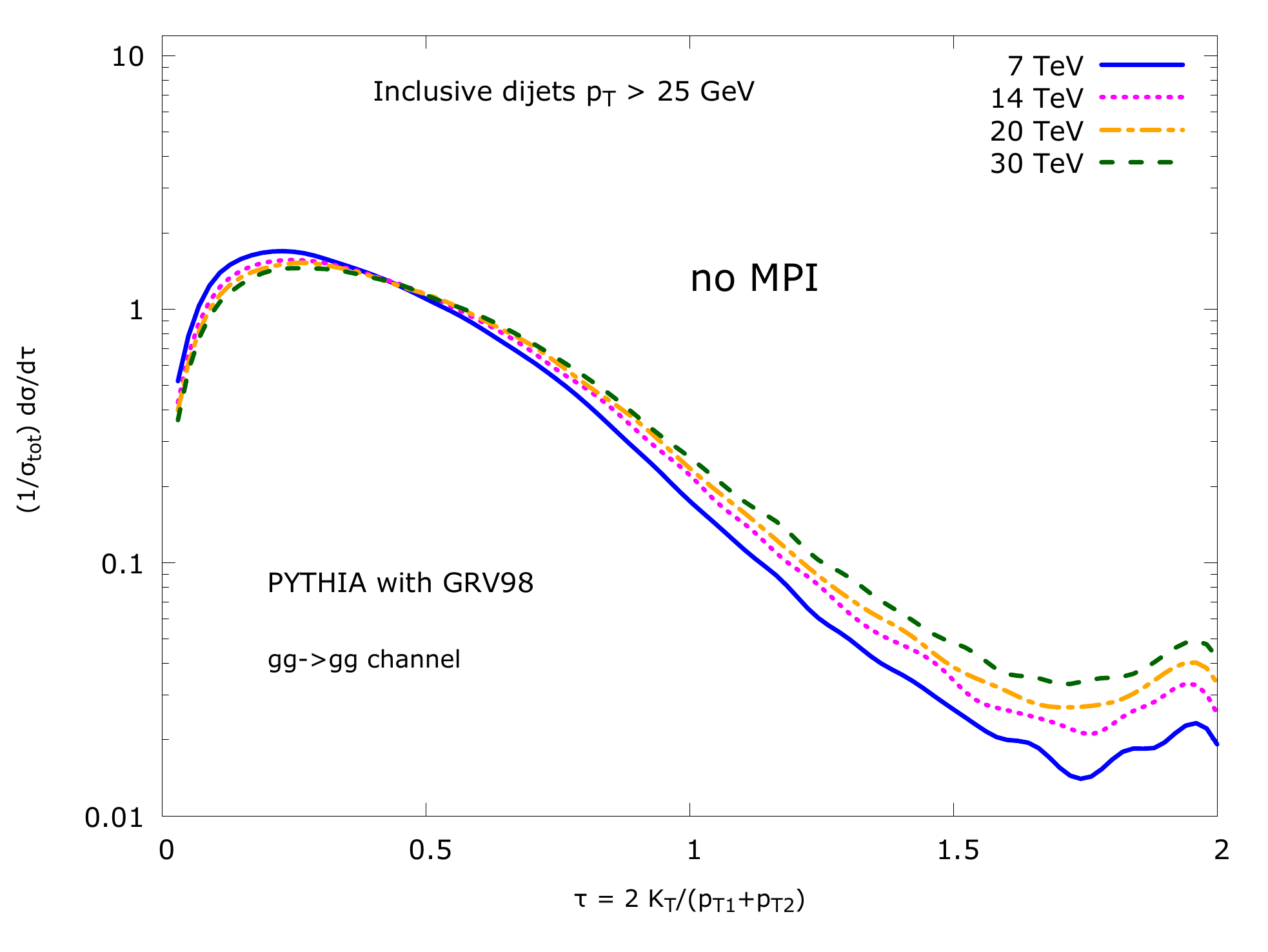}}$\,\,\,\,\,\,\,$\parbox{0.49\textwidth}{B)\\\includegraphics[width=0.49\textwidth]{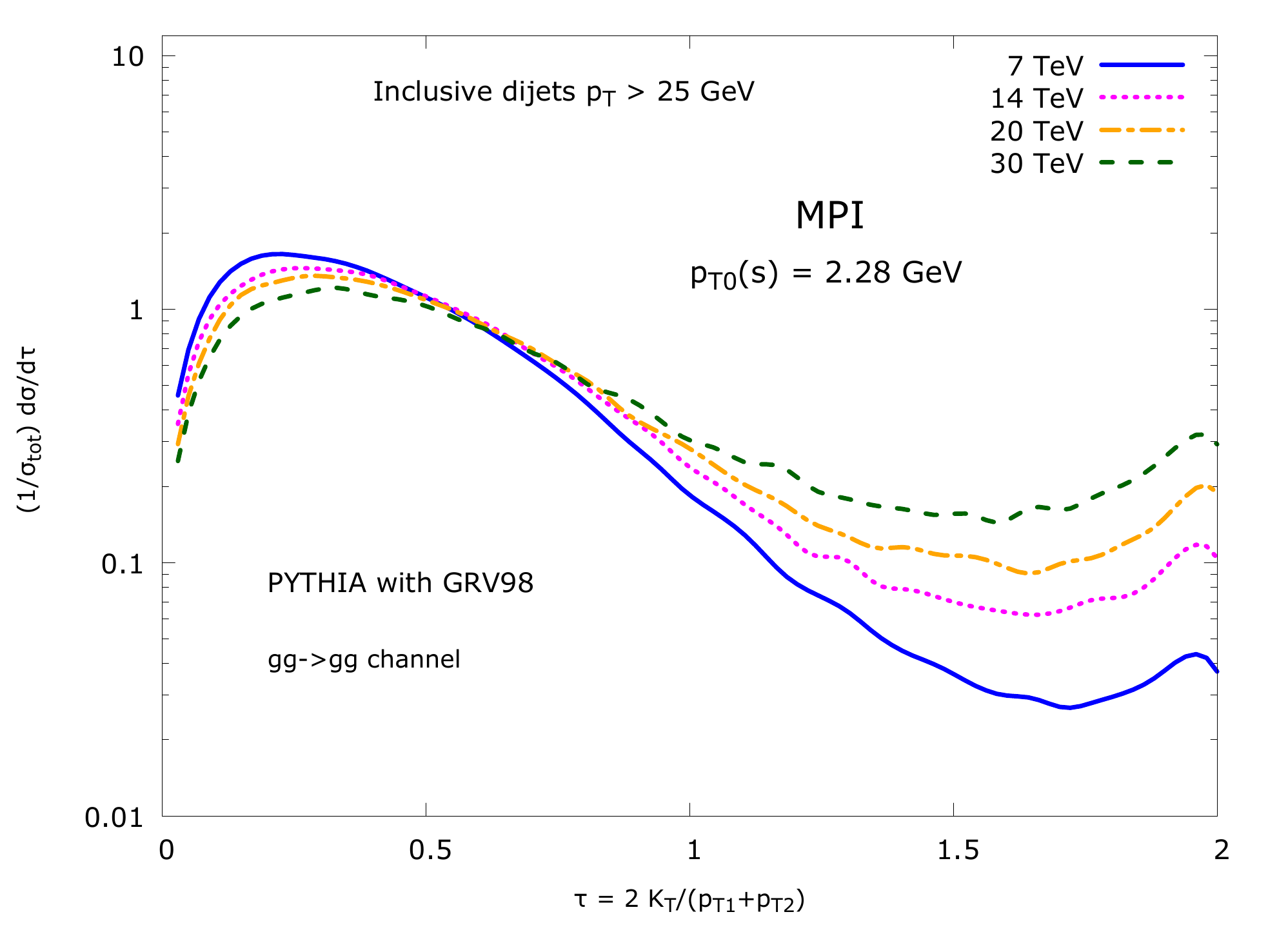}}
\par\end{centering}

\begin{centering}
\parbox{0.49\textwidth}{C)\\\includegraphics[width=0.49\textwidth]{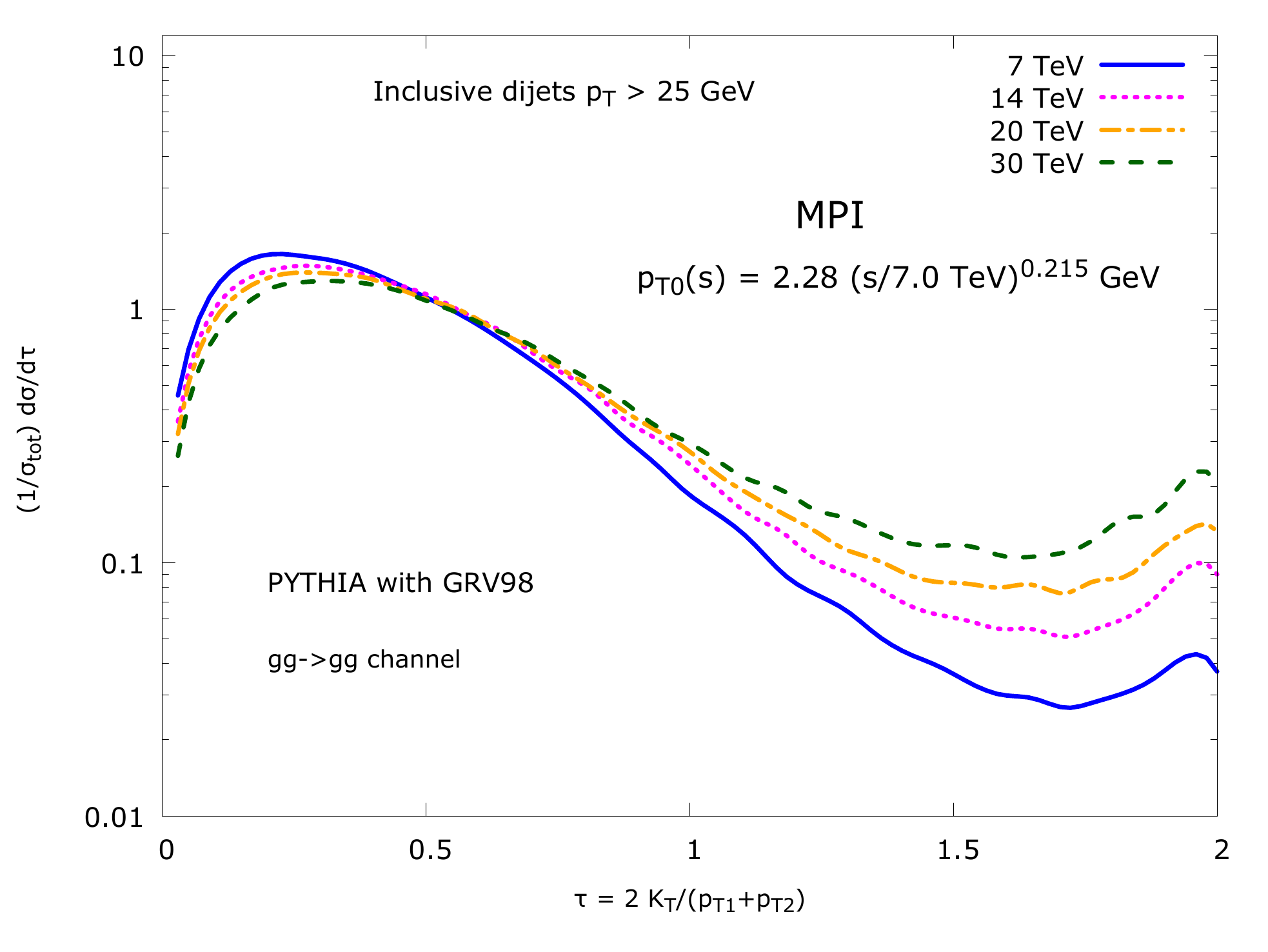}}$\,\,\,\,\,\,\,$\parbox{0.49\textwidth}{D)\\\includegraphics[width=0.49\textwidth]{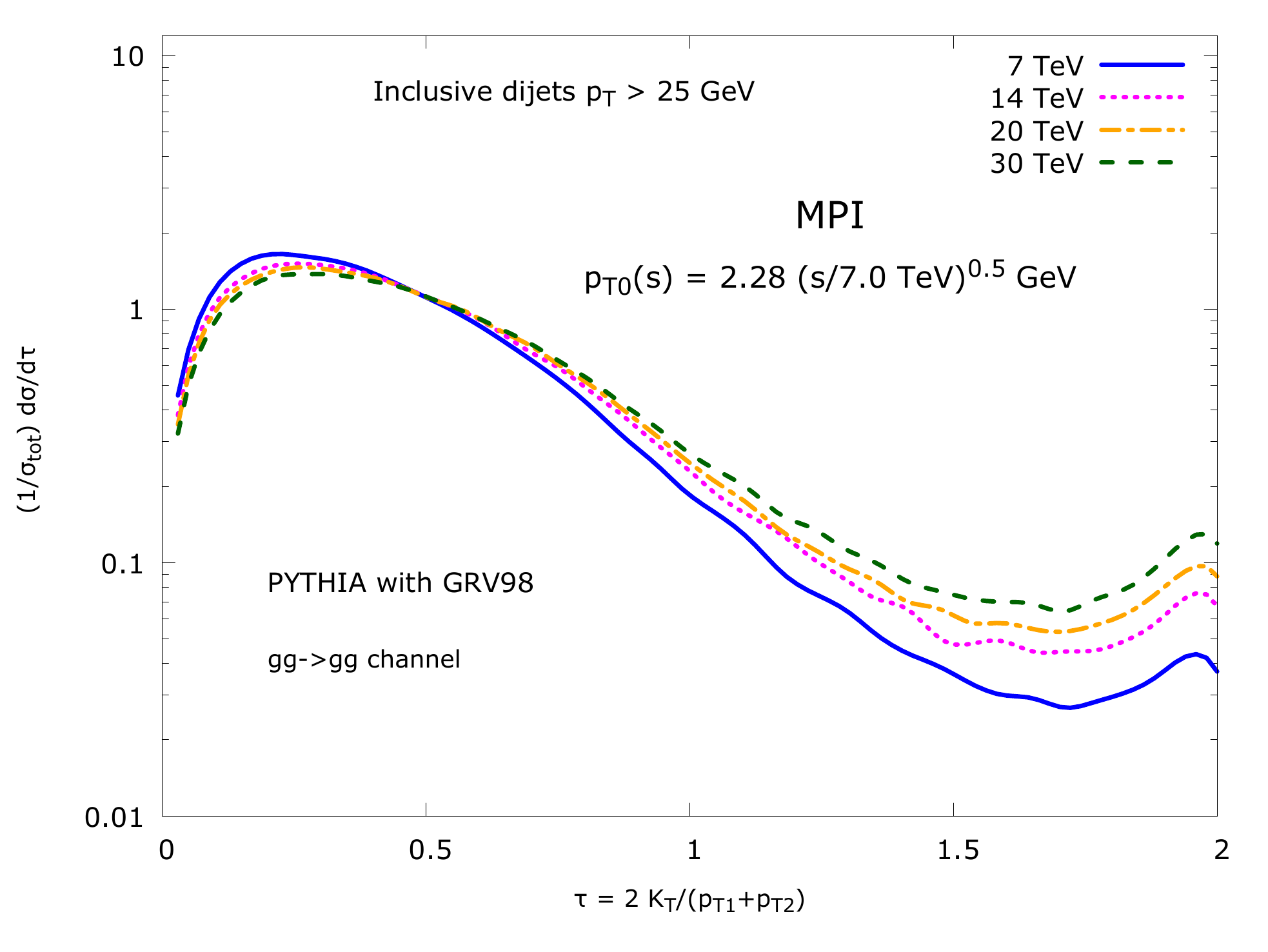}}
\par\end{centering}

\caption{Spectra of the variable $\tau=2K_{T}/\left(p_{T1}+p_{T2}\right)$
in $\mathsf{pythia}$ with parton showers, no hadronization and with
several choices of MPI model parameters. A) MPI is switched off, B)
the $p_{T0}\left(S\right)=\mathrm{const.}$, C) the standard choice
of (\ref{eq:pT0(s)}), D) the choice (\ref{eq:pT0(s)}) but with the
exponent approximately as twice as big. \label{fig:pythia_twist}}
\end{figure}

\begin{figure}
\begin{centering}
\parbox{0.49\textwidth}{A)\\\includegraphics[width=0.49\textwidth]{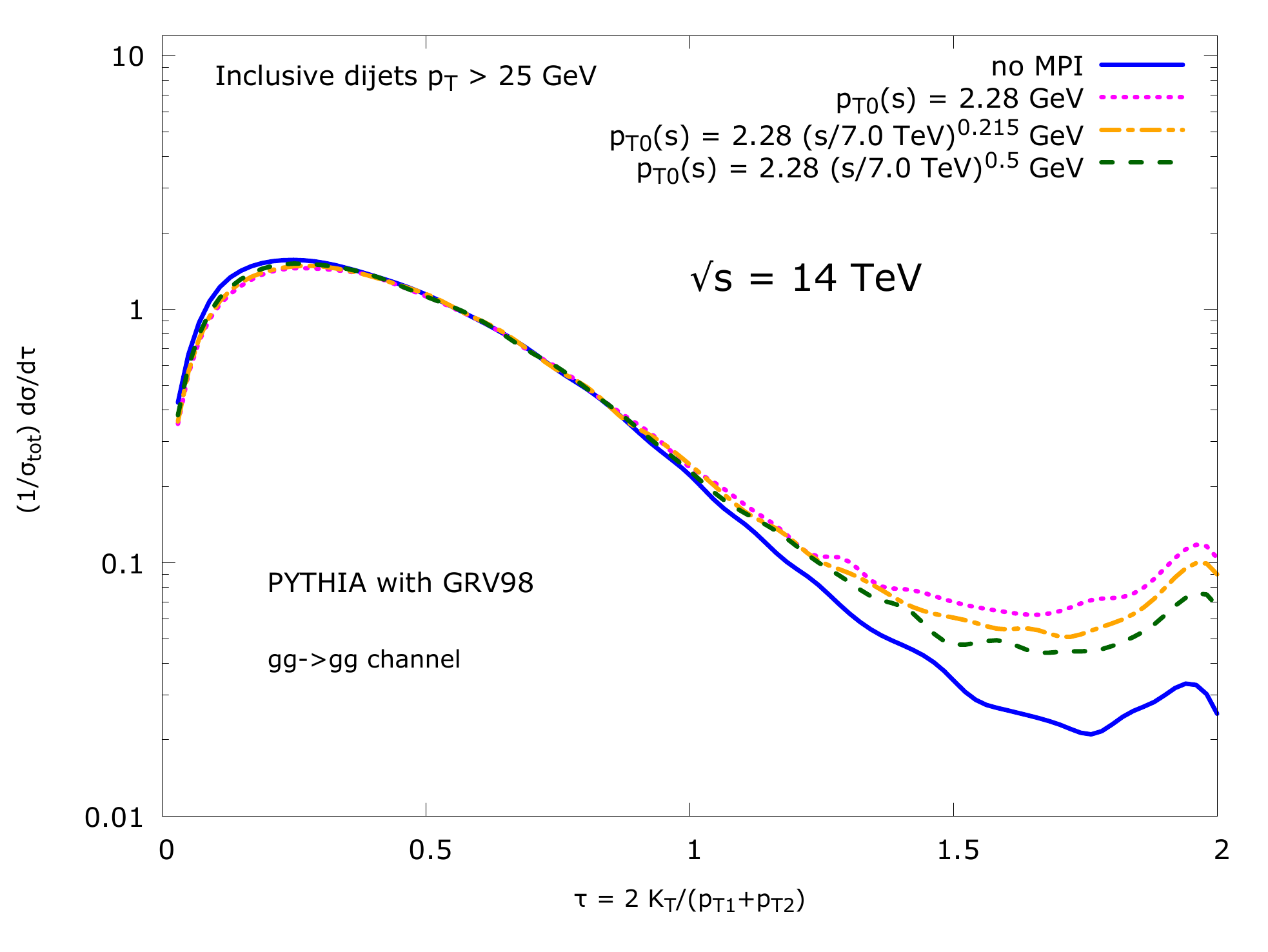}}$\,\,\,\,\,\,\,$\parbox{0.49\textwidth}{B)\\\includegraphics[width=0.49\textwidth]{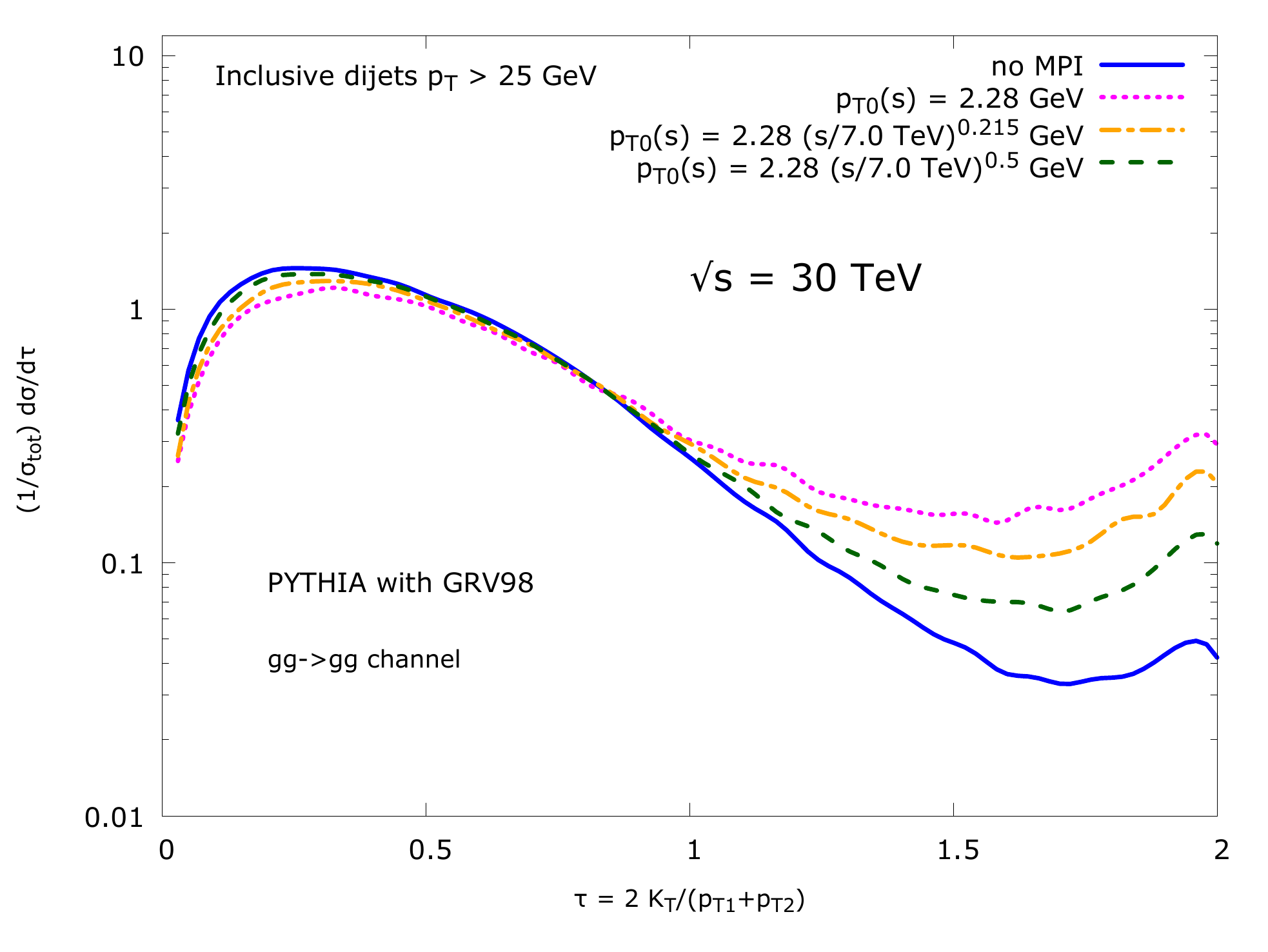}}
\par\end{centering}

\caption{
Similar to Fig.~\ref{fig:pythia_twist} but here we compare calculations with different 
MPI model parameter and fixed CM energy: A) 14 TeV, B) 30 TeV. \label{fig:pythia_twist_energy}}
\end{figure}

\begin{figure}
\begin{centering}
\parbox{0.49\textwidth}{A)\\\includegraphics[width=0.49\textwidth]{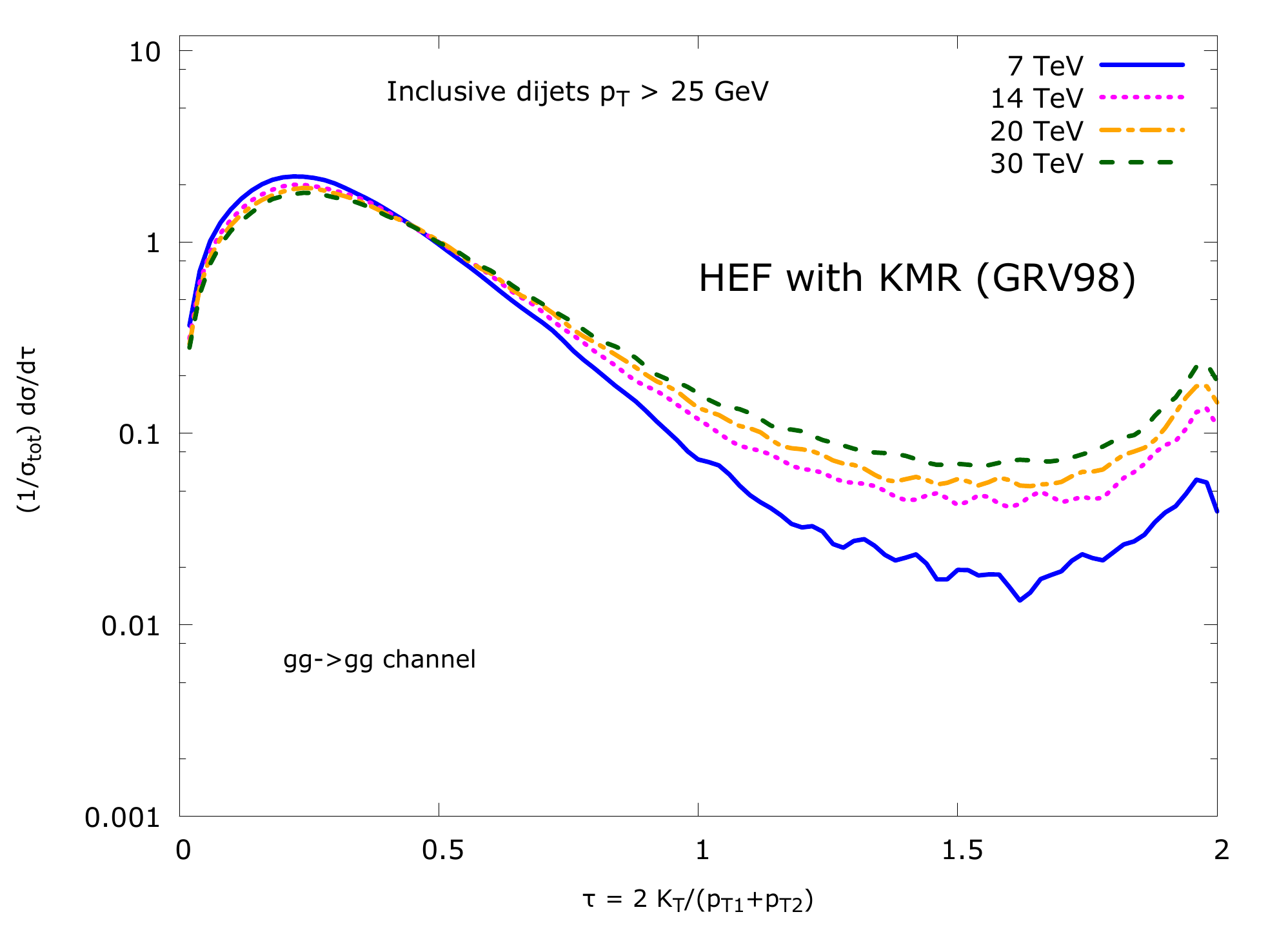}}$\,\,\,\,\,\,\,$\parbox{0.49\textwidth}{B)\\\includegraphics[width=0.49\textwidth]{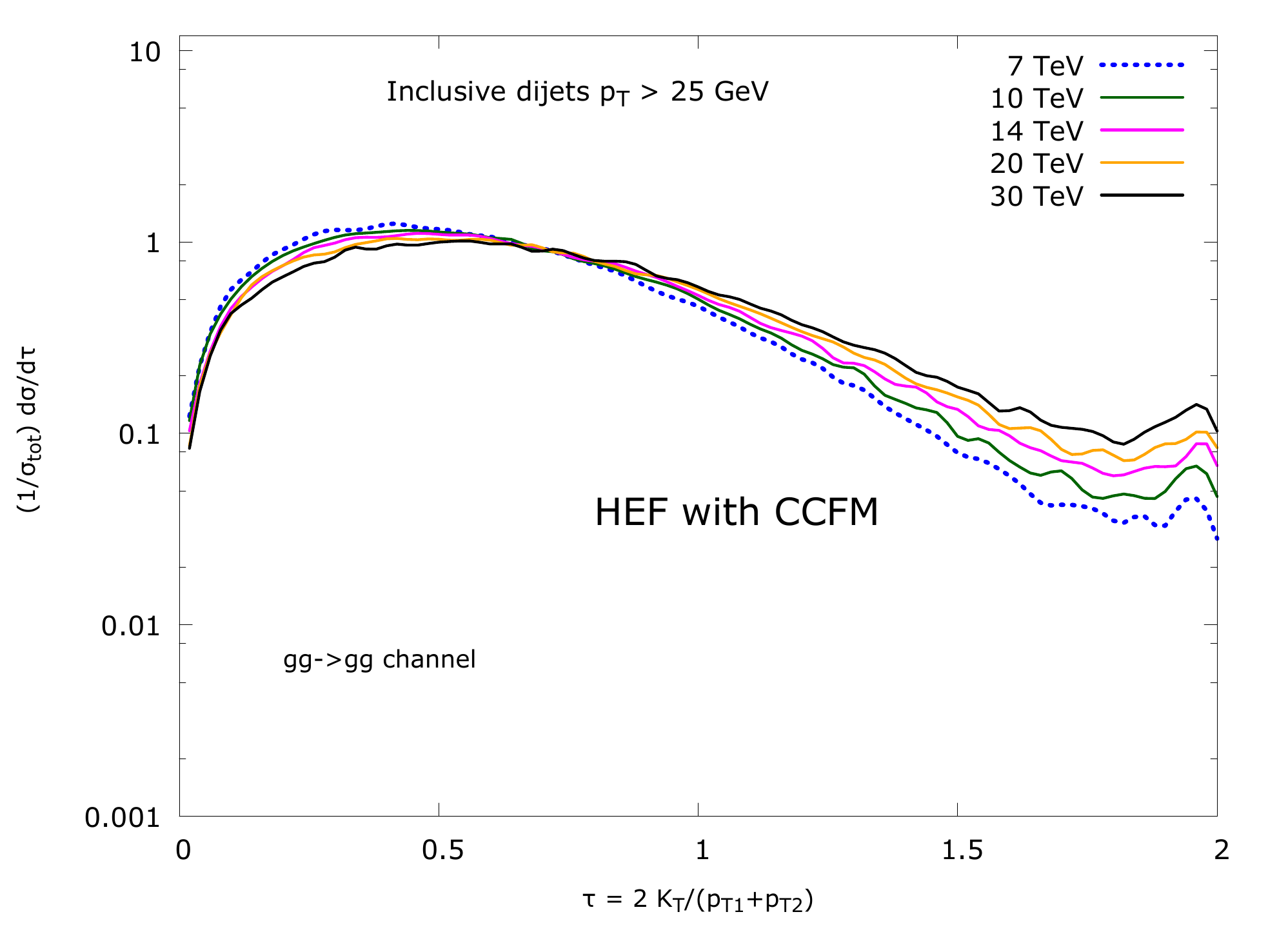}}
\par\end{centering}

\begin{centering}
\parbox{0.49\textwidth}{C)\\\includegraphics[width=0.49\textwidth]{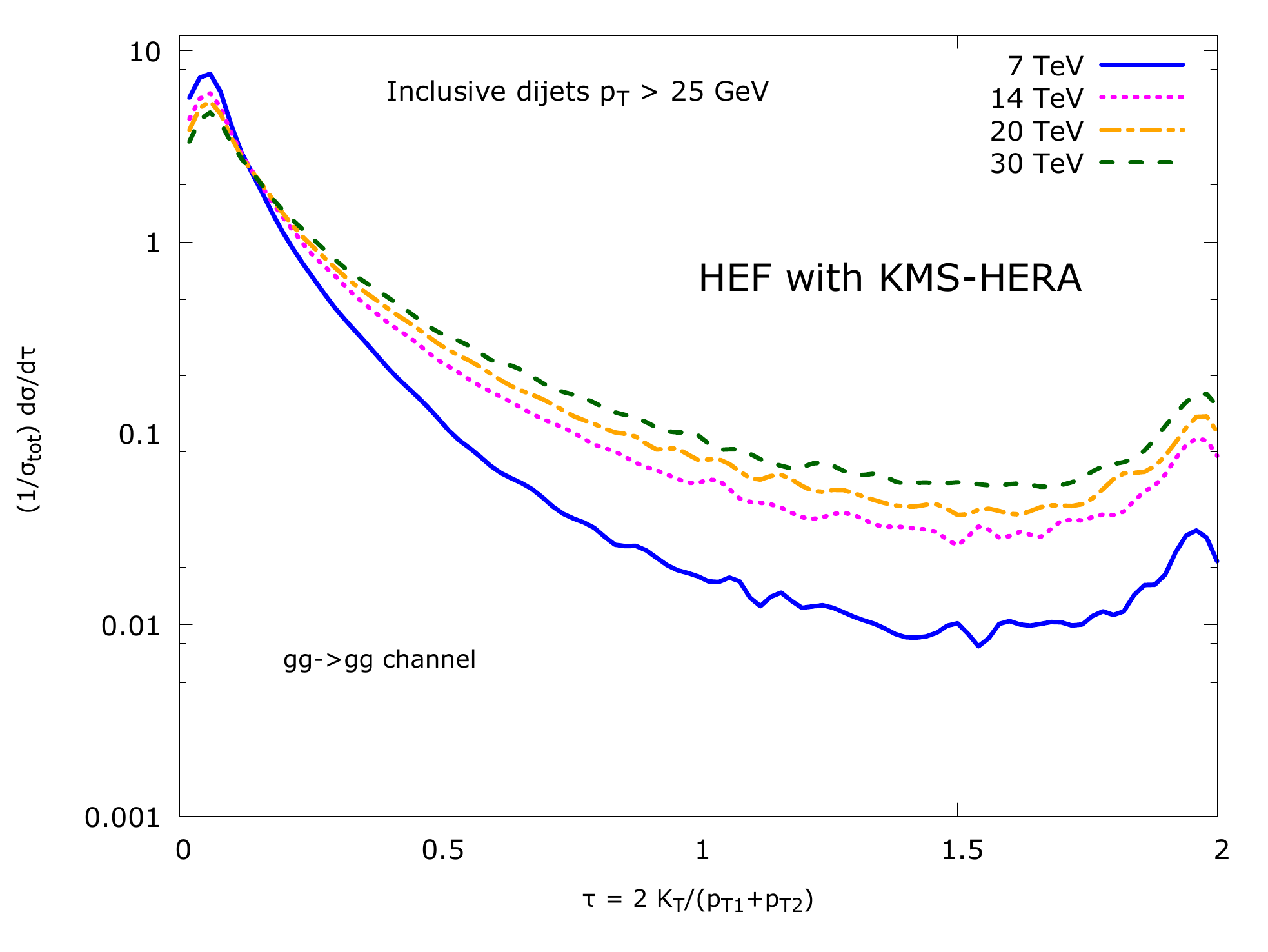}}$\,\,\,\,\,\,\,$\parbox{0.49\textwidth}{D)\\\includegraphics[width=0.49\textwidth]{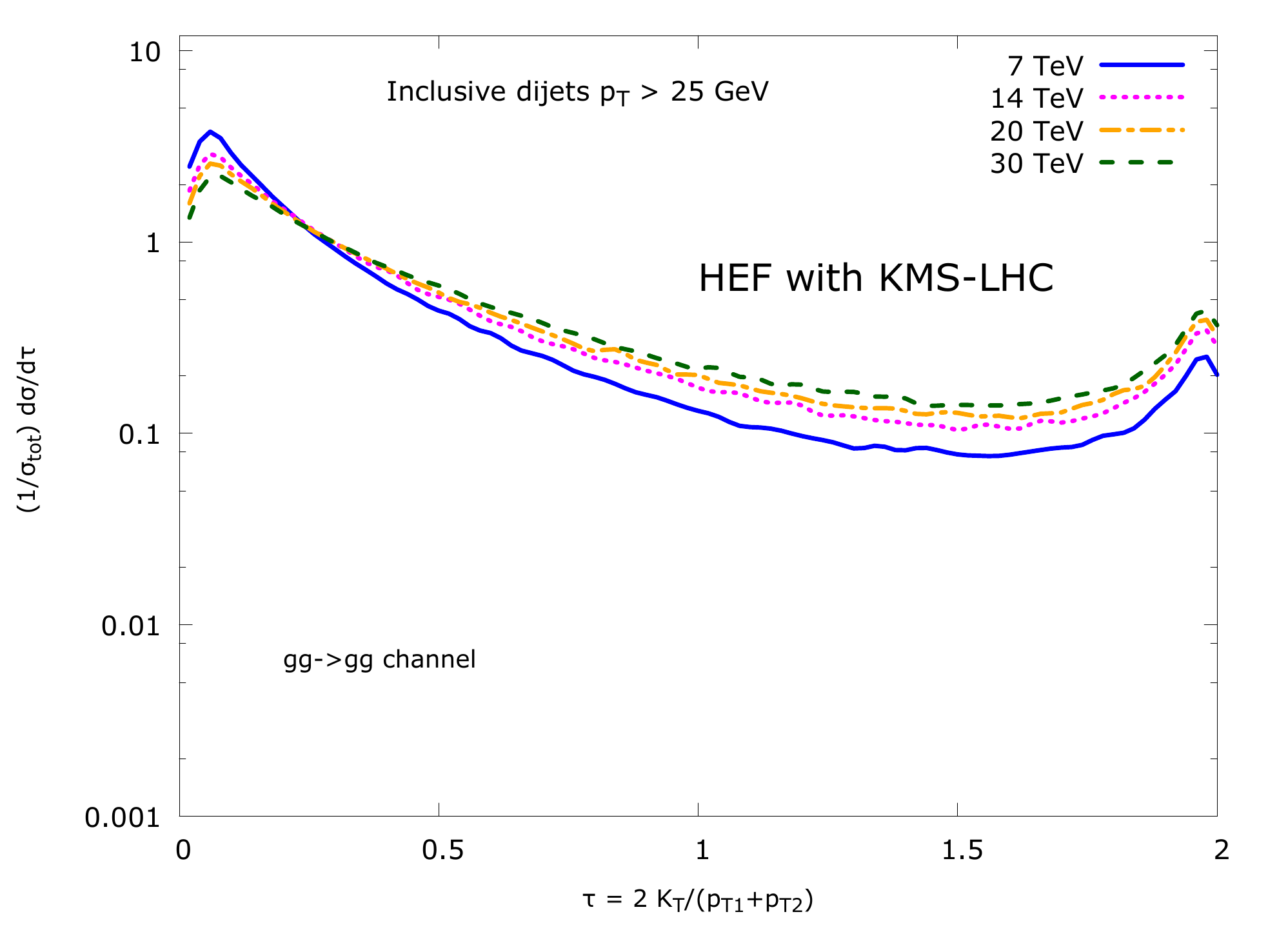}}
\par\end{centering}

\caption{Spectra of the variable $\tau=2K_{T}/\left(p_{T1}+p_{T2}\right)$
in HEF with various UGDs: A) KMR with GRV98, B) the CCFM, C) the KMS-HERA,
D) the KMS-LHC. \label{fig:HEF_twist}}
\end{figure}

In Fig.~\ref{fig:pythia_twist} we show the results for the CM energy
range $7-30\,\mathrm{TeV}$ calculated in $\mathsf{pythia}$ for a
few choices of the parameters in the parametrization of $p_{T0}\left(S\right)$.
Namely, we consider the following scenarios: (A) no MPI interactions,
(B) constant $p_{T0}=2.28\,\mathrm{GeV}$, (C) the standard implementation
given by (\ref{eq:pT0(s)}), (D) the choice (\ref{eq:pT0(s)}) with
 the exponent taken to be around twice as big. In Fig.~\ref{fig:pythia_twist_energy} we compare
 these scenarios for two fixed energies $14\,\mathrm{TeV}$ and $30\,\mathrm{TeV}$.
 In a similar manner we calculate
the spectra in $\tau$ using HEF in Fig.~\ref{fig:HEF_twist}. We
use the following UGDs described in Section~\ref{sub:setup}: (A)
KMR based on GRV98, (B) the CCFM, (C) the KMS-HERA and (D) the KMS-LHC.

\begin{figure}
\begin{centering}
\parbox{0.49\textwidth}{A)\\\includegraphics[width=0.49\textwidth]{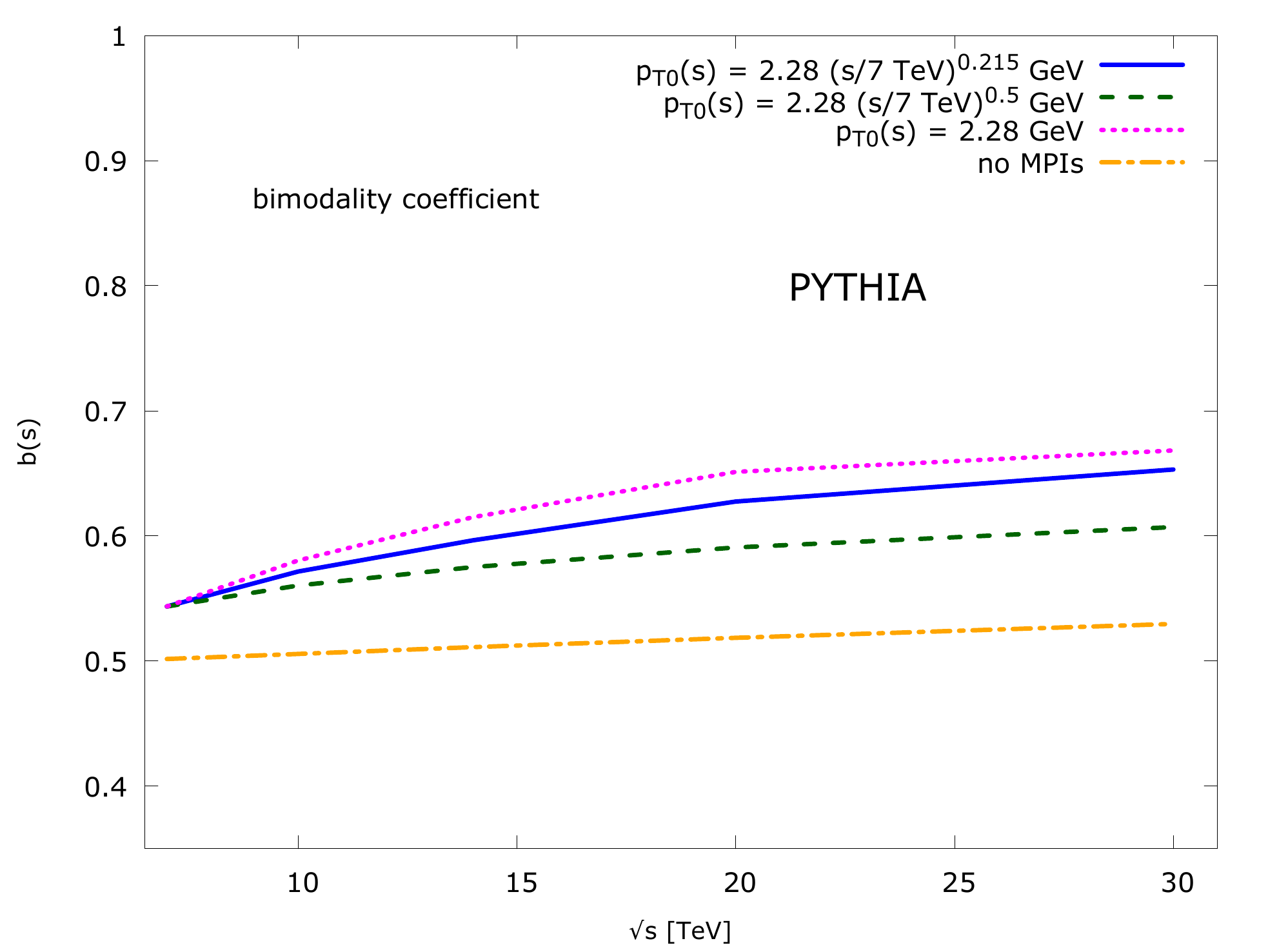}}$\,\,\,\,\,\,\,$\parbox{0.49\textwidth}{B)\\\includegraphics[width=0.49\textwidth]{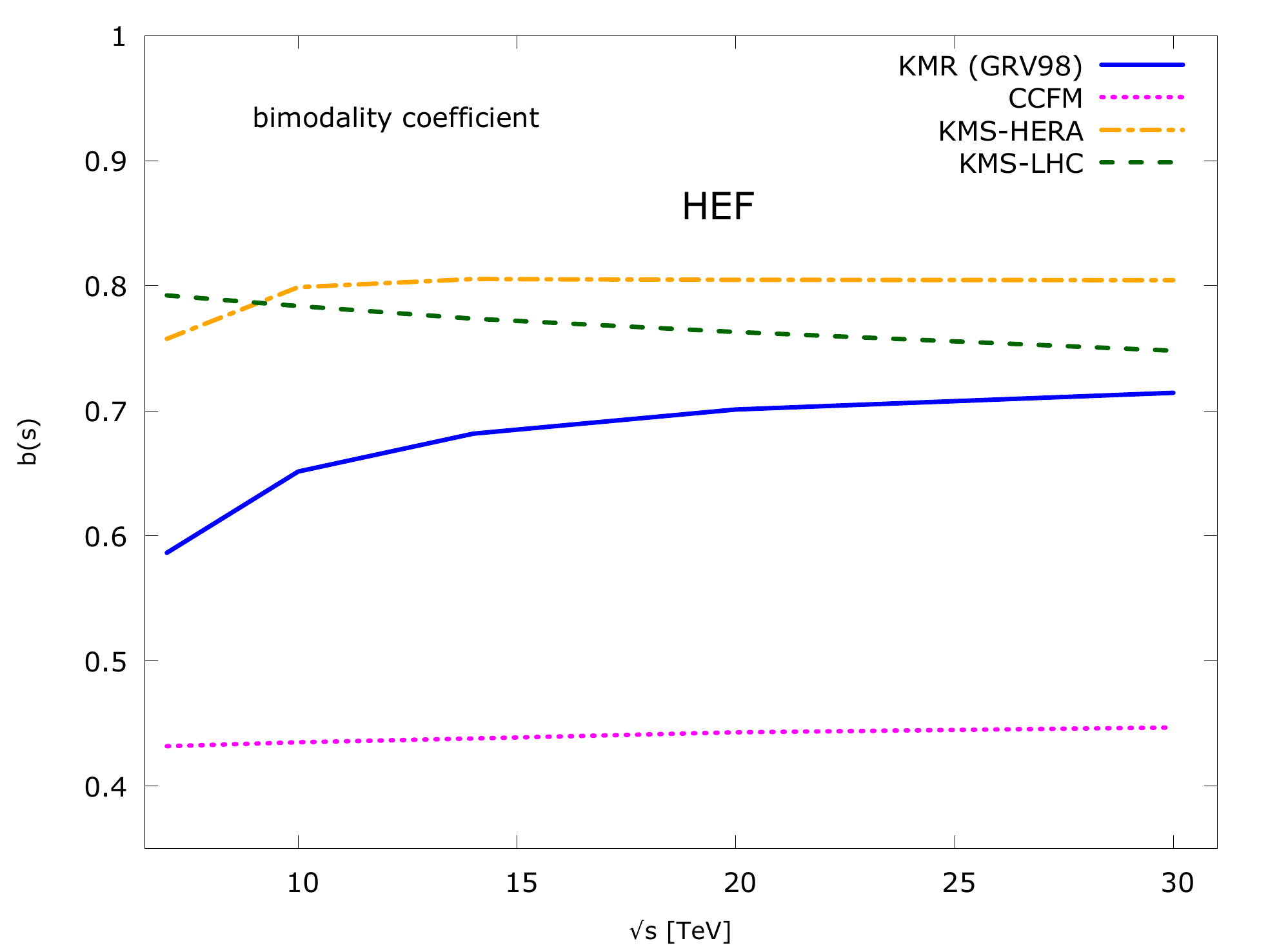}}
\par\end{centering}

\caption{The bimodality coefficient defined in (\ref{eq:bimodality_coeff})
for various approaches as a function of energy. A) $\mathsf{pythia}$
with different minijet suppression , B) HEF with various UGDs. \label{fig:bimodality_1}}
\end{figure}

Let us discuss first the spectra obtained from $\mathsf{pythia}$.
We see, that the distributions have a  \textit{bimodal} character,
i.e. they have two peaks, one close to $\tau=0$ and the second close
to $\tau=2$. The second peak (at large $\tau$) is much weaker than the leading peak
and its strength depends on the amount of MPIs present in the model:
the more MPIs the stronger the second peak. This is seen when comparing
the plots without suppression of minijets (Fig.~\ref{fig:pythia_twist}B),
throughout the increasing suppression (Figs.~\ref{fig:pythia_twist}C-D),
up to the `infinite' suppression (i.e. no MPIs, Fig.~\ref{fig:pythia_twist}A).
This is even more visible from Fig.~\ref{fig:pythia_twist_energy} where we 
compare these models for fixed energies.
In order to investigate the energy dependence of the second peak we
shall use the bimodality coefficient defined as
\begin{equation}
b=\frac{\gamma^{2}+1}{\kappa}\,,\label{eq:bimodality_coeff}
\end{equation}
where the skewness $\gamma$ and the kurtosis $\kappa$ are defined
as 
\begin{equation}
\gamma=\frac{\mu_{3}}{\sigma^{3}}\,,\,\,\,\,\kappa=\frac{\mu_{4}}{\sigma^{4}}\,,
\end{equation}
with $\mu_{n}$ being the $n$-th central moment and $\sigma$ the
standard deviation. We calculate $b$ as a function of energy in Fig.~\ref{fig:bimodality_1}A.
We see that the bimodality coefficient reflects (to some extent) the
energy dependence of the minijets contribution. When there are no
MPIs the coefficient is perfectly linear with energy. When we switch
on MPIs $b$ jumps to a higher value and then the increase is dictated
by the amount of suppression of minijets. Thus we may conclude, that
the bimodality coefficient of the normalized spectra in $\tau$ is
a reasonable measure of the minijet contribution as a function of
energy \textit{when a hard process is present}. 

\begin{figure}
\begin{centering}
\includegraphics[width=0.6\textwidth]{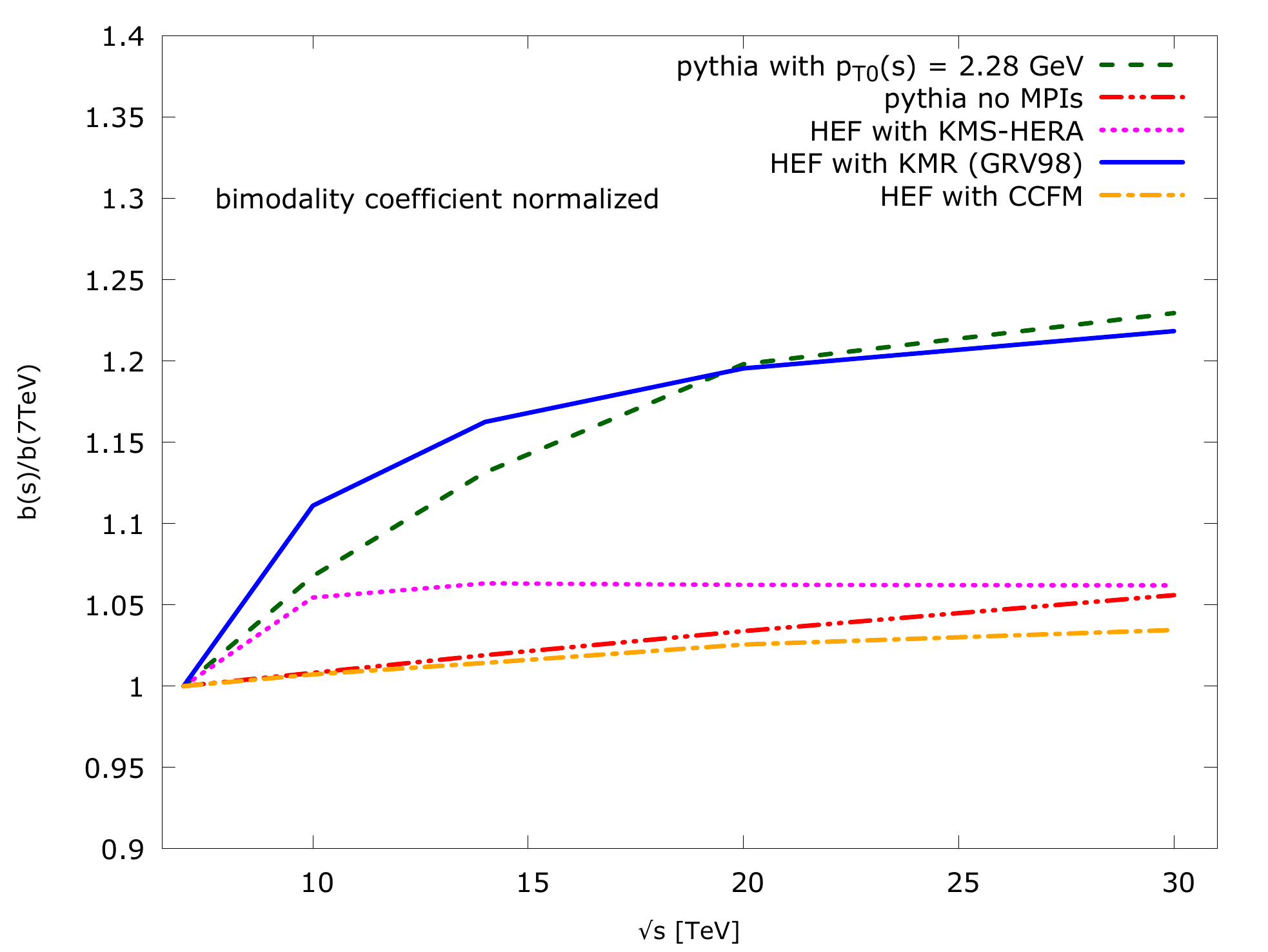}
\par\end{centering}

\caption{Comparision of energy dependence of the bimodality coefficient normalized
to the value at $7\,\mathrm{TeV}$ for some of the curves from Fig.~\ref{fig:bimodality_1}A-B.
\label{fig:bimodality_2}}
\end{figure}

Let us now turn to HEF and make a similar analysis. First we calculate 
the spectra in $\tau$ shown in Fig.~\ref{fig:HEF_twist}. We see
that KMR with GRV98 (Fig.~\ref{fig:HEF_twist}A) produce superficially similar
spectra to those of $\mathsf{pythia}$ with MPIs. The
CCFM (Fig.~\ref{fig:HEF_twist}B) looks more flat with the second peak only very slowly varying with energy. The KMS distribution which does not have the
hard scale evolution (i.e. the Sudakov resummation) produce very different
shapes. They are much more peaked near $\tau=0$ (Figs.~\ref{fig:HEF_twist}C-D).
Spectra for both versions of KMS also differ considerably with respect
to the second peak. 
To compare the energy evolution of the second peak 
 let us now extract the bimodality coefficient
from these spectra. The result is presented in Fig.~\ref{fig:bimodality_1}B.
First, we see that the normalizations vary significantly  for different
models. This is because the normalization is sensitive to the first
peak, which is different across  models. Second, looking at
the energy dependence, we see that the KMR with GRV98 has a similar (but not the same)
tendency to $\mathsf{pythia}$ with MPIs and the suppression parameter
$p_{T0}\left(S\right)=\mathrm{const}$. 
The rise with energy is slightly slower, but not as slow as the model (\ref{eq:pT0(s)}).
It is better
seen in Fig.~\ref{fig:bimodality_2} where we collect the $\mathsf{pythia}$
results with only extreme settings for clarity and some of the HEF
results. In this plot we normalize the bimodality coefficient by its
value at $7\,\mathrm{TeV}$ to compare the energy dependence. 
The conclusion from this plot is as follows. The HEF can render power corrections which
definitely can show similar energy evolution 
to the one from  MPIs in the event generator (due to the evolution of $p_T(S)$).
Here it is satisfied by KMR and KMS-HERA UGDs.
It seems however that the energy dependence they give flattens earlier than $\mathsf{pythia}$ minijet models.
The initial rise is also more rapid.

There is a comment in order. The bimodality coefficient from
$\mathsf{pythia}$ models depends on the jet radius $R$ as does the calculated $\tau$ distribution for large $\tau$.
 It is clear that by decreasing the jet radius we will reconstruct more jets
which will eventually start to balance each other. This sensitivity of large  $\tau$ to $R$ 
is a natural feature. 
In this regime the dijets are accompanied by large `underlying event' activity.
From LHC data  \citep{Aad2012} it is known 
that the underlying event observables are sensitive to $R$.
The sensitivity mechanism of $\tau$ distribution on $R$ will become more clear from the discussion below.

\begin{figure}
\begin{centering}
$\!\!\!\!\!$\parbox{0.5\textwidth}{A)\\\includegraphics[width=0.6\textwidth]{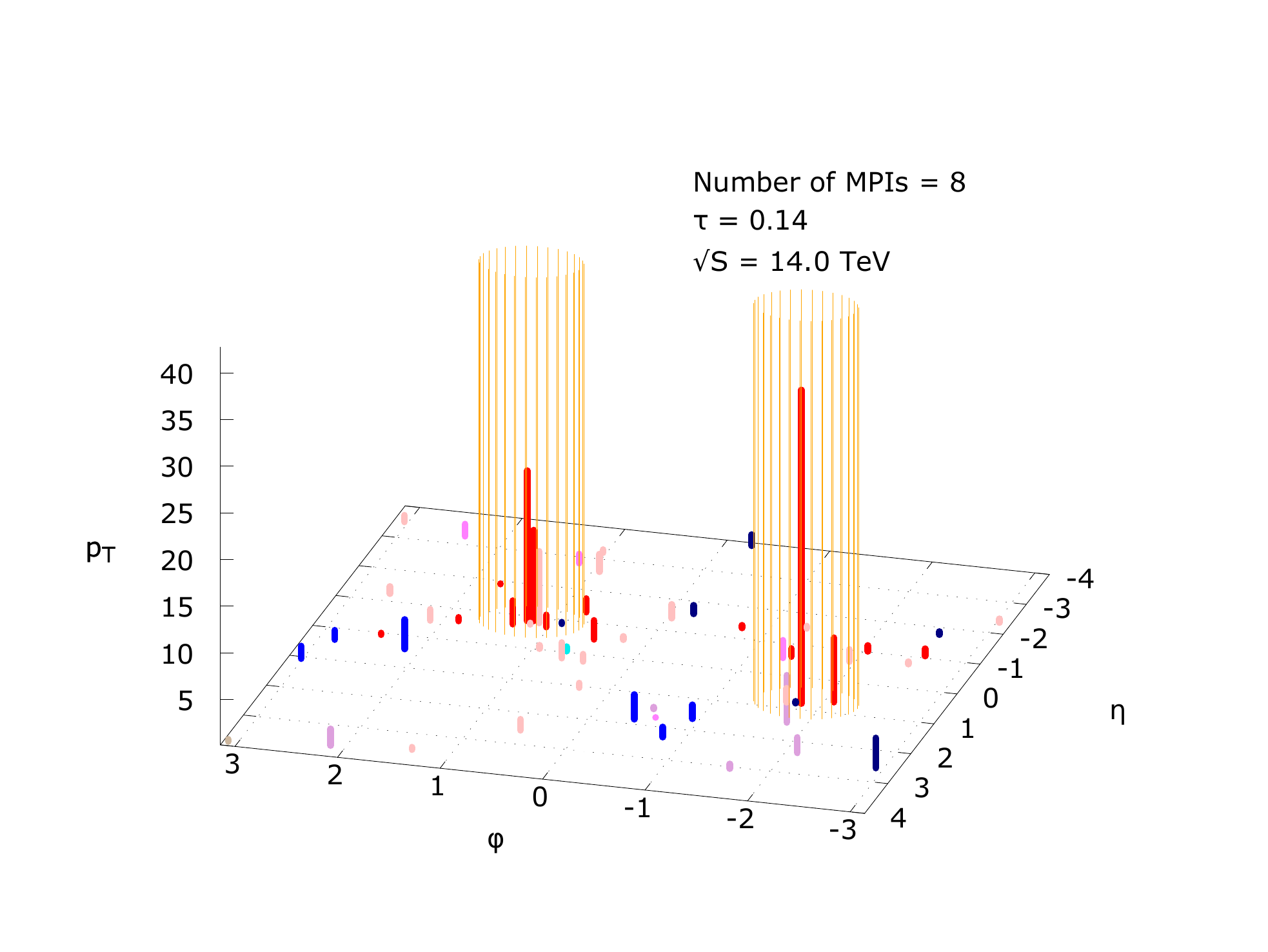}}$\,\,$\parbox{0.5\textwidth}{B)\\\includegraphics[width=0.6\textwidth]{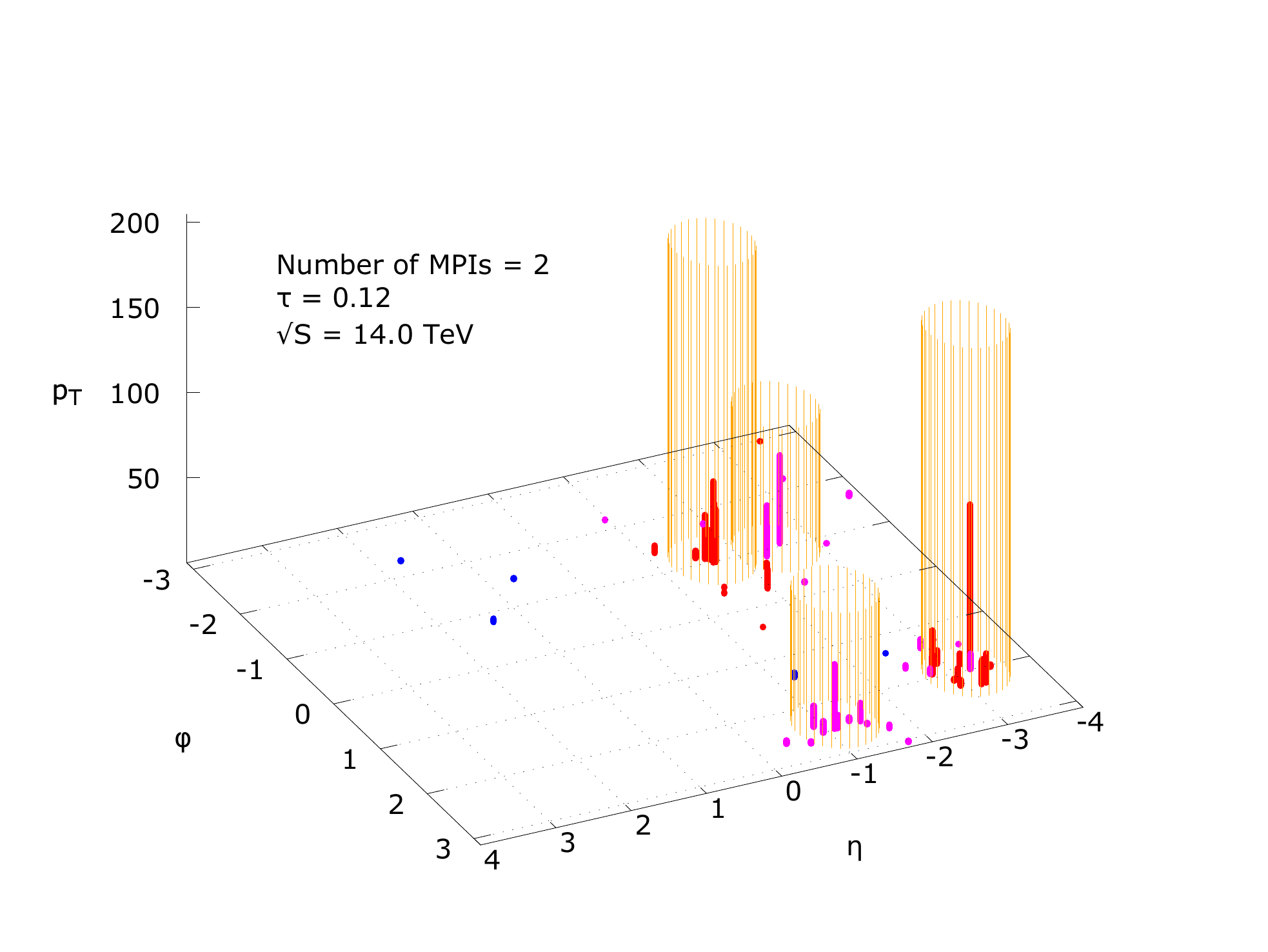}}
\par\end{centering}

\begin{centering}
$\!\!\!\!$\parbox{0.5\textwidth}{C)\\\includegraphics[width=0.6\textwidth]{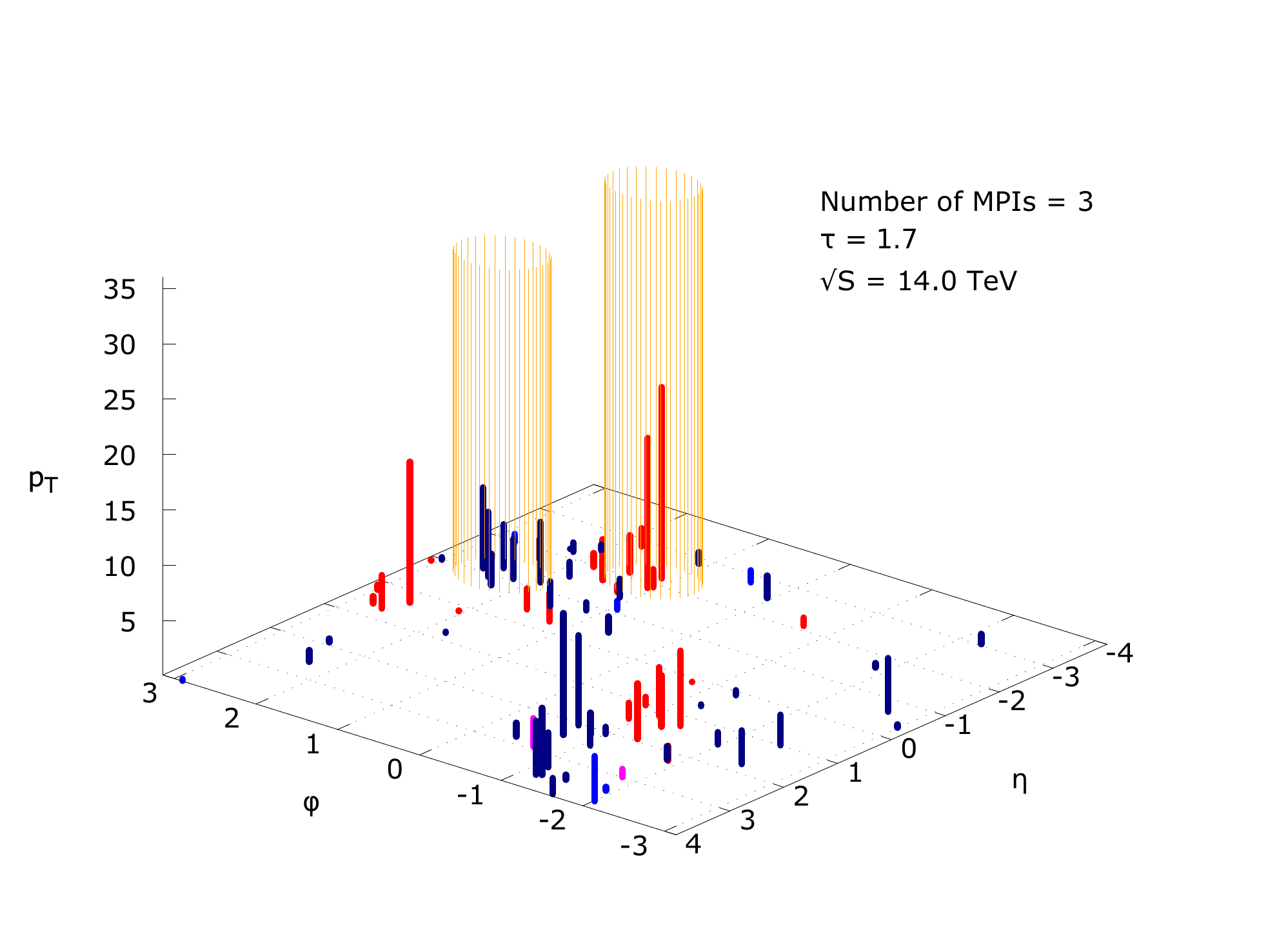}}$\,\,$\parbox{0.5\textwidth}{D)\\\includegraphics[width=0.6\textwidth]{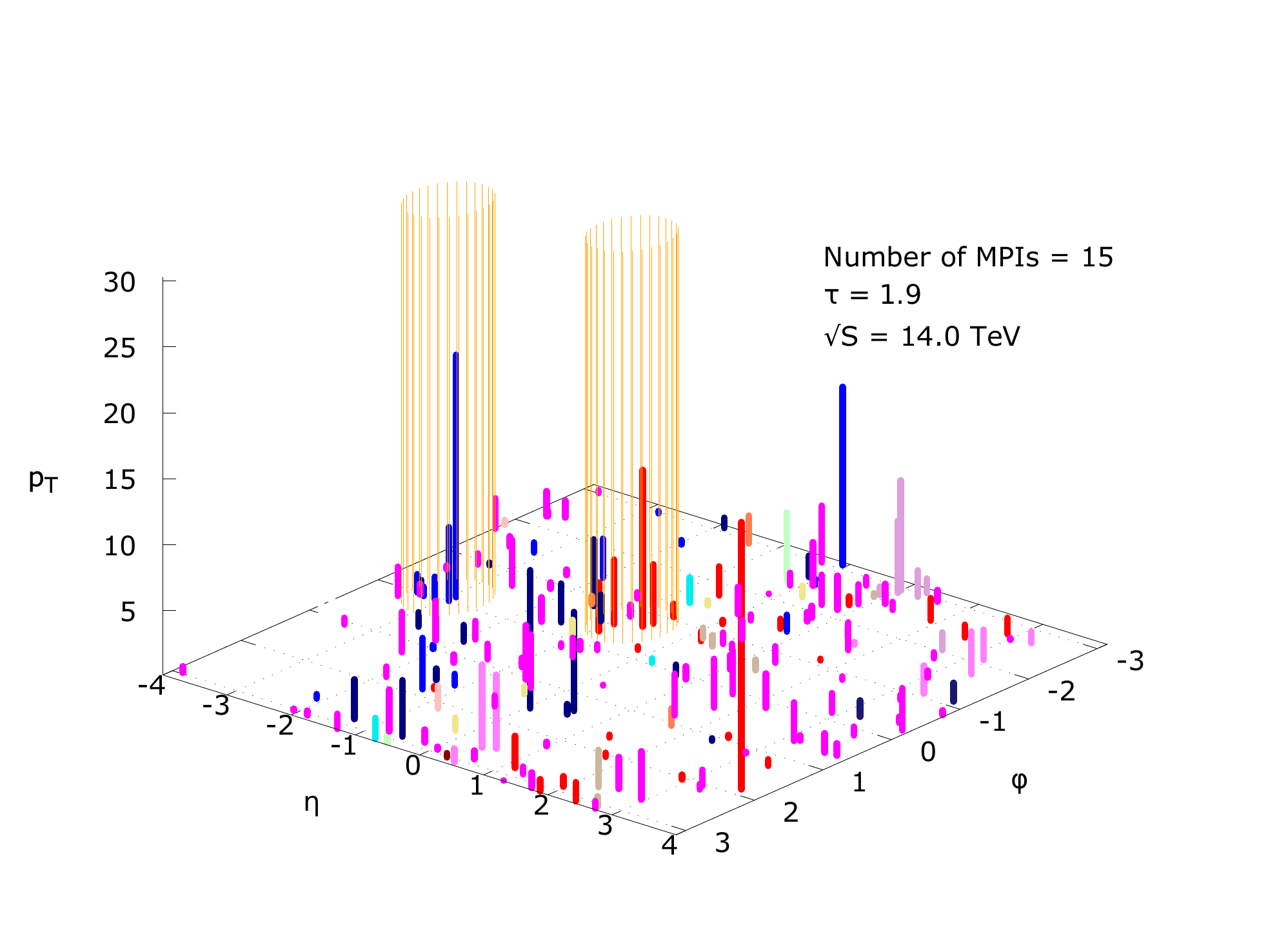}}
\par\end{centering}

\caption{Events in $\left(\phi,\eta,p_{T}\right)$ space from $\mathsf{pythia}$
where the final state particles and their MPI origin are shown. Particles
originating in various hard processeses (above certain $p_T$ threshold) are denoted using different
colors. The reconstructed leading jets are symbolically represented
by  cylinders with radius 0.5. The beam remnants are also displayed. The top row (A-B) represents events with
small relative momentum disbalance $\tau$ (relative to the hard scale).
The bottom row (C-D) represent events where the relative disbalance
$\tau$ is large. We see that the leading jets originate from different
subprocesses here. \label{fig:events}}
\end{figure}

To see more directly how the large $K_{T}$ disbalance
can be created due  to MPIs, we display in Fig.~\ref{fig:events} several events in $\left(\phi,\eta,p_{T}\right)$  space obtained from $\mathsf{pythia}$ with
MPI and parton showers. We have traced the origin of the final state
particles that later form the jets; particles originating in different
hard collision are denoted using different colors. The resulting jets
are displayed as cones with radius $R=0.5$. The two top plots present
two events with small disbalance relative to the hard scale (small $\tau$). We
see that the jets are reconstructed from the particles originating
in the same hard collision. The bottom plots show two events which
contribute to large disbalance to hard scale ratio $\tau$, thus to
power corrections. We see that the leading jets are reconstructed
from final state partons originating from different hard collisions.
Moreover, we see that a mixture of partons from different hard collisions
may enter a jet.

Let us now make some comments on the energy dependence of the power corrections from UGDs other then KMR, that is undergoing the BFKL evolution and its extensions. From Fig.~\ref{fig:bimodality_2} it seems that they give much weaker energy dependence than Sudakov-based KMR approach. In particular it is surprising for the  CCFM gluons.  Let us stress however, that different CCFM sets differ considerably
depending on the particular implementation
(we are not concerned in comparing various CCFM distributions in
this paper). Moreover, all BFKL-based gluons have been fitted to experimental data on the structure functions in Deep Inelastic Scattering. 
This process is dominated by small $k_T$ and thus
 the large $k_T$ tails of these distributions are burdened with rather sizeable errors. 
It is thus important to keep in mind these restrictions.

%%%%%%%%%%%%%%%%%%%%%%%%%%%%%%%%%%%%%%%%%%%%
\section{Summary and conclusions}
\label{sec:Conclusions}

Our work can be summarized as follows. We have performed a comprehensive
analysis of the  minijet cross section and its crucial component -- the $p_{T}$
cutoff. Despite the fact that in event generators minijets are described by the collinear formula, the
kinematic domain ventures out of the leading power approximation.
Therefore we attempted to explain the cutoff using various forms of
$k_{T}$-factorization for inclusive dijet production: the High Energy
Factorization (HEF) with two off-shell gluons in the initial state and an
extension of the Dokshitzer-Dyakonov-Troyan formula beyond the leading
power (IDDT). Both approaches involve unintegrated gluon distributions (UGD) which inject nonzero transverse
momentum into the hard process and thus there is a potential mechanism for a dynamical cutoff on small $p_{T}$.

 We have performed two analyses: (i)  direct calculations
of $p_{T}$ spectra in the low-$p_{T}$ region ($p_{T}>2\,\mathrm{GeV}$)
to see if the cutoff is generated, (ii) calculations for relatively
hard inclusive dijets with $p_{T}>25\,\mathrm{GeV}$ and analysis
of subleading effects in search for a patterns of minijets.

As far as the direct study (i) is concerned we find that the suppression which is generated is small and has a `kinematic' origin
and thus the opposite energy dependence than in the MC models. It is in fact something one should expect as the leading contribution to the
cross section in HEF/IDDT with $2\rightarrow 2$ hard process comes from the very small internal transverse momenta. 
The results are similar to the ones obtained from $\mathsf{pythia}$ when the `hard QCD' block is used.

In the study (ii) we use differential cross section in a variable $\tau$ defined to be the ratio of dijet disbalance to the average $p_T$. We observe, that when it is calculated in  $\mathsf{pythia}$ with MPIs it has a bimodal character with one bump located close to $\tau\sim 0$ and the second bump (much smaller) located close to $\tau\sim 2$. The second bump is sensitive to the $p_{T}$ cutoff in MPI model. The same observable calculated within HEF reveals similar feature. We investigated the energy dependence of the bimodality coefficient $b$ which characterizes the relative magnitude of the two peaks. We find that the energy dependence of $b$ calculated from   $\mathsf{pythia}$ resembles the energy evolution of the  $p_T$ cutoff in MPI model.
Thus by studying $b$ in HEF we could obtain information about minijets constituting emissions which lead to the dijet disbalance.
We found that the UGD constructed according to the prescription  which uses collinear gluon PDF and Sudakov form factor (proposed by Kimber, Martin and Ryskin and before that indirectly by Diakonov, Dokshitzer and Troyan) produces minijets which are only slightly suppressed with CM energy (for large energies). The UGDs with explicit BFKL kernel present give stronger suppression. None of the models recovers exactly the minijet suppression from the 
$\mathsf{pythia}$ event generator.

We note that while it is practically impossible to measure the minijets directly, where by `directly' we mean a measurement of  $p_T$ spectrum around $p_T\sim 2\, \mathrm{GeV}$ with reconstructed jets (although charged particle jets could be possible), our study (ii) is feasible with the current detectors operating at the LHC. 
This could supplement the measurements of underlying event (e.g. \citep{CMSCollaboration2015}) as a main source of restricting MPI model parameters \citep{Seymour2013}.

In the present work we did not discuss the gluon saturation \citep{Gribov1983} issues. It is clear that at some point for very high energies the nonlinear effects  in the gluon density should come into play, especially at small $p_T$. In a naive study, where one would just use UGD with nonlinear evolution of the Balitsky-Kovchegov type \citep{Balitsky1996,Kovchegov:1999yj} the situation would not change significantly, unless large saturation scale $Q_s\sim p_T$ is used. The point is however, that the HEF is not correct in the saturation domain and more complicated approach involving several UGDs is needed \citep{Kotko:2015ura,VanHameren2016}. Whether such improved factorization can generate a sizeable cutoff with the right energy dependence is still open, especially since the problem with the cutoff persists for large impact parameters where gluon densities are not too large \citep{Rogers2008,Rogers2010}.

\section*{Acknowledgements}

We would like to thank L.~Frankfurt, Y.~Dokshitzer and A.~Moraes
for useful discussions. The work was supported by the Department of
Energy Grants No. DE-SC-0002145, DE-FG02-93ER40771 and by the National Science Center, Poland, Grant No. 2015/17/B/ST2/01838, 

\bibliographystyle{JHEP}
\bibliography{library}

\end{document}